\documentclass{article}

\usepackage{calrsfs}
\usepackage{graphicx}
\usepackage{enumerate}
\usepackage[dvipsnames]{xcolor}
\usepackage{color}
\usepackage{url}
\usepackage{hyperref}
\usepackage{bm}
\usepackage{bbm}
\usepackage{float} 
\expandafter\let\csname endequation*\endcsname=\relax
\usepackage{amsmath,amsfonts,amssymb}
\usepackage{mathtools}
\usepackage[outdir=./]{epstopdf}
\usepackage{lineno,hyperref}
\modulolinenumbers[5]
\usepackage{subcaption}
\graphicspath{{../../Figures/}}

\newcommand{\av}[1]{\langle\, #1 \, \rangle}
\newcommand{\set}[1]{\left\{#1\right\}}
\newcommand{\Neigh}[1]{{\cal B}_{#1}}
\newcommand{\bra}[1]{\left[\, #1 \, \right]}
\newcommand{\pare}[1]{\left(\, #1 \, \right)}

\newcommand{\setZ}{\mathbbm{Z}}

\begin{document}


\title{The non linear dynamics of retinal waves}
\author{\href{https://orcid.org/0000-0003-1523-4187}{Bruno Cessac}\\
Universit\'e C\^ote d'Azur, INRIA, France\\
Biovision team and Neuromod Institute\\
Correspondence: bruno.cessac@inria.fr\\
\and
\href{https://orcid.org/0000-0001-5924-7823}{Dora Matzakou-Karvouniari}\\
Eurecom, Sophia-Antipolis, France\\
}

\maketitle

\begin{abstract}
We investigate the dynamics of stage II retinal waves via a dynamical system, grounded on biophysics, and analysed with bifurcation theory. We show how the nonlinear cells coupling and bifurcation structure explain how waves start, propagate, interact and stop. Especially, we analyse how the existence of a small region in parameter space, where dynamics returns in a recurrent way, gives rise to a very rich dynamics. In this context, we propose a non linear transport equation characterizing the waves propagation and interaction.
\end{abstract}

%


\section{Introduction}\label{Sec:Intro}

The visual system is an important part of the central nervous system. Although its functionality seems effortless, it carries out complex tasks including the reception of light and the formation of visual representations, the identification and categorization of visual objects, computing distances to and between objects and guiding body movements in relation to the environment \cite{marr:82,chalupa-werner:04,daw:12}. How the visual system acquires such abilities during development, pre- and post-natal, is a fascinating question.

Vision starts at the retina, a light-sensitive tissue at the back of the eye that covers about 65 percent of its interior surface \cite{besharse-bok:11}. The overwhelming capacity of the retina to convert complex visual scenes into spike trains - that send information to the visual cortex - is largely due to its layered structure and to dynamical interactions between retina-specific neurons, resulting in a complex, stimulus driven network dynamics. It is known that part of the shaping of the visual system during development is due to such network dynamics. Indeed, before retina is responsive to light, in early development, a wave activity is observed. The so-called retinal waves, reported in many vertebrate species - chicks \cite{sernagor-eglen-etal:00}, ferrets \cite{feller-butts-etal:97}, mice \cite{maccione-hennig-etal:14}, turtles \cite{sernagor-grzywacz:99}, macaques \cite{warland-huberman-etal:06}, are spontaneous bursts of activity propagating in the developing retina and playing a fundamental role in shaping the visual system (retinotopy, binocular vision) and retinal circuitry. 

Certainly, there are genetic instructions organizing development and, in particular, visual system shaping and retinal waves occurrence. However, these instructions could either be quite detailed, tightly organizing each step of development, or, they could be less constrained, just setting up the main steps. In this perspective, retinal waves would emerge due to purely dynamical mechanisms broadly controlled by a genetic program. Then, the prevalence of retinal waves showing very similar patterns across many different species and developmental stages, would suggest that they are generated by common, generic, collective, non-linear mechanisms that still needs to be unravelled. Developing mathematical models constitutes a way to extract these putative underlying mechanisms, a strategy that has been applied over the last twenty years for retinal waves.

Indeed, several models have been proposed to describe this phenomenon, mainly in the stage of development called stage II, mediated by Starbust Amacrine Cells (SACs) coupled via the acetylcholine neurotransmitter and nicotinic receptors
\cite{feller-butts-etal:97,butts-feller-etal:99,godfrey-swindale:07,godfrey-eglen:09,hennig-adams-etal:09,lansdell-ford-etal:14},  (see section \ref{Sec:Biophys} for more detail).
For an extended review, see also \cite{godfrey-eglen:09,gjorgjieva-eglen:11}. 
All these models, mostly based on numerical simulations, have been able to reproduce experimental characteristics of retinal waves such as their size, duration, speed and frequency. However, their approaches mainly lie on capturing phenomenological features of waves and most of them do not 
analyse mathematically the model dynamics and mechanisms responsible for retinal waves generation, propagation and termination. In addition they require fine tuning of some parameters.
 A natural question is what happens if these modelling parameters are modified? Addressing this issue requires a theoretical analysis of the model's structural stability (stability of a behaviour with respect to parameters variations). This is actually a salient question as physiological parameters
such as the reversal potential for GABA receptors or the efficacy of cholinergic synapses \cite{zheng-lee-etal:04} are known to change during development, thereby inducing potential changes in wave dynamics. 

Especially, in \cite{zheng-lee-etal:04}, Zheng et al. have experimentally shown that the spontaneous stage II retinal waves are mediated by a \textit{transient} network of SACs, connected through excitatory cholinergic connections \cite{zheng-lee-etal:06}, which are formed only during a developmental window up to their complete disappearance. Especially, the intensity of the acetylcholine coupling is \textit{monotonously decreasing} with time (see Fig. 3B in \cite{zheng-lee-etal:04} and table \ref{Tab:AchEVolution} below). How do stage II retinal waves dynamics depend on this parameter? How do their characteristics (e.g. size, duration, speed) evolve when acetylcholine coupling decreases? What could be the impact on visual system development? 
In the present paper, we investigate the question of the dynamic origin and structure of stage II retinal waves, their distribution, and the stability of their behaviour, during the stage II developmental window, upon a decreasing acetylcholine coupling compatible with experimental findings. 

This paper is the continuation of a study starting with \cite{karvouniari-gil-etal:19} and fully exposed in the PhD thesis \cite{karvouniari:18} aiming to: (1) propose a detailed biophysical modeling of the cellular mechanisms of the spontaneous activity in immature SACs, addressing questions directly accessible and linked to pharmacological manipulations on retinal waves; (2) develop a mathematical analysis of this model, based on dynamical systems and bifurcation theory so as to propose generic mechanisms for wave initiation, propagation and stop. In addition, such analysis allows to
question the structural stability of the model; (3) validate the working hypothesis that SACs produce a wide spectrum of waves duration, size, interwave period, across species, thanks to a single fact: they are close to specific bifurcations generating what Izhikevich calls Type 1 excitability \cite{izhikevich:00,izhikevich:07}. Although point (2) and (3) might look contradictory we will show here that this is not the case. \\
\indent The paper \cite{karvouniari-gil-etal:19} was devoted to the study of single, isolated cells dynamics. The detailed modelling of individual SACs dynamics as autonomous, rhythmic bursters and the mathematical analysis of our dynamical system using bifurcation theory helped us identify the key parameters which control bursting in immature SACs. Especially, we exhibited that few biophysical parameters regulating calcium and potassium activity control bursting and we proposed a testable experimental prediction on the role of voltage-dependent potassium channels on the transitory excitability properties of SACs along development. We also proposed an explanation on how SACs can exhibit a large variability in their bursting periods across different species , as observed experimentally, yet based on a simple, unique, mechanism.\\
\indent In the present paper, extending the single neuron dynamics, we model in detail the mutual cholinergic synaptic connections between SACs, ending up exploring the mechanisms of SACs synchronization and waves. In section \ref{Sec:Model}, we give a brief account of retinal waves properties, for non expert readers, before introducing a multi-dimensional model of SACs featuring their intrinsic dynamics (bursting and hyperpolarization) and their non linear coupling via acetylcholine. We argue that the dynamics of SACs waves is essentially controlled by two parameters slowly evolving in time: one, $G_A$, controlling the excitatory cholinergic cells coupling, and the other, $G_S$, controlling cell's hyperpolarisation and refractoriness. Thanks to a bifurcation diagram in the space $\set{G_S,G_A}$, we analyse, in section \ref{Sec:WavesDyn}, how waves start, propagate, interact and stop. Mostly, the variety observed in waves dynamics comes from the fact that they start in a tiny region in the space $\set{G_S,G_A}$ delimited by bifurcation lines, where waves initiation is quite sensitive to perturbations such as noise, which is a feature of type 1 excitable systems \cite{izhikevich:00,izhikevich:07}. Although this region is tight, the slow dynamics returns to it in a recurrent way, regenerating the potentiality to trigger new waves sensitive to perturbations with a wide variation in waves characteristics. In addition, this scenario holds on an interval of acetylcholine coupling, explaining the apparent contradiction between a variability induced by closeness to bifurcations and stability to variations in developmental parameters.
On these tracks we numerically investigate, in section \ref{Sec:VaryingAch}, the effect of varying acetylcholine coupling during development, in a range of parameters extracted from the experimental literature. Simulations are done in a 1-dimensional lattice, with nearest neighbours interactions. To extend our analysis to larger dimensions we derive, in section \ref{Sec:Transport}, transport equations for $G_S,G_A$, now considered as propagating fields shaping waves dynamics. From this, we are able to compute the wave speed as a function of acetylcholine coupling, as well as to show the existence of a critical value of this coupling, below which no wave can propagate. As we argue, these transport equations bare interesting analogies with the Kardar-Parisi-Zhang (KPZ) equations of interface growth \cite{kardar-parisi-etal:86} on one hand, and Self-Organized Criticality \cite{bak-tang-etal:87} on the other hand, opening up perspectives for future research. 

\section{Stage II retinal waves model}\label{Sec:Model}

\subsection{The biophysics of retinal waves}\label{Sec:Biophys}
Retinal waves are bursts of activity occurring spontaneously in the developing retina of vertebrate species, contributing to the shaping of the visual system organization \cite{wong-meister-etal:93, firth-wang-etal:05, sernagor-hennig:13, ford-feller:12}. They are characterized by localized groups of neurons becoming simultaneously active, initiated at random points and propagating at speeds ranging from $100$ $\mu m/s$ (mouse, \cite{singer-mirotznik-etal:01}, \cite{maccione-hennig-etal:14}) up to $400$ $\mu m/s$ (chick, \cite{sernagor-eglen-etal:00}), with changing boundaries, dependent on local refractoriness \cite{ford-feller:12,ford-felix-etal:12}. This activity, slowly spreading across the retina, is an inherent property of the retinal network \cite{zheng-lee-etal:06}. 
More precisely, the generation of waves requires three conditions \cite{ gjorgjieva-eglen:11,ford-feller:12}: 
\begin{enumerate}[(C1)]
\item A source of depolarization for wave initiation (\textit{"How do waves start?"}). Given that there is no external input (e.g. from visual stimulation in the early retina), there must be some intrinsic mechanism by which neurons become active \footnote{Recent experimental works by E. Sernagor lab, suggest that some specific cells, whose type has not yet been identified, act as pacemaker triggering waves \cite{montigny-krishnamoorthy-etal:19}. We shall not consider this aspect in our work although it could be included in the proposed formalism.}.
\item  A network of excitatory interactions for propagation (\textit{"How do waves propagate?"}). Once some neurons become spontaneously active, how do they excite neighboring neurons?
\item  A source of inhibition that limits the spatial extent of waves and dictates the minimum interval between them ("\textit{How do waves stop ?}"). 
\end{enumerate}
 Wave activity begins in the early development, long before the retina is responsive to light. It emerges due to several biophysical mechanisms, which change during development, dividing retinal waves maturation into 3 stages (I, II, III) \cite{sernagor-hennig:13}. Each stage, mostly studied in mammals, is characterized by a certain type of network interaction (condition C2): gap junctions for stage I; cholinergic transmission for stage II; and glutamatergic transmission for stage III. In this work, we focus on stage II. 

During this period, the principal mechanism for transmission is due to the neurotransmitter acetylcholine (Ach) with nicotinic receptors \cite{feller-butts-etal:97,zheng-lee-etal:06,sernagor-hennig:13}. The first functional cholinergic connections are formed around birth, firstly at the level of Starburst Amacrine Cells (SACs). The emergence of waves depends on cellular mechanisms studied by \cite{zheng-lee-etal:06} (for rabbits), where it is found that Starburst Amacrine Cells emit spontaneous intrinsic calcium bursts (condition C1). When bursting a SAC emits acetylcholine thereby increasing the level of excitability of its neighbours. When the bursts occurring in a neighbourhood of a quiet SACs are synchronized, they eventually lead it to burst, inducing a wave propagation. Stage II waves are therefore spatiotemporal phenomena resulting from the spatial coupling of local bursters (SACs) via acetylcholine (condition C2). In addition, a strong after hyperpolarization current (sAHP)  induces a long refractory period for the recently active neurons, preventing the propagation of a new wave for a period of order of a minute, in the region where a wave has recently spread  \cite{zheng-lee-etal:06}. More generally, sAHP plays a prominent role in shaping waves periodicity and boundaries (condition C3), as we discuss in this paper. 

\subsection{The model}\label{Sec:ModelDef}

Among the 3 stages of retinal waves, stage II is the one which received the most attention from modellers  \cite{feller-butts-etal:97,butts-feller-etal:99,godfrey-swindale:07,godfrey-eglen:09,hennig-adams-etal:09,gjorgjieva-eglen:11,lansdell-ford-etal:14} (see \cite{kahne-rudiger-etal:19} for a recent model on stage I, and \cite{choi-zhang-etal:14} for a model of stage III). The model we propose is inspired by the work of \cite{hennig-adams-etal:09} and \cite{lansdell-ford-etal:14}, with strong modifications, in the model itself and in the methods used for study. It was introduced in \cite{karvouniari-gil-etal:19} at the level of single cells (no network) where it was thoroughly justified on biophysical grounds, leading to experimental predictions. The approach to study its dynamics, based on a bifurcation analysis of an individual SAC dynamics was original too. Here, we extend the study to a network of SACs, with an analysis still grounded as well on bifurcation theory. 
In this section, we define the equations ruling the model's dynamics.   

Note that SACs display a regular tiling of the retina with a disk-shaped dendritic tree \cite{tauchi-masland:84}. Their average distance is  $a \simeq 50 \, \mu m$. We thus define a network of Starburst Amacrine Cells distributed on a regular lattice in $\setZ^d$, $d=1,2$, with lattice spacing $a$, where sites/neurons are labelled with an index $i=1 \dots N$.
The state variables characterizing neuron $i$ are: $V_i(t)$, the membrane potential, $N_i(t)$, the gating variable for fast $K^+$ channels, $R_i(t)$ and $S_i(t)$, the gating variables for slow $Ca^{2+}$-gated $K^+$ channels, $C_i(t)$, the intracellular $Ca^{2+}$ concentration, $A_i(t)$, the extracellular acetylcholine concentration emitted by neuron $i$.  We note $\Neigh{i}$ the set of SACs which are in synaptic contact with SAC $i$. A SAC is not connected to itself, ($i  \, \not \in \, \Neigh{i}$). 
The structure of neighbours is nearest neighbours. This is mainly because it allows to write explicit transport equations in term of a Laplacian (see section \ref{Sec:Transport}). More complex structures have been considered in \cite{karvouniari:18} or \cite{karvouniari-gil-etal:16} for this model. We come back to this point in the discussion section.

The parameters values are carefully calibrated with respect to biophysics, found in the literature or fitted from experimental curves in \cite{abel-lee-etal:04}, \cite{zheng-lee-etal:04} and \cite{zheng-lee-etal:06} (see \cite{karvouniari:18} for detail on the fitting procedures). Parameters values and units are given in appendix \ref{AppendixParameter}. \\

Dynamics has several time scales, from fast (a few milliseconds) to slow (a few seconds) to very slow (about one minute). These multiple scales are at the source of the dynamic complexity and the core of the present analysis, so we present the equations ruling the evolution of SACs taking into account this time scale separation.
First, the membrane potential of SAC $i$ obeys:
\begin{equation}\label{eq:Voltage}
\begin{array}{lll}
C_m \frac{dV_i}{dt}&=&  I_{L_i} + I_{C_i} + I_{K_i} + I_{S_i} + I_{A_i};
\end{array}
\end{equation} 
where $C_{m}$ is the membrane capacitance, $I_{L_i}=-g_L \, \pare{V_i - V_L}$ is a leak current, with leak conductance $g_L$ and leak reversal potential $V_L$. The current:
\begin{equation}\label{eq:calcium}
I_{C_i}=-g_C M_\infty(V_i)(V_i-V_C),
\end{equation}
is a calcium current of the Morris-Lecar form \cite{morris-lecar:81},
with maximal calcium conductance $g_C$, calcium reversal potential $V_C$ and $M_\infty(V)=\frac{1}{2} [1+\tanh (\frac{V-V^{(1)}}{V^{(2)}})]$. 
The \textit{fast} potassium current $I_{K_i}$ takes the form $I_{K_i}(V_i,N_i)=-g_K  \, N_i \, (V_i-V_K)$,
where the evolution of the voltage-gated $K^+$ channel gating variable $N_i(t)$ is given by:
\begin{equation}\label{eq:N}
\tau_N \, \frac{d N_i}{d t}=\Lambda(V_i) \, \pare{N_{\infty}(V_i)-N_i},
\end{equation}
with $\Lambda(V_i)=\cosh(\frac{V_i-V^{(3)}}{2 V^{(4)}})$,
and $N_{\infty}(V_i)=\frac{1}{2} [1+\tanh (\frac{V-V^{(3)}}{V^{(4)}})]$.
Here $\tau_N$, the characteristic time of the activation variable $N$, is of order $5$ ms. $I_{K_i}$ is therefore quite fast, in contrast to the original Morris-Lecar paper, calibrated in order to capture the frequency of the fast repetitive firing of SACs, (around $20$ Hz \cite{zheng-lee-etal:06}). $V^{(1)},V^{(2)},V^{(3)},V^{(4)}$ are tuning parameters whose value is given in the appendix \ref{AppendixParameter}.

The slow After Hyperpolarization (sAHP) current $I_{S_i}=-G_{S_i} \, \pare{V_i - V_K}$
is a calcium-gated potassium current, whose effective conductance is:
\begin{equation}\label{eq:GS}
G_{S_i}=g_S \, R_i^4,
\end{equation}
$g_S$ being the maximal sAHP conductance, evolves very slowly (time scale of order one minute). This conductance is controlled by a cascade of mechanisms involving the entrance of calcium in the cell, and, thus, the increase of the intra-cellular calcium concentration, $C_i$, upon membrane potential increase.
Increasing $C_i$ increases the fraction, $S_i$, of saturated calmodulin that binds to slow calcium-gated potassium receptors. This increases the fraction, $R_i$, of bounded terminals in the corresponding potassium channels, thereby increasing $G_{S_i}$.
There is a power $4$ because $4$ bound terminals are needed to open a $Ca^{2+}$-gated $K^+$ channel. This large exponent, $4$, is actually quite important for
the waves dynamics, as developed below.

The corresponding equations are \cite{karvouniari-gil-etal:19}:
\begin{equation}\label{eq:RsAHP}
\left\{
\begin{array}{lll}
\tau_C \, \frac{d C_i}{d t}&=-\frac{\alpha_C}{H_X} \,C_i+C^{(0)} + \delta_C \, I_{C_i}(V_i);\\
\tau_S \, \frac{dS_i}{dt} &= \alpha_S C_i^4 (1-S_i) - S_i;\\
\tau_R \, \frac{d R_i}{d t}&= \alpha_R \, S_i \, (1-R_i)-R_i;
\end{array}
\right.
\end{equation}
where $I_{C_i}$ is given by \eqref{eq:calcium}. The value of the tuning parameters $\alpha_C,C^{(0)},H_X,\delta_C,\alpha_S,\alpha_R$ used in this paper  are given in the appendix \ref{AppendixParameter}.
The time scale $\tau_C$ is of order $2$ s whereas $\tau_S,\tau_R$ are of order one minute in real SACs.
Actually, in the simulation displayed in section \ref{Sec:AchDecays} we will consider shorter time scales, of order $10$ s, to reduce the computational time. \\

All the terms in \eqref{eq:Voltage} considered up to now correspond to local, uncoupled, dynamics of SACs. In addition, SACS are coupled via the Ach current:
\begin{equation}\label{eq:IA}
I_{A_i} =  -G_{A_i} \, (V_i-V_A),
\end{equation}
with:
\begin{equation}\label{eq:GA}
G_{A_i}=g_A \, \sum_{j \in \Neigh{i}}\, \frac{A_j^2}{\gamma_A+A_j^2},
\end{equation}
where $g_A$ is the maximal Ach conductance and $\gamma_A=K_d^2$, $K_d$ is the half-activation constant of Ach production ($\sim 1 - 4$ nM). The effective Ach conductance depends on the Ach concentration produced by SACs $j$ connected to $i$. The exponent $2$ comes from the physiology of nicotinic receptors: two Ach molecules have to bind to the receptor to open the corresponding ion channel (permeable to sodium and potassium, resulting in a reversal potential $V_A$ close to $0$ mV). The Ach concentration $A_j$ depends on the pre-synaptic cell's voltage via the differential equation:
\begin{equation} \label{eq:Achprod}
\frac{d A_j}{d t}=-\mu_A \, A_j \, + \, \beta_A T_A(V_j),
\end{equation} 
where $\mu_A$ is a degradation coefficient (of order $2 \, s^{-1}$), $\beta_A$ is the maximal production rate of $A$ (in $nM s^{-1}$), and the production term is:
\begin{equation}\label{eq:tach}
T_A(V_j)=\frac{1}{1+e^{-\kappa_A (V_j-V_A)}}.
\end{equation}
The parameters $\mu_A,\kappa_A,\beta_A,V_A$ have been fitted from experiments \cite{zheng-lee-etal:06,karvouniari:18}. They are uniform on the SACs population (independent of $i$). 
The set of equations \eqref{eq:Voltage}-\eqref{eq:tach} constitutes a non linear dynamical system of dimension $7 \times N$ on a spatial lattice of $N$ SACs. All parameters value are reported in appendix \ref{AppendixParameter}. 

Note that the sAHP conductance, $G_{S_i}$, depends on cell $i$ \textit{intrinsic} activity, whereas the Ach conductance, $G_{A_i}$, depends on the activity of cells \textit{connected} to cell $i$. Also note the notations: $g_S,g_A$ are physiological parameters (maximal conductance), whereas $G_S,G_A$
are dynamical variables depending non linearly on the system evolution. The time scales of evolution are rather different: $G_A$ has a time scale of order seconds (due to the evolution of $A$, eq. \eqref{eq:Achprod}), $G_S$ has a time scale of order a minute (due to the evolution of $R$). Finally, later in the paper (section \ref{Sec:AchDecays}) we are considering the evolution of $g_A$ corresponding, physiologically, to a time scale of order a day (see Tab. \ref{Tab:AchEVolution}).

\subsection{Bifurcation analysis}\label{Sec:BifAnalysis}

\begin{figure}
\centerline{
\resizebox{\textwidth}{0.4\textheight}{
\includegraphics[]{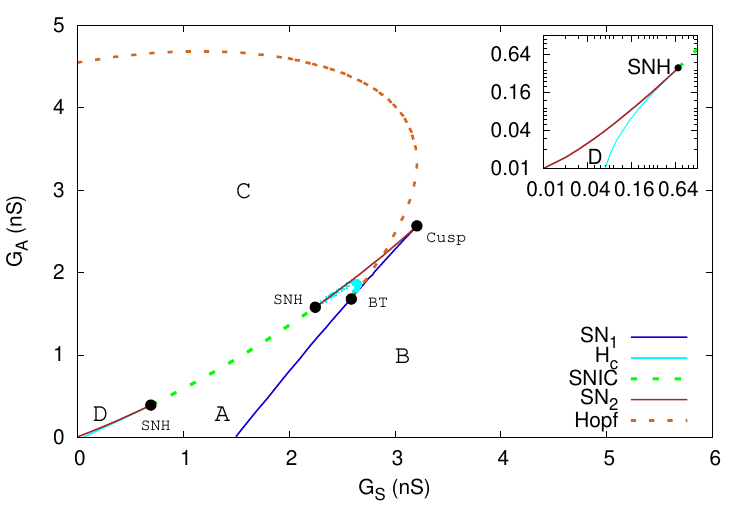} 
}
}
\vspace{0.2cm}
\centerline{
\resizebox{0.4\textwidth}{0.15\textheight}{
\includegraphics[]{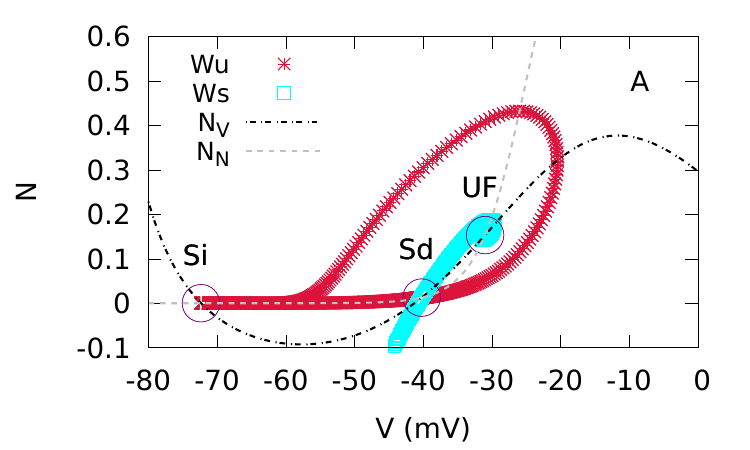} 
}
\resizebox{0.4\textwidth}{0.15\textheight}{
\includegraphics[]{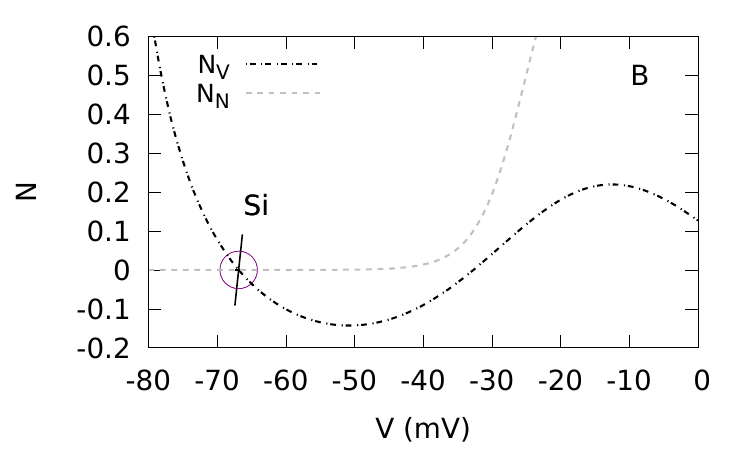}
}
}
\centerline{
\resizebox{0.4\textwidth}{0.15\textheight}{
\includegraphics[]{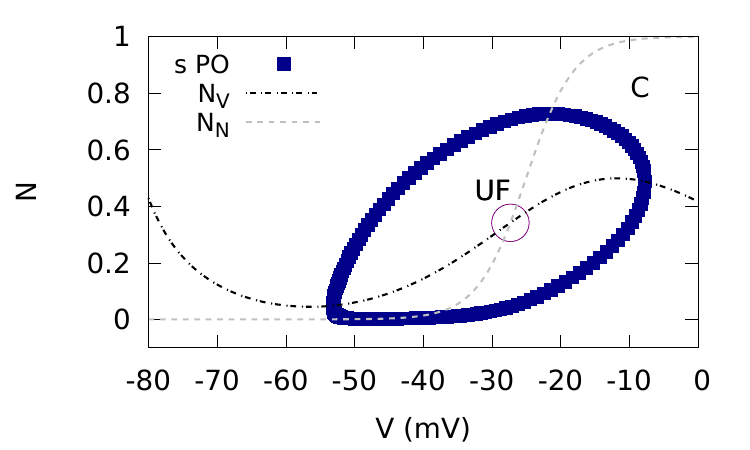}
}
\resizebox{0.4\textwidth}{0.15\textheight}{
\includegraphics[]{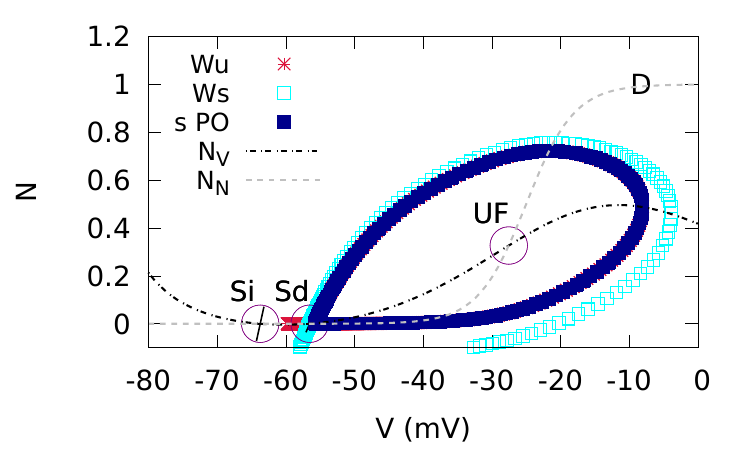}
}
}
 \caption{\scriptsize\textbf{1. Top. Bifurcation diagram} of the fast dynamics \eqref{eq:Voltage}-\eqref{eq:N} in the plane $G_S,G_A$. "SN" stands for "Saddle-Node bifurcation", "Hc" for "Homoclinic bifurcation", "Hopf" for "Hopf bifurcation", "SNIC" for "Saddle-Node on an Invariant Cycle","SNH" for Saddle-Node-Homoclinic point (also called saddle-node separatrix loop\cite{schecter:87}, SNSL), "BT" stands for Bogdanov-Takens. At the SNH codim-2 bifurcation point, there is a Hc bifurcation line
connecting to the BT point with a SNIC emerging from the opposite
side of the SNH in which the $SN_2$ and Hc are tangent. At the BT point, there is, on one side the $SN_1$ line and, on the other side the Hc, the $SN_1$ and a Hopf line, tangent at the BT point (see \href{https://team.inria.fr/biovision/bifurcations-map-of-the-retinal-waves-model/}{https://team.inria.fr/biovision/bifurcations-map-of-the-retinal-waves-model/} for a zoom on this region, explanations and movies).
  \textbf{Inset:} Zoom on region $D$, in log scale. The Hc and $SN_2$ line meet tangentially at the SNH point and a SNIC line emerges on the other side. \textbf{2.  Phase portraits} of regions $A,B,D,C$ with nullclines ($N_V$, in black, is the $V$ nullcline, $N_N$, in grey, is the $N$ nullcline), stable manifold of hyperbolic fixed points ($W_s$ in cyan), unstable manifold of the saddle-fixed points ($W_u$ in red), stable periodic orbit (in blue). "Si" means "Sink", "$Sd$" means "Saddle", "UF" means "Unstable Focus", "SPo" means "Stable Periodic orbit". 
 \label{Fig:BifurcationDiagram_GS_GA}}
 \end{figure}

The key observation affording the model analysis is that the evolution of the effective conductance $G_{S_i}$ (sAHP) and $G_{A_i}$ (Ach) of a SAC, $i$, is slow compared to the fast dynamics of $V_i, N_i$. The idea is thus to consider $G_{S_i},G_{A_i}$ as slow parameters tuning the evolution of the fast dynamics \eqref{eq:Voltage},\eqref{eq:N}. 
In addition, we will see that the pattern of slow-time evolution of $G_{S_i},G_{A_i}$ during a wave is pretty similar for all cells. So, the central idea is to consider first equations \eqref{eq:Voltage}-\eqref{eq:N} with tunable parameters, the sAHP conductance, $G_S$, and the Ach conductance, $G_A$, independent of the cell.  The core of this analysis is then displayed in the bifurcation diagram of Fig. \ref{Fig:BifurcationDiagram_GS_GA}.
We explain the diagram in this section. Later, we will let $G_S,G_A$ slowly evolve in time for each cell, so as to describe the dynamics of waves initiation, propagation and stop. 

Let us now comment about Fig. \ref{Fig:BifurcationDiagram_GS_GA}. More details are given in the legend. Additional information, not necessary for the understanding of the paper, are available on the web page  \href{https://team.inria.fr/biovision/bifurcations-map-of-the-retinal-waves-model/}{https://team.inria.fr/biovision/bifurcations-map-of-the-retinal-waves-model/}. Especially, the region in the interval $\sim \pare{G_S,G_A} \in [2.2,3.2[ \times [1.3,2.7[$, containing the SNH, BT, and SNH points, is not discussed in the paper, for a simple reason: the dynamics of our model, with its parameters grounded on experiments, never enters in this region (see Fig \ref{Fig:Trajectory_GS_GA}). A zoom on this region, with the corresponding bifurcation diagram, can nevertheless, be found at the aforementioned web page.  Let us also mention that, in this region, dynamics ought to be very close to the model studied by Moreno-Spiegelberg et al \cite{arinyo-i-prats-moreno-spiegelberg-etal:21,moreno-speigelberg-arinyo-i-prats-etal:22} dealing with traveling pulses in Type 1 excitable media with some important differences though: (1) the presence of noise in our model; (2) the fact that our parameters $G_S,G_A$ are slowly evolving according to the fast dynamics; (3) the presence of long lasting refractory regions shaping waves propagation. Although it could actually be interesting to see what happens in our model when entering this region, note that this would require values of $g_A$ about $10$ times higher than what we have inferred from experimental data (see Table \ref{Tab:AchEVolution}).

Excluding this region, Fig. \ref{Fig:BifurcationDiagram_GS_GA} reveals that the parameter space $G_S,G_A$ is divided into four regions, $A,B,C,D$. In region $A$, there is a stable fixed point (Sink, noted Si), corresponding to a rest state, and an unstable focus (UF), separated by a Saddle point (Sd) whose stable ($W_s$, in cyan) and unstable ($W_u$, in red) manifolds are shown.
These manifolds are important as they trigger the homoclinic bifurcation discussed below. The corresponding phase portrait is plotted in Fig. \ref{Fig:BifurcationDiagram_GS_GA}, middle, left. 
In region $B$, there is a unique stable fixed point (Si), a rest state, whose voltage $V$ and activation variable $N$ depend on $G_S,G_A$ (Fig. \ref{Fig:BifurcationDiagram_GS_GA}, middle, right). In region $C$, there is a stable periodic orbit (sPO) corresponding to fast oscillations in the voltage and activation variable (Fig. \ref{Fig:BifurcationDiagram_GS_GA}, bottom, left).

Finally, in region $D$, which lies in between the cyan $H_c$ and the brown $SN_2$ lines in the bifurcation diagram, with $G_S \in [0,0.7]$ (see inset), there is a stable fixed point Si, corresponding to a rest state, and a stable periodic orbit (Fig. \ref{Fig:BifurcationDiagram_GS_GA}, bottom, right). These two attractors are separated by a Saddle point (Sd). Region $D$ is very narrow in the parameters space and hardly visible on the diagram at this scale. That's why, in the inset zooming on this region, we use log scale. \textit{This is nevertheless an essential zone as discussed below.} 
The web page \href{https://team.inria.fr/biovision/bifurcations-map-of-the-retinal-waves-model/}{https://team.inria.fr/biovision/bifurcations-map-of-the-retinal-waves-model/} actually proposes several movies showing the evolution of the phase portrait when moving in the plane $\set{G_S,G_A}$ along several pathways. 

When varying $G_S,G_A$ the transition between regions correspond to the bifurcation lines displayed in Fig. \ref{Fig:BifurcationDiagram_GS_GA} top. "SN" refers to Saddle-Node, "Hc" to homoclinic, "Hopf" to Hopf bifurcation, and "SNIC" to Saddle-Node on an Invariant Cycle 
(see \cite{guckenheimer-holmes:83} or \cite{ermentrout-terman:10} for a classification of bifurcations). One goes from region $A$ to region $B$ by the saddle-node bifurcation $SN_1$ where the Sd and UF of region $A$ coalesce and disappear. 
For the transition from $A$ to $C$ there are two situations: (1), crossing region $D$. Here, the stable and unstable manifold of Sd give rise to a limit cycle by homoclinisation ($H_c$), then, Si and Sd coalesce by the saddle-node bifurcation $SN_2$, leaving a unique attractor, the stable periodic orbit of region $C$. (2) Direct transition from $A$ to $C$ by a SNIC bifurcation (green line in Fig. \ref{Fig:BifurcationDiagram_GS_GA}).
Movies illustrating these bifurcations are available on the web page \href{https://team.inria.fr/biovision/bifurcations-map-of-the-retinal-waves-model/}{https://team.inria.fr/biovision/bifurcations-map-of-the-retinal-waves-model/}. 

The equation characterizing the bifurcation line of the transition between region $D$ and $C$ via the $SN_2$ bifurcation will be used later. It is given by  \cite{karvouniari-gil-etal:19}:
\begin{equation}\label{eq:SN2}
-G_A \pare{V_{Si}-V_A}  - G_S \pare{V_{Si}-V_K}=I_{SN},
\end{equation} 
where $V_{Si}<0$ is the rest voltage on the sink in region $D$. $V_{Si}$ is actually a function of $G_S,G_A$ given by eq. \eqref{eq:VSi} below. The current $I_{SN} \sim 0.3$ pA \cite{karvouniari-gil-etal:19}. A similar equation holds for the homoclinic transition where $I_{SN}$ is replaced by $I_{Hc}$, with a value quite close to $I_{SN}$.


In region $D$ there is bi-stability  between a rest
state and an excited oscillatory state (in contrast to region A where there is bistability between two stationary states, one with low and the other with high voltage).  Noise allows to switch from rest to oscillatory state, as developed below. This corresponds to what Izhikevich call Type 1 excitability \cite{izhikevich:00,izhikevich:07}. Now, in the bifurcation diagram, 
which has been represented in large range of parameters values $G_S,G_A$ for completeness, region $D$ looks quite small and may appear as probably irrelevant, but we show that it is not. Recall first that $G_S$ corresponds, in the model, to $g_S \, R^4$ where $g_S$ is the maximal sAHP conductance, whereas $R \in [0,1]$ is a probability.
For our parameters value, in the absence of waves, cells are at rest with $R \sim 0.3$ (see Fig. \ref{Fig:CRS}), so that $G_S \sim 0.0162$ nS for $g_S=2$ nS, and $G_S \sim 0.081$ nS for  $g_S=10$ nS. In addition, at rest too, ($V \sim -70$ mV),  $T_A(V) \sim 0$ (eq. \eqref{eq:Achprod}), so that the Ach production is essentially zero, therefore $G_A \sim 0$. This implies that, at rest, cells are in region $D$. More generally, we argue below that there is long period, at the end of sAHP time course, where the cell lies in region $D$ (see also  Fig. \ref{Fig:Trajectory_GS_GA}). As we will see, this has a big impact on wave initiation and propagation.


\subsection{The mechanism of bursting}\label{Sec:Bursting}

Bursting, the succession of fast oscillations and rest state, is a well known phenomenon, widely studied in the literature of mathematical neuroscience \cite{izhikevich:07,ermentrout-terman:10}.
Here we describe its mechanism in our model (for more detail, see \cite{karvouniari-gil-etal:19}). Consider a cell, at rest (non oscillating), thus either in region $A,B$  of the bifurcation map or in the rest state (Si) of region $D$. This cell can be lead to an excited state with fast oscillations (stable limit cycle in region $C$ or $D$) from different scenarios discussed in the next section. 
Here, we are just interested in what happens when the cell starts to produce these fast oscillations.  

In this situation, the average voltage is high ($\sim  -30$ mV, see in Fig. \ref{Fig:BifurcationDiagram_GS_GA}, the phase portraits of region C, D). The rising in voltage increases the intracellular calcium concentration and triggers the generation of an increasing sAHP current, according to eq. \eqref{eq:RsAHP}. In the bifurcation diagram this corresponds to an increase in $G_S$, which eventually leads the cell to region $A$ or $B$ where no stable limit cycle exists. Thus, fast oscillations stop and the cell is feeling a strong, slow, after hyperpolarization current that slowly decreases its voltage. As a consequence, its intracellular calcium production decreases, thus sAHP, leading the cell slowly back to its rest value. This process takes around one minute \ during which the cell is hyper-polarized. This is how bursting is produced (see Fig. 3 in \cite{karvouniari-gil-etal:19}). 

\begin{figure}
\centerline{
\resizebox{0.8\textwidth}{0.3\textheight}{
\includegraphics[]{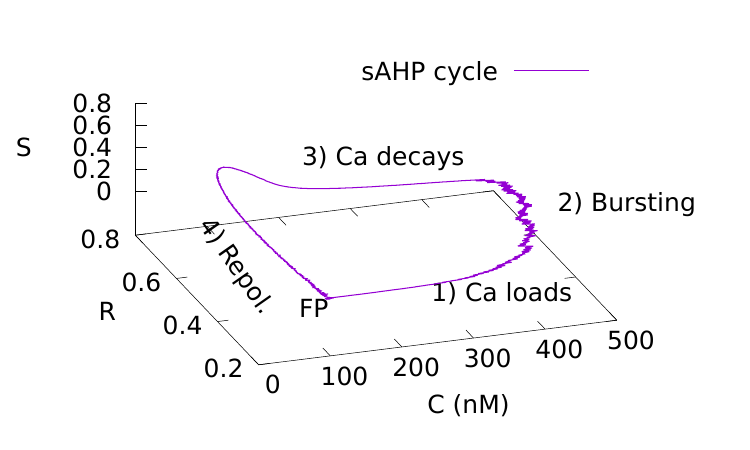}
}
}
 \caption{\footnotesize\textbf{Typical course of the variables $C,R,S$ for a cell.} in the presence of noise. "FP" means "Fixed Point", "Repol" means "Repolarization". Here, we have plotted one piece of trajectory corresponding to a single bursting period. In general, successive burst correspond to non superimposing trajectories, due to the effect of coupled cells.  
  \label{Fig:CRS}
}
 \end{figure}

Note that this corresponds to a pseudo-cycle, shown in Fig. \ref{Fig:CRS}, in the subspace of variables $C,R,S$, controlling the sAHP production. The following remark is important. The values taken by $R$ during the cycle, $R \in [0.3,0.7]$, correspond, for $g_S=10$ nS, to $G_S \in [0.081,2]$. As already mentioned, $R$ is quite slow compared to Ach. So, during the re-polarization phase, $G_A$ has a very low value close to $0$ (see e.g. Fig. \ref{Fig:Trajectory_GS_GA}).
Thus, when getting back to rest with a low $G_A$ value the SAC successively
crosses region $B,A$ then $D$.  When a cell is in region $B$ or $A$ the only possibility to excite it is to shift it to region $C$, requiring that $G_A$ is large enough, namely, that neighbours cells are bursting and $g_A$, the maximal Ach conductance is very large. Otherwise, it is not possible to excite it. When a cell is in region $D$ it can be excited, either by increasing $G_A$ and shifting it to region $C$, or, by jumping from the rest state (Si) to the period orbit, e.g. by a small amount of noise as developed in section \ref{Sec:NIB}. All in all, this implies that, on its way back to rest after hyperpolarization, a cell undergoes two sub-phases. An \textit{absolute refractory} (AR) phase where it is not possible to excite it and a \textit{relative refractory phase} (RR) where it can be excited again. In particular, in region $D$, cells are in the RR phase. Actually, the cell spends a relatively long time in the relative refractory phase. Although it can be excited anywhere in this phase, the level of excitation required to make it enter in the fast oscillations strongly depends on its location in region $D$ (see Fig. \ref{Fig:PBurst}), i.e. on the actual value of $G_S$ and $G_A$. 

\section{From single cell dynamics to waves} \label{Sec:WavesDyn}

From the previous analysis we have learned that a SAC can either be: (1) at rest (region $D$); (2) excited with fast oscillations (region $C$ or $D$); (3) hyperpolarized, either in the absolute refractory period  where it is not possible to excite it (region $A$ or $B$), or, in the relative refractory phase (region $D$, stable fixed point). When excited, each SAC undergoes the  excitation-hyperpolarization  cycle (Fig. \ref{Fig:CRS}) before eventually returning to rest.

We now analyse how this individual dynamics, coupled via Ach, induces retinal waves and controls waves initiation, propagation and stop.

\subsection{The mechanism of wave initiation} \label{Sec:WavesInit}

Let us consider a situation where all cells are at rest so that they cannot burst without an external influence.
By assumption, no cell is excited and no external influence can come from other cells. Actually, as the dynamics \eqref{eq:Voltage}-\eqref{eq:tach} is deterministic, a SAC at rest would therefore stay at rest indefinitely, without a possibility to trigger a wave. However, there is always noise in real neurons due e.g. to fluctuations in ionic currents from the random opening of ionic channels. We model this by a white noise $\xi(t)$,  whose amplitude is constant with time and controlled by a parameter $\eta$. 
In our formulation $\eta \, \xi(t)$ has the physical dimension of a current, hence the white noise $\xi(t)$ has a physical dimension of $ms^{-\frac{1}{2}}$, $\eta$ is in $pA \, ms^{\frac{1}{2}}$. The influence of this noise is
studied in detail in the paper \cite{karvouniari-gil-etal:19} for single and isolated neurons. Here, we outline the main consequences on retinal waves. Another scenario of wave generation, not requiring noise, is discussed in the discussion section \ref{Sec:SB}.

\subsubsection{Noise Induced Bursting}\label{Sec:NIB}

In region $D$, neurons are bistable: a neuron in the rest state (Si) can go to the excited oscillatory state (sPO) provided it receives a kick (here, noise) allowing its state, in the fast-dynamics phase space $V-N$,  to jump above the Sd barrier and to be trapped in the attraction basin of the stable periodic orbit (Fig. \ref{Fig:BifurcationDiagram_GS_GA}). Crossing the saddle barrier leads to the stable periodic orbit where the neuron stays a certain time, producing fast oscillations, returning eventually back to the Si attraction basin. So, in region $D$ we observe periods of fast oscillations punctuated by return to rest.
Note that, in region $A$, it is possible to observe isolated spikes, with fast return to rest, because noise can make the trajectory going beyond the saddle barrier. As there is no attractor beyond this barrier, the trajectory eventually returns back to the attraction basin of Si, following the unstable manifold of Sd, which is attracted to Si. This takes a bit of time though. In the trajectory, this effect is manifested by one or a few short spikes in the voltage. We want to differentiate these two situations focusing on bursting cells. Indeed, to anticipate on waves propagation, we need SACs to be in the excited state long enough to produce Ach emission so as to eventually excite the neighbours. As the characteristic time of Ach growth is of order $1-2s$ (a few $\mu_A^{-1}=0.53$ s), the cell must be excited for a duration of this order. This does not happen in region $A$. In region $D$ it does happen, although it depends on the noise intensity.

A good, numerical and automatic indicator to ensure that a SAC is excited long enough, is to fix a threshold $\theta$ on calcium concentration (here $\theta=4 \, C^{(0)}$, where $C^{(0)}$ is the baseline calcium concentration). We consider that a burst arises at time $t$ when $C(t) > 4 \, C^{(0)}$. As we checked, this condition is a good criterion to ensure that a sAHP current is generated.  Note that the characteristic time for calcium rising, $\tau_C=2$ s (eq. \eqref{eq:calcium}) is of the same order as the characteristic time for Ach production, so this criterion is also valid to ensure that SAC is excited long enough to produce Ach. 

\begin{figure}[H]
\begin{center}
\includegraphics[width=0.45\textwidth,height=0.25\textheight]{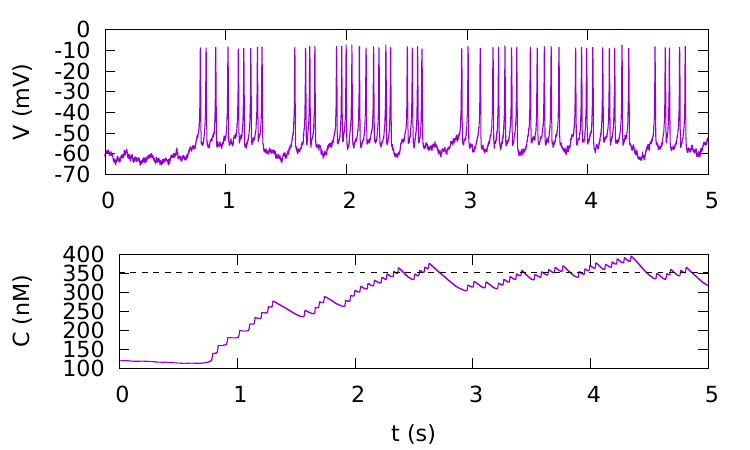}
\hspace{0.2cm}
\includegraphics[width=0.45\textwidth,height=0.25\textheight]{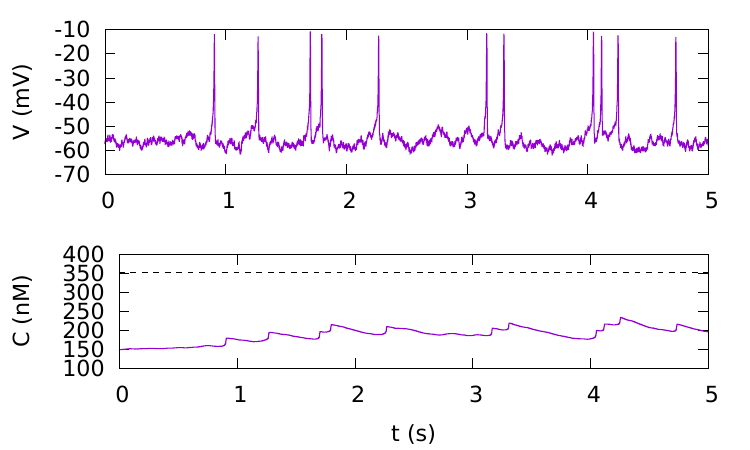}
%
\end{center}
\caption{\scriptsize\textbf{Noise induced bursting.} Both figures show the join evolution of voltage (upper trace) and calcium concentration (lower trace) for $\eta=10 \, pA \, ms^{\frac{1}{2}}$. \textbf{Left.} Region $D$. \textbf{Right.} region $A$. In region $D$ the trajectory stays long enough onto the stable periodic orbit to trigger a high increase in calcium concentration.
 In region $A$ noise can make the trajectory go in the region delimited by the saddle and its unstable manifold, leading to a short spike in $V$, not long enough to let the calcium concentration grow above the threshold $\theta$ (black line).
}
\label{Fig:NIB}
\end{figure}

\subsubsection{Bursting probability map}\label{Sec:BurstingProbabilityMap}

Here we want to compute the probability that calcium dynamics crosses the threshold $\theta$ ensuring the generation of Ach and sAHP for the full dynamics. As in the previous section, we consider $G_S,G_A$ as constant. 

The rest state corresponds to $\frac{dV}{dt}=0$ and $\frac{dN}{dt}=0$ (fast dynamics) in the absence of noise, and for fixed values of $G_S,G_A$. This corresponds to a voltage:
\begin{equation}\label{eq:VSi}
V_{Si} = \frac{g_L \, V_L + g_C \, M^\ast \,V_C +g_K N^\ast V_K + G_S \, V_K+G_A V_A}{g_L  + g_C +g_K  + G_S+G_A}
\end{equation}
where $M^\ast \equiv M_\infty\bra{V_{Si}}$, $N^\ast \equiv N_\infty\bra{V_{Si}}$.
This is thus, in general, a non linear function of the parameters $G_S,G_A$. 

Now, due to noise, the voltage $V$ has fluctuations $\delta V$ around the rest state, $V=V_{Si}+\delta \, V$. Assuming that $\delta V$ is small enough so that $M_\infty(V) \sim M_\infty(V_{Si}) \, + \, \delta V \, M'_\infty(V_{Si})$, $N_\infty(V) \sim N_\infty(V_{Si}) \, + \, \delta V \, N'_\infty(V_{Si})$, the fluctuations are given, to the first order in $\delta V$, by the stochastic differential equation $\frac{d \delta V}{dt} = -\frac{1}{\tau(G_S,G_A)} \, \delta V + \frac{\eta}{C_m} \xi_t$, where the characteristic time $\tau \equiv  \tau(G_S,G_A)=\frac{C_m}{G^\ast(G_S,G_A)}$ with:
\begin{equation} \label{eq:Gast}
G^\ast(G_S,G_A)=
g_L + g_C \, \pare{ M^\ast+ M^{'\ast} \pare{V_{Si}-V_C}}+g_K \pare{ N^\ast+ N^{'\ast} \pare{V_{Si}-V_K}} + G_S + G_A,
\end{equation}
where
$M^{'\ast}= M'_\infty(V_{Si})$ and the same for $N^{'\ast}$. Thus, $\delta \, V$ is, in this first order approximation, a Ornstein-Uhlenbeck process with mean zero, covariance $C(t,t') =\frac{\eta^2}{2 \, C_m^2} \, \tau \, \pare{e^{-\frac{t-t'}{\tau}}- e^{-\frac{t+t'}{\tau}}}$, for $t \geq t'$, and variance
%
$\sigma^2(t)
=\frac{\eta^2}{2 \, C_m \, G^\ast}  \, \pare{1- e^{-\frac{2 \, t}{\tau}}}$.
%
When $t$ is large enough, $t \gg \tau$, this variance converges to a constant $\sigma^2 \equiv \sigma^2\pare{G_S,G_A}$ with:
\begin{equation}\label{eq:sigma}
\sigma(G_S,G_A) = \frac{\eta }{\sqrt{2 \, C_m \, G^{\ast}(G_S,G_A)}}.
\end{equation} 

\begin{figure}
\resizebox{\textwidth}{0.25\textheight}{
\includegraphics[]{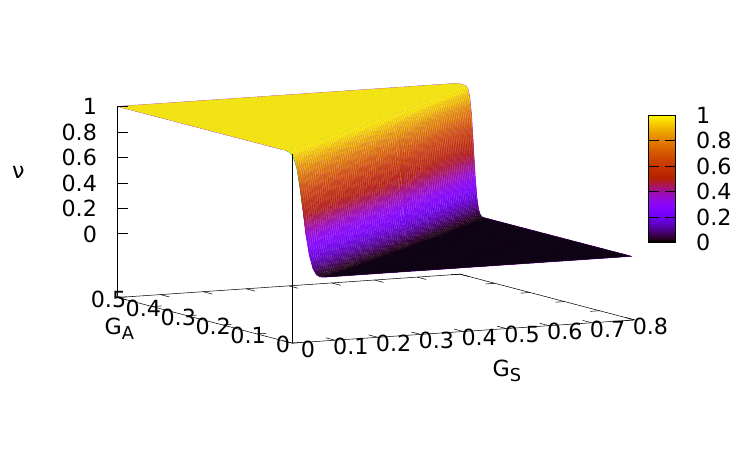}

\hspace{0.5cm}
\includegraphics[]{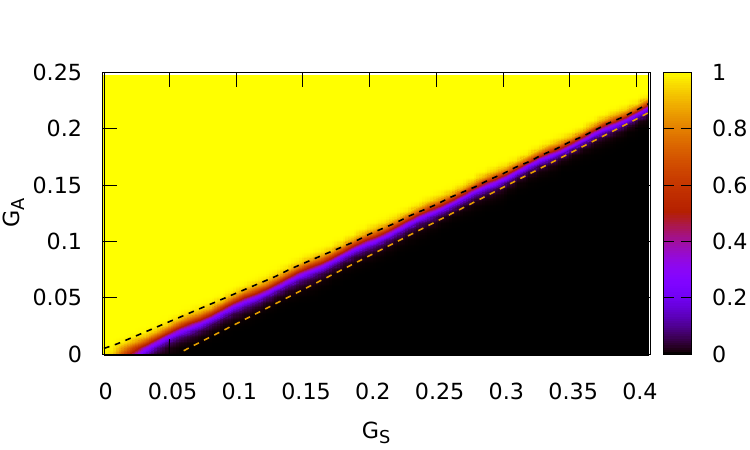}
}
\caption{\textbf{Probability} that a cell enters in the bursting regime where calcium exceeds the threshold $4 \, C^{(0)}$, in the space $G_S,G_A$, for $\eta=10 \, pA \, ms^{\frac{1}{2}}$. \textbf{Left.} $3$ D view of the probability with color map. \textbf{Right.} Color map of the bursting probability, plotted together with the Homoclinic bifurcation line (dashed orange) and the $SN_2$ bifurcation line (dashed black). This is a zoom of region $D$ (which extends further in the directions $G_S$ and $G_A$). Black corresponds to a $0$ probability while yellow corresponds to a probabilty $1$. The intermediate scales are shown in the color bar. 
}
\label{Fig:PBurst}
\end{figure}

Let us now call $V_{Sd}$ the voltage of the saddle point in region $D$.
A necessary condition for bursting to occur is that, at some time $t$, $V_{Si} +  \delta V(t) > V_{Sd}$ corresponding to a probability $1 - \Pi\pare{\frac{V_{Sd}-V_{Si}}{\sigma(t)}}$,
where $\Pi(x)=\int_{-\infty}^x \frac{e^{-\frac{y^2}{2}}}{\sqrt{2 \pi}} \,dy$.
This probability is a sigmoidal function, where the inflexion point depends on $G_S,G_A$ (via $G^\ast$), as well as the slope (via $\sigma(G_S,G_A)$) (see Fig. \ref{Fig:PBurst}).
More generally, the density and the cumulative distribution of the first passage time can be computed exactly \cite{borodin-salminen:02,alili-patie-etal:05b}. In the simplest case, it is given by an erf function, and more generally, by Hermitte polynomials.

However, this is not enough for what we want here. We want to compute the probability, $\nu(G_S,G_A)$, that, for given values of $G_S,G_A$ the \textit{calcium concentration} crosses the threshold $\theta$. In terms of $V$, this means we want to estimate the probability that, starting from $V=V_{Si}$ at time $0$, $V(t)$ crosses the $Sd$ barrier and \textit{stays long} enough beyond this barrier to let calcium rise and cross the threshold. We have not been able to compute analytically this probability. This could may be analytically done using Girsanov-Freidlin-Wentsel \cite{freidlin-wentzell:98} estimates (considering the stable limit cycle, an $\omega$-limit set, as an equipotential set), but this is beyond our technical reach. 

We made therefore numerical simulations to compute this probability in the plane $\set{G_S,G_A}$. We proceeded as follow. For each value of $G_S,G_A$ on a grid with resolution $10^{-2}$, one generates $N_{samp}=100$ trajectories where, for each trajectory, one starts from the rest state,  iterates the dynamics with noise over a  maximal time duration $T=20$ s, and seeks the occurrence of an increase in the calcium conductance above the threshold $\theta$. The estimated probability $\nu(G_S,G_A)$ is the number of occurrence of such long bursts divided by $N_{samp}$.

The result is shown in Fig. \ref{Fig:PBurst}. As expected, this is a sigmoid function, well approximated by a form: 
\begin{equation}\label{eq:nu}
\nu(G_S,G_A) \sim f\pare{\frac{G_A-G_{A_c}(G_S)}{\sigma(G_S,G_A)}}
\end{equation}
(Fig. \ref{Fig:PBurst}, top) where $f$ is a sigmoid. Here, $G_{A_c}$ is a threshold value of $G_A$, depending on $G_S$ and constrained by the 
$SN_2$ and Homoclinic bifurcation line (eq. \eqref{eq:SN2}).   Likewise, $\sigma(G_S)$, the inverse slope of the sigmoid function, is constrained by the Homoclinic  and $SN_2$ bifurcation lines.
$\sigma(G_S)$ depends on $\eta$, the noise intensity and tends to zero as $\eta \to 0$. Also, it decays to zero like the distance between the $SN_2$ and homoclinic bifurcation line. Thus, beyond the point where 
$SN_2$ and $H_c$ meet (at the SNH point), the transition is zero-one, corresponding actually to direct transition from region $A$ to region $C$ by a SNIC.

Let us make a few remarks to conclude this section. First, on the role of noise intensity. When $\eta \to 0$ (deterministic case), $\nu \to 0$: there is no possibility to burst in the absence of noise. For $\nu > 0$ the probability is a sigmoid function with a transition zone delimited by the 
boundaries of region $D$. When noise level is low, it takes a long time to the cell to jump to the stable periodic orbit, but when it is there, it stays sufficiently long to produce Ach dynamics, and, for calcium dynamics to reaches the threshold $\theta$. When $\eta$ increases, the time to jump on the sPO decreases, but the time where the trajectory stays on it decreases too. If $\eta$ becomes too large the cell cannot stay long enough on the stable periodic orbit to produce enough Ach to excite its neighbours. 
This implies that there is an \textit{interval} of $\eta$ values where retinal waves can actually propagate. 

Second, the probability to trigger a burst is highly sensitive to the value of $G_S,G_A$ because $\nu$ is quite sharp. Consider now a SAC on its way back to rest in the cycle of Fig. \ref{Fig:CRS}, somewhere in region $D$, thus in the relative refractory phase. Its level of $G_S$ depends on its position in the cycle, its level of $G_A$ depends on the state of its neighbours. And a tiny change in these value dramatically change the probability that this cell becomes excited. Especially, taking again into account the time scale separation between hyperpolarization and cholinergic coupling, and considering the $G_A$ dynamics for a constant $G_S$, a tiny change in the level of excitation of its neighbours dramatically impacts the probability the SAC has to become excited. This has a strong impact on waves propagation and waves distribution.


\subsection{The mechanism of waves propagation and stop}\label{Sec:WavesProp}

When a cell is in the excited, oscillating, state, it produces Ach, 
 according to \eqref{eq:Achprod}, \eqref{eq:tach}, on a time scale of order $2$ seconds, increasing the Ach current received by its post-synaptic neighbors. This increases the $G_A$ conductance of its neighbours cells. We investigate what happens to a cell  receiving an Ach current from its neighbours.
 There are actually $3$ cases, depending where the cell is located in the bifurcation map.
 
\begin{figure}
\centerline{
\resizebox{\textwidth}{0.35\textheight}{
\includegraphics[]{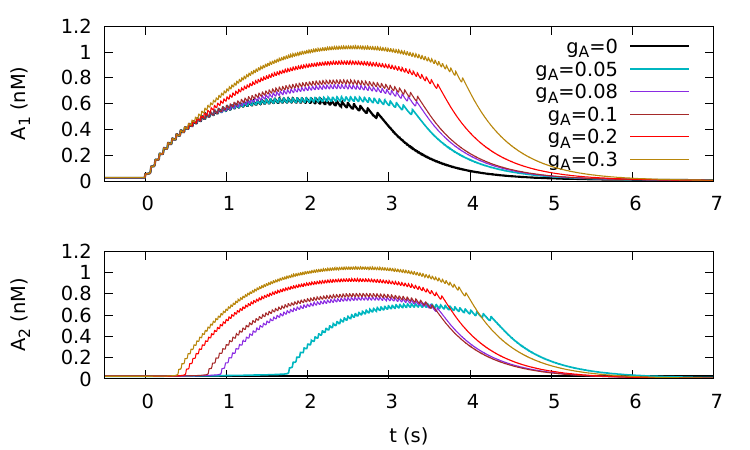}
\hspace{0.6cm}
\includegraphics[]{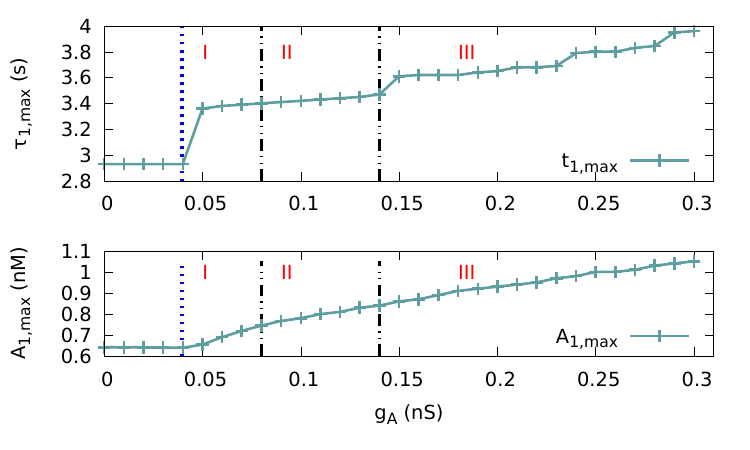}
}
}
 \caption{ \label{Fig:AchProfile_NIB_2cells}. \textbf{Left. Evolution of Ach concentration for 2 coupled cells,} during a burst, for different values of $g_A$ in nS. Row $1$ corresponds to cell $1$ and row $2$ to cell $2$. Here, cell 1 bursts first due to a small pulse in V while cell $2$ is driven by cell $1$: it bursts due to the Ach current generated by $1$. 
 \textbf{Right. Burst duration, $\tau_{1,max}$, and Ach peak, $A_{1,max}$, of cell $1$ under the feedback of cell $2$.} Region I,II,III are explained later in the paper. The blue dashed line correspond to a value of $g_A$, computed in section  \ref{Sec:ThresholdgA}, eq. \eqref{eq:g_0}, below which cell $1$ can not trigger a burst in cell $2$.
 These two figures illustrate the mutual non linear interaction of two bursting cells. 
Especially, the Ach current generated by $2$ prolongs the burst of $1$. Compare, indeed, the black trace for cell $1$ (no coupling so that cell $2$ stays at $A=0$), to e.g. the blue trace of cell 1 ($g_A=0.05$) where the burst is longer. 
Both cells are then mutually coupled and this mutual coupling reinforces their synchrony.
As shown on the right, the effect of this coupling on burst duration $\tau_{1,max}$ depends on the regime, I,II,III.  
%
}
\end{figure}

 \begin{enumerate}
\item\textbf{The cell is already excited} (region $C$ or sPO in $D$). Then, it produces Ach thereby exciting the cells which are exciting it, inducing a strong non linear feedback coupling mechanism, enhancing the Ach production, until the sAHP production takes place. This is illustrated, for two cells, in Fig. \ref{Fig:AchProfile_NIB_2cells}. This mutual coupling increases the Ach conductance $G_A$ of coupled cells, corresponding to moving vertically in the bifurcation map of Fig. \ref{Fig:BifurcationDiagram_GS_GA}. So, a cell initially on the sPO of region $D$ (e.g. excited by noise) is eventually lead to region $C$ where its oscillations are now essentially insensitive to noise. 
\item\textbf{The cell is hyperpolarized, in the absolute refractory regime,} (region $B$ or $A$ too far from region $C$ to be lead in $C$ by an increase of $G_A$). Here, the cell cannot be excited no matter how it is perturbed.
SACs in this state constitute therefore an \textit{impassable barrier where no wave can go through}. 
\item\textbf{The cell is hyperpolarized, in the relative refractory regime.} Here, the cell can actually be excited thanks to different modalities, which can combine. First, because the cell feels an increasing $G_A$ it moves vertically in the bifurcation map and can be eventually lead to region $C$. Second, it can be excited by noise, when it is in region $D$, either in the rest state or on its way back to rest in the  excitation-hyperpolarisation  cycle of Fig. \ref{Fig:CRS}. This last case is the most interesting. Indeed, as discussed in the previous section, when a cell is moving in region $D$ the probability to become excited due to noise depends dramatically on its position in $D$ which is moving since the value of $G_S,G_A$ slowly evolve in time, with different time scales. Wave propagation in this regime is essentially driven by noise.
 \end{enumerate}

From this simple description we anticipate that there is a minimal value $g_{A_0} \equiv g_{A_0}\pare{G_S,\eta}$ for $g_A$, the maximum Ach conductance, below which a wave cannot propagate. This value is computed in section \ref{Sec:Transport}.

We have here the core mechanism of waves propagation and stop. When a SAC starts to burst it excites its post-synaptic neighbours. This corresponds, for those excited neighbours,  to an increase in $G_A$, which either increases their probability to be excited by noise in region $D$ or, if $g_A$ is large enough, leads them to region $C$. This way, one can generate a chain reaction, a retinal wave. Now, due to the sAHP activation, excited SACs eventually leave the bursting phase and go to an hyperpolarized state where they cannot be excited any more during a few tens of seconds. 
Thus, a wave leaves, behind it, an \textit{absolute refractory region where the forthcoming waves cannot go through for a long period}. This induces spatio-temporal waves interactions where each wave has to propagate into a transient landscape of excitable/absolute refractory/ relative refractory cells and can collide. Before addressing the detail of waves propagation let us finish this section with two important remarks.

\subsection{Link to forest fires}\label{Sec:ForestFires}
This dynamics bares some analogy with forest fires \cite{bak-chen-etal:90,chen-bak-etal:90,drossel-schwabl:92,grassberger:02}, as discussed in \cite{hennig-adams-etal:09,lansdell-ford-etal:14}. However, there is fundamental difference, illustrated in Fig. \ref{Fig:AchProfile_NIB_2cells}: the mutual non linear coupling between cells prolongs the bursts, a feature which is definitely not present in forest fire models. Actually, this prolongation effect has been reported in Zheng et al experiments \cite{zheng-lee-etal:06}. It is possible to block Ach receptors with a chemical cocktail. In the presence of this cocktail the burst duration is reduced. Another consequence of this non linear coupling, observed as well in experiments, is to increase the sAHP duration, hence, the persistence of refractive zones, with strong consequences on waves propagation. Therefore, due this non linear interaction, retinal waves do not reduce to "classical" paradigm of forest fire as proposed e.g. in \cite{bak-chen-etal:90,drossel-schwabl:92} or in \cite{hennig-adams-etal:09,lansdell-ford-etal:14} for retinal waves.

 However, the so-called "fire-diffuse-fire" model, appearing for intracellular calcium waves and introduced in \cite{dawson-keizer-etal:99} actually present some important features of what we observe in our model: bi-stability, continuous versus saltatory propagation depending on a threshold for a characteristic parameter. 
 The mathematical study presented in \cite{thul-smith-etal:08}, exhibits conditions to have this "fire-diffuse-fire" in a spatial landscape with bidomains and, in particular, the wave speed is computed. More generally, intracellular calcium waves share many properties with retinal waves: the necessity of an initial nucleus to seed the wave, the spread of excitability and the emergence of refractory domains, the effect of stochasticity (which renders calcium waves functionally robust and adaptive to changing environmental conditions \cite{thurley-skupin-etal:12}).
In contrast to these works, focusing on Ca dynamics modelled by phenomenological equations of reaction-diffusion type, in our work Ca dynamics is part of a process involving bursting at fast time scale and Ach coupling at medium scale: cholinergic waves generate calcium waves which are related to the existence of refractory sAHP domains. 
Therefore, a possible modelling would be to couple our Ach dynamics to calcium waves shaping the sAHP landscape. This opens up a way to interesting studies at least on the numerical side, the mathematical analysis being seemingly (at least to us) out of reach.

\subsection{The typical pathway of a SAC in the $G_S,G_A$ plane depends on $g_A$}\label{Sec:Pathway}

\begin{figure}
\resizebox{1.2\textwidth}{0.25\textheight}{
\includegraphics[]{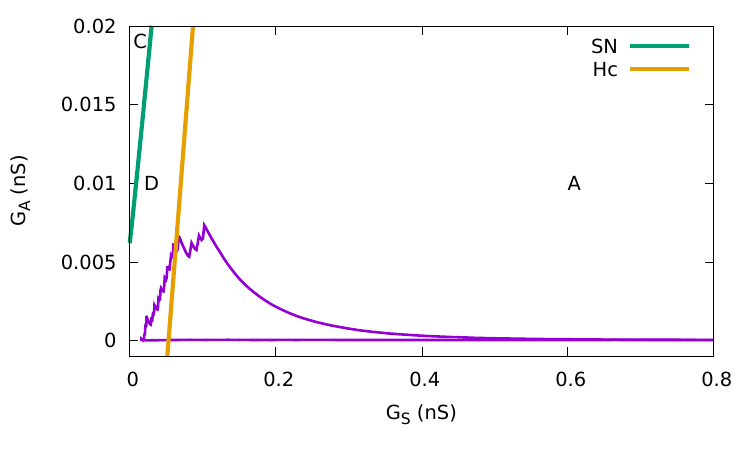} 
\hspace{0.05cm}
\includegraphics[]{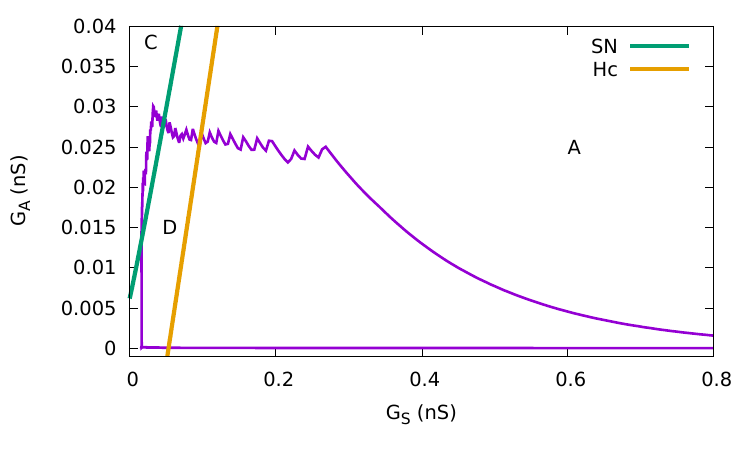}
\hspace{0.05cm}
\includegraphics[]{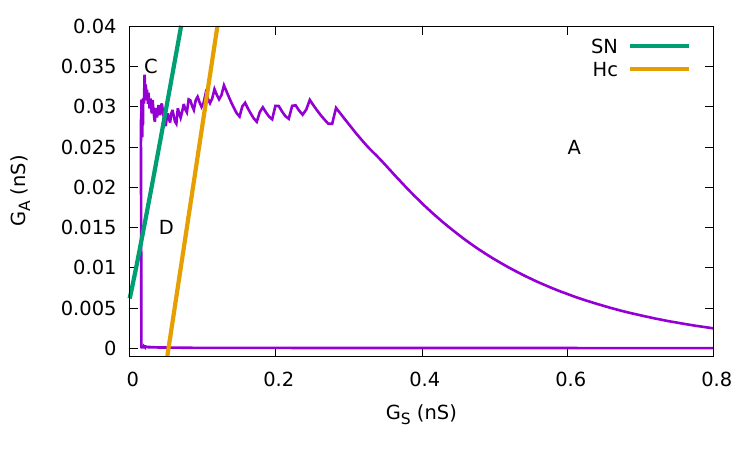}
\hspace{0.05cm}
\includegraphics[]{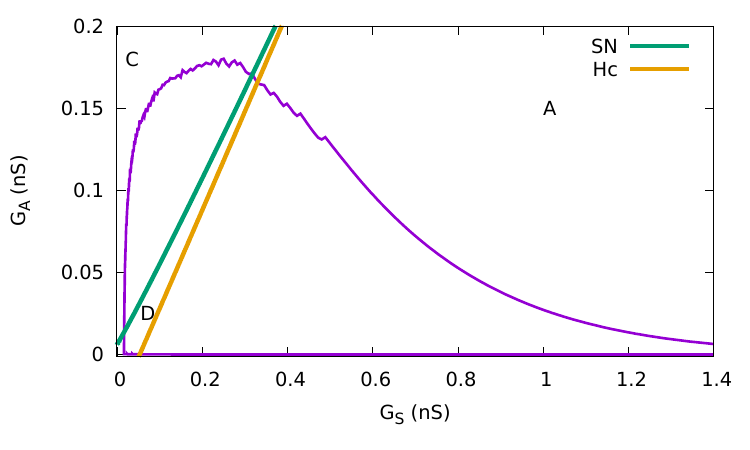}
}
\resizebox{1.2\textwidth}{0.25\textheight}{
\includegraphics[]{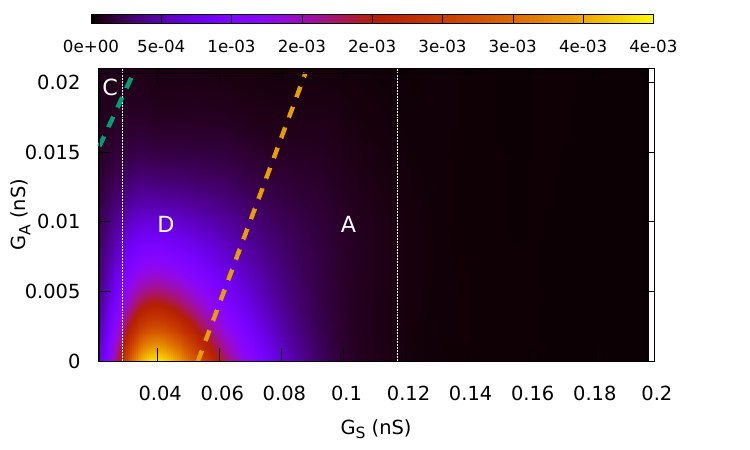}
\hspace{0.05cm}
\includegraphics[]{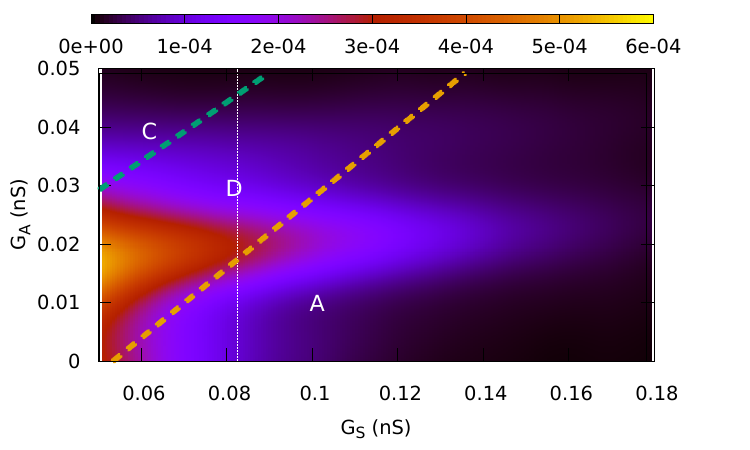}
\hspace{0.05cm}
\includegraphics[]{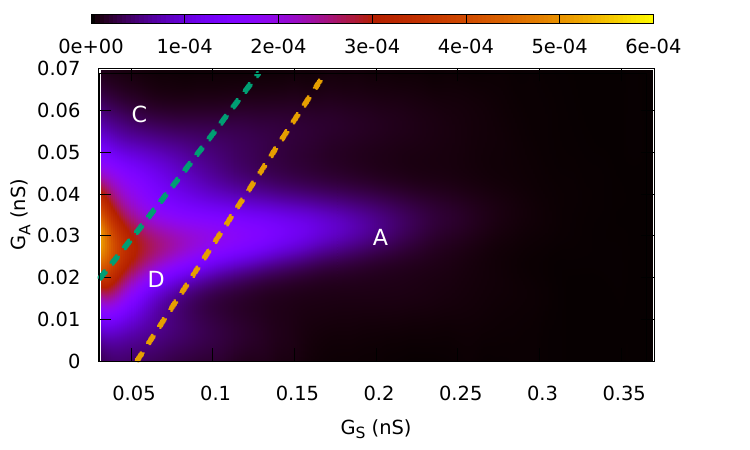}
\hspace{0.05cm}
\includegraphics[]{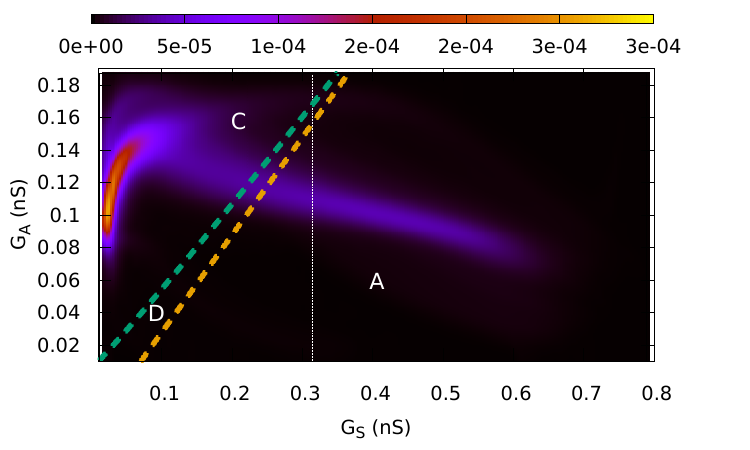}
}
\caption{\scriptsize\textbf{Top.} Example of trajectory of a bursting cell in the plane $\set{G_S,G_A}$ for, from left to right, $g_A=0.04,0.08,0.1,0.2$ nS. Note the oscillations in $G_A$ coming from the bursting of neighbours even if the cell itself is not bursting (e.g. region $A$). Green and orange lines respectively correspond to SN and Hc bifurcation.  \textbf{Bottom.} Probability density
of cells state in the plane $\set{G_S,G_A}$ for the same $g_A$ values as in Fig. \ref{Fig:Propagation}. Color bar on top indicates the corresponding value of probabilities (estimated on boxes with size $1 \, pS^2$ in the plane $\set{G_S,G_A}$).  Note that the scales are different in the different figures. Green and orange lines respectively correspond to SN and Hc bifurcation.  
\label{Fig:Trajectory_GS_GA}
}
\end{figure}

A SAC involved in a wave undergoes a pseudo-cyclic pathway in the plane $\set{G_S,G_A}$,  driving it from rest (in region $D$ of Fig. \ref{Fig:BifurcationDiagram_GS_GA}) to burst (limit cycle of region $D$ or of region $C$) to region $A$ or $B$ during hyperpolarization. Then, the cell returns back to region $D$ where it is likely to be reactivated by noise, with a probability depending on its location in region $D$ (Fig. \ref{Fig:PBurst}). The important remark now is that the time spent by the pseudo-cyclic pathway in the different regions depend on $g_A$, the maximal Ach conductance, as shown in Fig. \ref{Fig:Trajectory_GS_GA} top.
At the bottom row of this figure we have displayed heat maps, showing the probability of occupancy of SACs in the plane $\set{G_S,G_A}$ during waves, with labels corresponding to regions $A,C,D$ and lines corresponding to bifurcations frontiers. 

We observe that for low values of $g_A$ (left-most column, corresponding to regime I introduced below), bursting cells are mostly located in region $D$. 
Here, bursting is induced by noise where two cells have little chances to burst at the same time. In contrast, in the central columns $2,3$ (corresponding to regime II introduced below) cells are in-between region $D$ and region $C$ where Ach production is high enough to provoke cells synchronization and waves. As $g_A$ grows, cells have a decreasing probability to be in region $D$ and wave propagation becomes less sensitive to noise. Finally, the right most column (regime III) corresponds to cells mostly located in region $C$, with high Ach production, inducing a high global activity.

\section{The effect of varying Ach coupling during development} \label{Sec:VaryingAch}

In this section we analyse numerically waves dynamics as $g_A$, the maximal Ach conductance, varies.

\subsection{Ach coupling decays during development} \label{Sec:AchDecays}

We extrapolate here the potential behaviour of $g_A$ during development from Fig 3B in Zheng et al, \cite{zheng-lee-etal:04}. 
 They puff Ach with a concentration $1$ $mM$ while they block synaptic transmission with $C_d^{2+}$. They monitor the induced Ach current and plot its maximum (arising after $\sim 2$s) as a function of the developmental time. They observe a clear decay. From this figure
 we extrapolated the table \ref{Tab:AchEVolution} below and the plausible behaviour of $g_A$. 

The Ach conductance of a nicotinic channel is proportional to $\frac{A^2}{\gamma_A + A^2}$ with $\gamma_A \sim 1 $ nM. Even if the Ach concentration resulting from the puff has been reduced when the maximum of the induced current is reached we may assume that the remaining Ach concentration is quite larger than $K_d$. This means that $\frac{A^2}{\gamma_A + A^2} \sim 1$ (saturation). Note that receptor desensitisation is quite
important during a puff, as the receptors are completely overwhelmed with transmitter, so this assumption seems reasonable.
Therefore, in our model the current resulting from the puff is $I_{puff}=-n g_A (V_{clamp} - V_A)$
where $n$ is the number of activated cholinergic synapses connecting a SAC, $V_{clamp}=-70$ mV (patch clamp in Zheng et al experiment) and $V_A=0$ mV. 
Thus $g_A= - \frac{I_{puff}}{n \, V_{clamp}}$. The number $n$ is difficult to estimate from experiments. From \cite{zheng-lee-etal:06} we may fix $n \sim 25-30$ so that  $g_A \sim \frac{I_{puff}}{2000}$, which gives us hints on the conductance per synapse (assuming actually that $n$ is constant during development).

This gives the following table of values for $g_A$, where we added, in the comments line, the corresponding stage of development (rabbit). Stage III corresponds to glutamatergic waves birefly commented in section \ref{Sec:Biophys}. What is important here is the huge decay of $g_A$ during development (2 orders of magnitude).
 \begin{table}[!htb!]
\begin{center}
    \begin{tabular}{ | p{2.6cm}| p{2.6cm} | p{2.6cm} |p{4cm} |}     \hline
    Date (days after birth) & $I_{puff}$ (pA) & $g_A$ (nS) & Comments \\ \hline
    -2 & $1500$ &  $0.765$ & Stage II \\ \hline
    -1 & $1150$ &  $0.587$ & Stage II\\ \hline
    1 & $900$ &  $0.459$ & Stage II \\ \hline
    2 & $800$ &  $0.408$ & Stage II\\ \hline
    3 & $700$ &  $0.357$ & Stage II \\ \hline
    4 & $500$ &  $0.255$ & Stage II \\ \hline
    5 & $220$ &  $0.112$ & Stage II \\ \hline
    8 & $120$ &  $0.061$ & Stage II \\ \hline
    10 & $80$ &  $0.041$ & Transition II-III \\ \hline
    11 & $50$ &  $0.025$ & Transition II-III \\ \hline
    14 & $50$ &  $0.025$ & Stage III \\ \hline
    17 & $30$ &  $0.015$ & Stage III, huge error bars \\  \hline
    18 & $20$ &  $0.01$ & Stage III, huge error bars\\  \hline
    19 & $20$ &  $0.01$ & Stage III, huge error bars\\  \hline
    21 & $10$ &  $0.005$ & Stage III, huge error bars\\  \hline
   \end{tabular}
\end{center}
\caption{Values of $g_A$ versus perinatal days, extrapolated from Fig. 3B in \cite{zheng-lee-etal:04}.}
\label{Tab:AchEVolution}
\end{table}

Beyond the important observation of a  rapid decay of the cholinergic synapses during maturation, this table provides us hints to explore the central topic of this paper: \textit{how could stage II retinal waves dynamics be impacted by a continuous change in the cholinergic coupling ?} In the next sections we address in more detail the mechanism of wave
propagation and interaction in a random, slowly evolving, landscape of $G_S,G_A$. We also study numerically the wave sizes and durations distribution. As we show, it strongly depends on the maximal intensity of Ach coupling $g_A$.
The values of $g_A$ displayed in Table \ref{Tab:AchEVolution} are actually only indicative on the modelling side, as we are considered, in simulations, a nearest neighbours topology different from the real connectivity. Nevertheless, we tune model parameters to have stage II waves in a similar range of values for $g_A$. Especially, the wave speed we observe are in agreement with experimentally measured values \cite{maccione-hennig-etal:14}, (see Fig. \ref{Fig:Propagation}) .

\subsection{The setting of numerical simulations} \label{Sec:NumSim}

Let us start with a few operational definitions used for numerical investigations. 
We consider a wave as a \textit{causal} event. The bursting of a cell at time $t$ is either spontaneous, or caused by bursting cells \textit{connected} to it. As a consequence, a wave is mathematically a list of events of type ($t_{start}$,$S_t$). $S_t$ is a subset of bursting cells at time $t$ defined by the following constraints: 
\begin{enumerate}[(1)]
\item The wave starts from a cell, or a group of connected cells, $S_{start}$ starting to burst at time $t_{start}$.
\item For $t>t_{start}$, $S_t$ must intersect $S_{t-dt} \cup \partial S_{t-dt}$, where $\partial S_{t-dt}$ is the neighbourhood of $S_{t-dt}$. This mathematically corresponds to causality and connectedness. 
\item The wave stops at the last time t where $S_t$ is non empty. This time is called $t_{end}$.
\end{enumerate}
Therefore, two disjoint clusters $S_t, S'_t$ (i.e. with no common cell even at the boundary) correspond to two \textit{distinct} waves. When two waves intersect we still consider them as distinct, they do not penetrate each other. Thus, from this perspective, they do not spread any more at the intersection points. 

We call "size" the total number of cells involved in a wave, i.e. $s=\sum_{t=t_{start}}^{t_{end}} \#S_t$, where $\#S_t$ is the cardinal of the set $S_t$. The sum is discrete because time is discrete in simulations. The duration of the wave is $D=t_{end}-t_{start}$.
The \textit{global activity,} noted $n$, is the \textit{total} number of active cells at a given time. 

All our simulations are limited to a one dimensional space. This is due to several reasons. First, considering the huge difference between the fastest and slowest time scales in the model, simulations in 2D, for reasonable sizes ($\sim 100 \times 100$  cells) are enormously time consuming. In addition, the definition of waves boundaries is easy in 1D (boundaries are points), while these are irregular frontiers in 2D, requiring sophisticated algorithms. As we show in the next section, it is actually important to separate waves when studying their statistics, instead of considering the total number of active cells per unit time, independently of their localisation. Finally, these simulations are intended to illustrate the main content of this paper, which is theoretical. 
We discuss about the effect of dimensionality and connectivity in section \ref{Sec:SOC}. %

In the simulations we considered characteristic times $\tau_S$ and $\tau_R$, which fix the time scale of sAHP, of order $10$ s, while, in experiments, they are rather of order a minute. We did that to reduce the computational time, especially, the refractory times between successive waves.

\subsection{Waves dynamics and scaling}\label{Sec:WavesInteraction}

We now illustrate the effect of increasing $g_A$ on waves dynamics in our model. We essentially observe $3$ regimes, already commented above and now  more precisely defined and illustrated by Fig. \ref{Fig:Regimes}, \ref{Fig:Propagation}, \ref{Fig:Scaling}. Note that we used the labelling I,II,III for these regimes, not to be confused with the retinal waves stages I,II,III. Our analysis sticks at stage II retinal waves.
\subsubsection{Quantitative description of the regimes}\label{Sec:QuantDescRegimes}

\begin{figure}
\centerline{
\resizebox{\textwidth}{0.3\textheight}{
\includegraphics[]{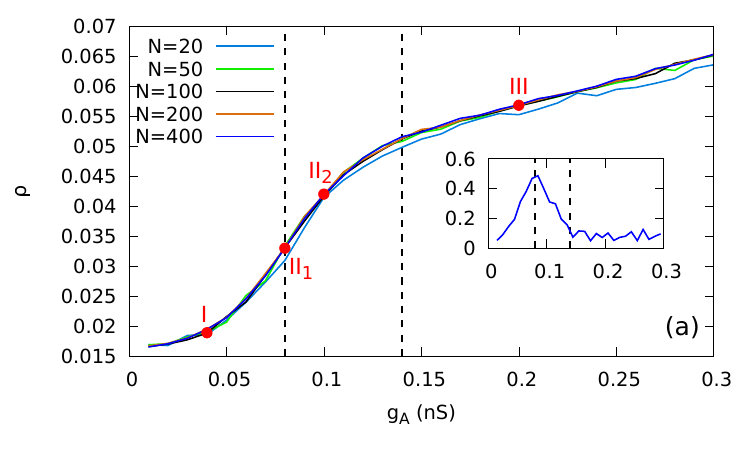}
\hspace{0.2cm}
\includegraphics[]{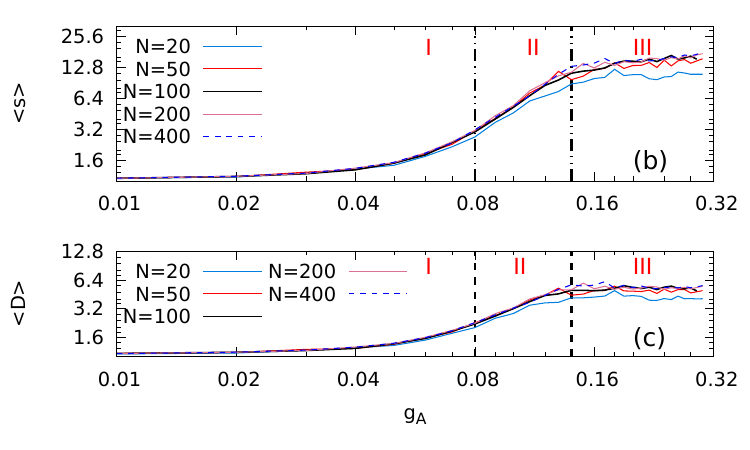}
\hspace{0.2cm}
\includegraphics[]{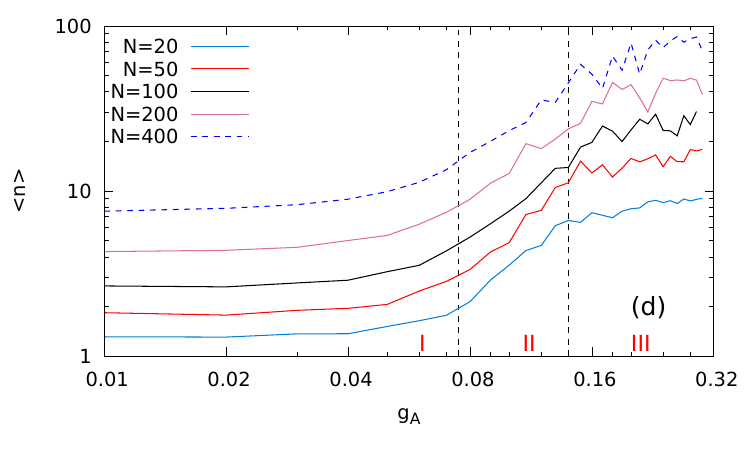}
}
}

\caption{\textbf{Left (a).} Probability $\rho$ that a cell is bursting at a given time as a function of $g_A$, for $N=20$, $50$, $100$, $200$, $400$. The inset shows the numerical derivative of $\rho$ for $N=400$. 
The maximum of the derivative defines the transition between regime I and II (leftmost dashed line). This transition is further explained in section \ref{Sec:Transport}. The second dashed line, roughly defines the transition between regime II and III and 
arises at $g_A \sim 0.14$ nS. Red points with labels correspond to specific values of $g_A$ : one in regime I, two in regime II ($II_1$ at the transition between I and II and $II_2$) and one in regime III, used for representative plots in the subsequent figures. 
\textbf{Center.} Average wave size, $\av{s}$, label (b), and duration, $\av{D}$, label (c), in seconds, as a function of $g_A$, for the same values of $N$ (in logscale). Vertical black lines correspond to the separation in $3$ regimes.
 \textbf{Right (d).} Mean activity, $\av{n}$, (see text for a definition) in the same conditions.
 Note that size and duration do not depend on $N$, the lattice size, in contrast to $n$, which, actually grows further for larger values of $g_A$ (not shown).
\label{Fig:Regimes}
}
\end{figure}

\begin{itemize}
\item[\textbf{Regime I.}] \textbf{Localized waves}. When $g_A$ is small, cells can be excited by noise but the Ach they produce does not increase sufficiently the probability of excitation of their neighbours to induce a large scale propagation.
\item[\textbf{Regime II.}] \textbf{Waves are competing through the sAHP landscape.} When $g_A$ further increases,  a bursting SAC increases enough the probability of this neighbours to be excited so that some neighbours become eventually excited. The mutual excitation drives the cells in a robust synchronisation regime, as illustrated in Fig. \ref{Fig:AchProfile_NIB_2cells}. Waves can propagate but they are blocked by the sAHP landscape left by previous waves. Here, therefore, waves size is dramatically dependent on the waves history. In this regime waves characteristics (size, duration, activity) are quite sensitive to $g_A$ and their probability laws have  a large tail. 
\item[\textbf{Regime III.}] \textbf{Waves are cooperating through Ach coupling}. When $g_A$ is large enough, waves may eventually spread through the whole lattice. In this regime, waves born at different times, from different sites, eventually collide with the effect of prolonging the burst durations. When $g_A$ increases further, one is progressively driven to a full synchrony regime where all cells fire synchronously with a period fixed by sAHP characteristic times ($\tau_S,\tau_R$).
\end{itemize}
These 3 regimes are first roughly indicated in Fig. \ref{Fig:Regimes}, where we plotted, in Fig.  \ref{Fig:Regimes}(a) (left), the probability $\rho$ that a cell is bursting at a given time, as a function of $g_A$ for $N=20$, $50$, $100$, $200$, $400$. We essentially see abrupt changes in the slope of $\rho$, allowing to (broadly, for the moment) locate these $3$ regimes (Inset of Fig. \ref{Fig:Regimes} (a)). The separation between regime I and II holds at the unique point where $\frac{\partial \rho}{\partial g_A}$ is maximum, that is the point where the probability $\rho$ that a cell bursts at a given time has the maximum sensitivity to variations of $g_A$. Regime III is the range of $g_A$ values where $\rho$ increases linearly. 
We have also indicated, in red, 4 points corresponding to the three regimes (with one point at the transition between regime I and II), used to characterize in more details waves behaviour. 
In Fig. \ref{Fig:Regimes}, center, we have plotted the average wave size (label (b)) and duration (label (c)) as a function of $g_A$, for the same values of $N$. Recall (section \ref{Sec:NumSim}) that we call "size" the total number of cells involved in a wave, where a wave is composed of \textit{connected} bursting cells. The global activity (total number of cells bursting at a given time) is plotted on the right of Fig. \ref{Fig:Regimes}.

Similar regimes have been reported in the literature, especially in \cite{hennig-adams-etal:09,lansdell-ford-etal:14}. What these studies neglected though is the role of closeness to bifurcations, non linear feedback, and their consequences on waves propagation. 

\subsubsection{Waves propagation in the different regimes}\label{Sec:WavesPropRegimes}

\begin{figure}
\resizebox{1.1\textwidth}{0.25\textheight}{
\includegraphics[]{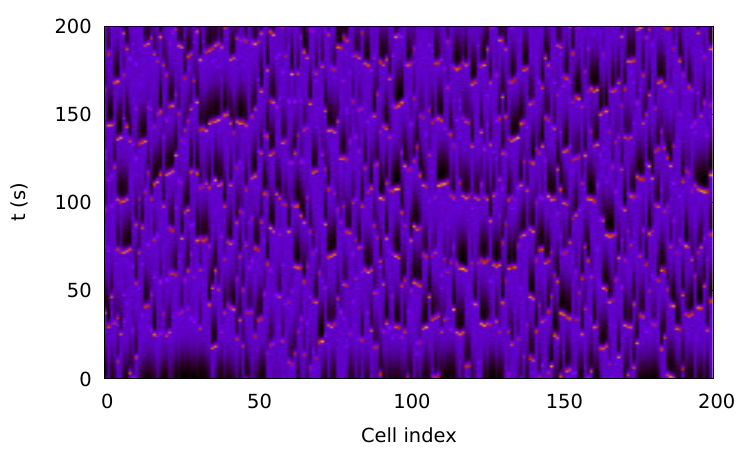}
\hspace{0.1cm}
\includegraphics[]{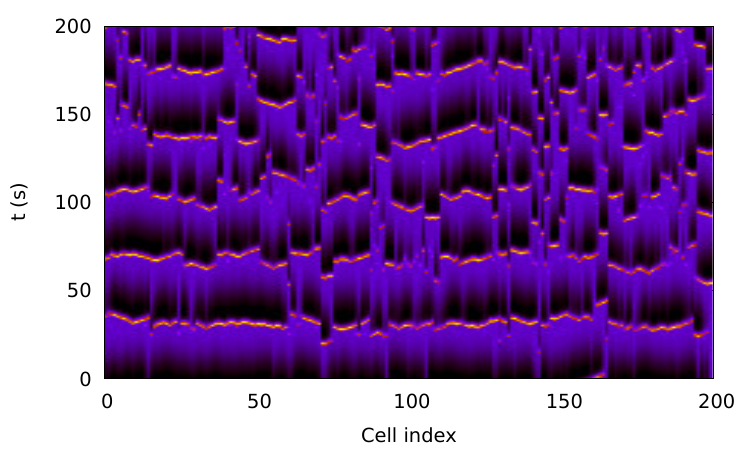}
\hspace{0.1cm}
\includegraphics[]{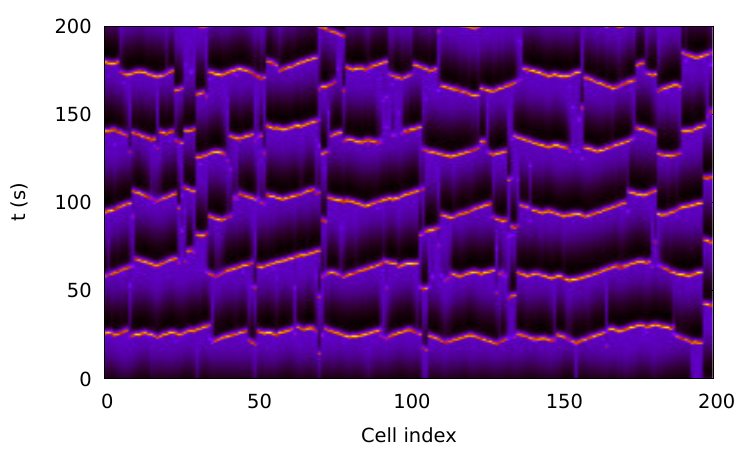}
\hspace{0.1cm}
\includegraphics[]{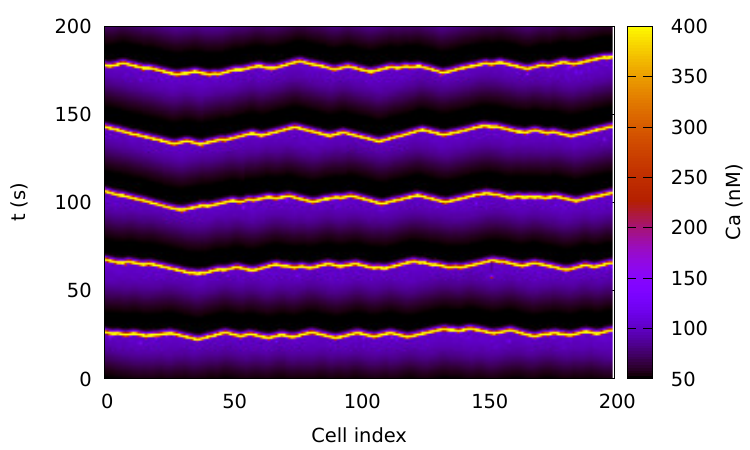}
}
\resizebox{1.1\textwidth}{0.25\textheight}{
\includegraphics[]{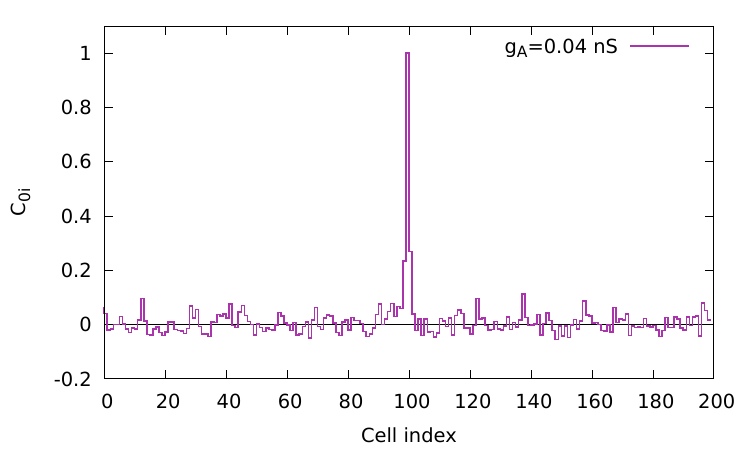}
\hspace{0.1cm}
\includegraphics[]{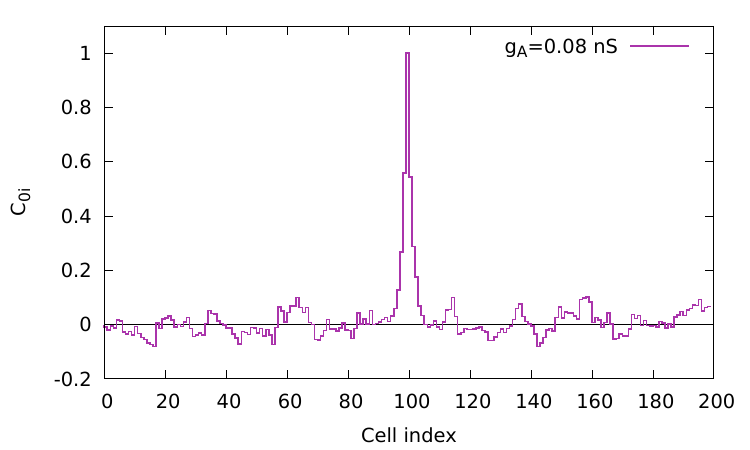}
\hspace{0.1cm}
\includegraphics[]{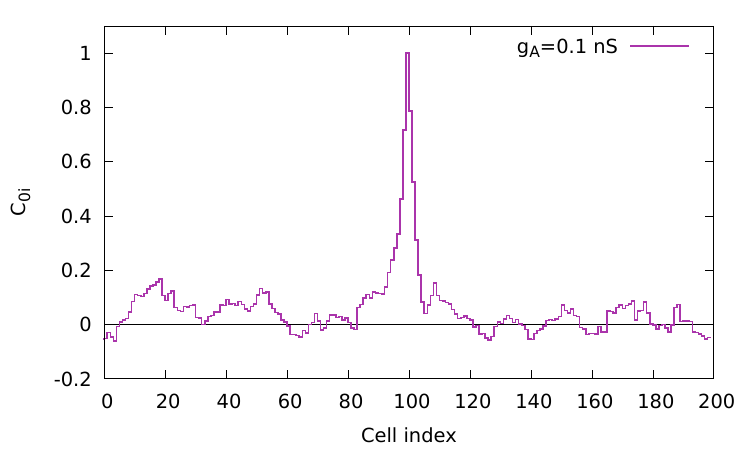}
\hspace{0.1cm}
\includegraphics[]{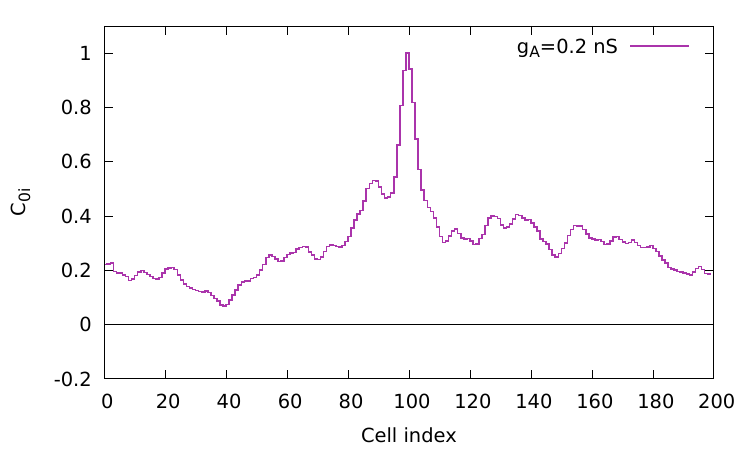}
}
 \caption{\textbf{Spatial characterization of the $3$ wave regimes. Top. Propagation of waves} in the presence of noise ($\eta=6  \, pA \, ms^{-\frac{1}{2}}$) for $200$ cells illustrated by the evolution of calcium concentration.   
\textbf{From left to right}: $g_A=0.04$ nS (regime I); $g_A=0.08$ and $g_A=0.1$ nS (regime II, point $II_1$, $II_2$ in Fig. \ref{Fig:Regimes}); $g_A=0.2$ (regime III). Hot zones (orange, yellow) are active (bursting) zones; black zones are absolute refractory zones; purple zones correspond to SACs in the rest state. Propagating waves are stopped by absolute refractory regions. 
\textbf{Bottom. Normalized calcium correlation} $\frac{{\cal C}(i)}{{\cal C}(i_c)}$ between the central cell and other cells for the same values of $g_A$. There are zones of positive correlation and negative correlations (see text).
 \label{Fig:Propagation}
 }
 \end{figure}

This propagation is illustrated in
Fig. \ref{Fig:Propagation}. From left to right each column corresponds respectively to regime $I$ ($g_A=0.04$ nS); regime $II$ ($g_A=0.08$ nS corresponding to the point $II_1$ in Fig. \ref{Fig:Regimes} (a) and $g_A=0.1$ nS, point $II_2$ in Fig. \ref{Fig:Regimes} (a)); regime $III$ ($g_A=0.2$ nS). The top row shows the time evolution (vertical axis) of calcium concentration in the lattice (horizontal axis) in these different regimes. We can see the initiation and propagation of waves.  The bottom row shows the
normalized correlation of calcium activity of the center cell (index $i_c$) with another cell in the lattice (index $i$).
The correlation is numerically computed as: 
\begin{equation}\label{eq:CorCa}
{\mathcal {C}}(i)=\frac{1}{T} \int_0^T C_i(t) \, C_{i_c}(t) \, dt - \frac{1}{T} \int_0^T C_i(t) \, dt \, \frac{1}{T} \, \int_0^T C_{i_c}(t) \, dt,
\end{equation}
where $T$ is the total duration of the simulation, $i_c$ is the label of the center cell, and  $i$ the label of a cell in the lattice. In Fig. \ref{Fig:Propagation}, bottom row, we have displayed the normalized correlation $\frac{{\mathcal C}(i)}{{\mathcal C}(i_c)}$ so that the maximum is $1$.\\

In regime $I$, (left column) the entrance of a SAC in the bursting regime initiates a propagation which ends very soon. This is because, for this regime of parameters, the Ach current, even at its maximum, is not sufficient to cross the $S_d$ barrier. Only noise affords it, with a low probability, depending on noise intensity and on $g_A$. As a consequence, $\rho$, $\av{s}$, $\av{D}$, $\av{n}$ grow slowly with $g_A$. In regime I the calcium correlation decays rapidly with the distance to the center cell: cells are essentially decorrelated.

On the opposite, in regime III, (right column), top, we see waves spreading through the whole lattice. Waves appear almost periodically with a frequency $\sim 35$ s, essentially constrained by sAHP dynamics (i.e. the return time to region $D$ dependent on the times $\tau_S,\tau_R$) and the time necessary to activate a cell and, thereby, restart a wave. In this regime $\rho$ grows slowly, and $\av{s}$, $\av{D}$, $\av{n}$ saturate. The saturation of $\av{s}, \av{D}$ is due to our modelling choice: when two waves intersect we consider them as distinct. Thus, they do not spread any more at the intersection points. As a consequence, these quantities do not depend on the lattice size. 
The plateau is actually determined by the probability to start a wave and the wave speed. 
In this regime, $n$, the total activity, behaves in a different way. One expects the total number of active cells to increase with the lattice size, but, as we see in figure \ref{Fig:Propagation}, all cells in the lattice are not necessarily simultaneously active. As $g_A$ further grows, however, one achieves a full synchronisation where all cells are active. These full synchrony regime is periodic and alternate of full lattice refractoriness.  
Finally,  the instantaneous calcium correlation is strictly positive throughout the lattice, confirming that, in this regime, the correlation length is the lattice size. 

 The intermediate regime II is shown in Fig. \ref{Fig:Propagation}, middle columns. 
On the top row, we see how waves propagation has to cope with the sAHP landscape left by the previous waves. Spreading waves are stopped by the sAHP trace left by previous waves.
Therefore, waves have a widespread distribution (see also Fig. \ref{Fig:Scaling}).
Calcium correlation also shows an interesting profile. It reveals that correlation alternate from positive to negative as the distance to the center cell increases. Positive correlations correspond to cells in the same state as the central cell $i_c$ (i.e. bursting when $i_c$ is bursting; hyperpolarized when $i_c$ is hyperpolarized), whereas negative correlations correspond to cells in the opposite states. Thus, calcium correlations is another way to reveal the existence of refractory/active domains shaping waves propagation and interactions. Increasing $g_A$ from regime I to regime III via regime II leads to an increase of the correlation length, which has the size of the lattice in region III. 

In these figures one clearly see what has been announced in the beginning of this section. In regime II waves interact via their sAHP landscape, thereby obstructing each other. In contrast, in regime III they have a direct interaction and the mutual Ach feedback prolongs the cells activity. The transition between these two regimes is reminiscent of a percolation transition, as anticipated in \cite{hennig-adams-etal:09}. However, the transition we observe holds at a specific $g_A$ value and is therefore not robust to changes of $g_A$ during development. This aspect is further discussed in section \ref{Sec:SOC}.


\subsubsection{Waves statistics}\label{Sec:WavesStats}
\begin{figure}
\resizebox{1.1\textwidth}{0.25\textheight}{
\includegraphics[]{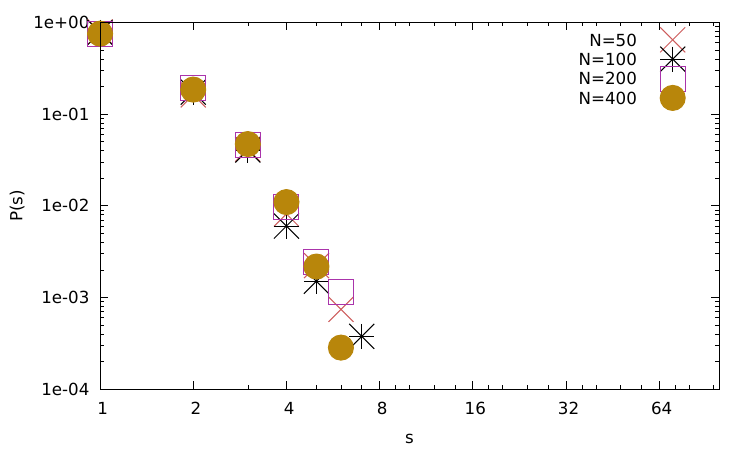}
\includegraphics[]{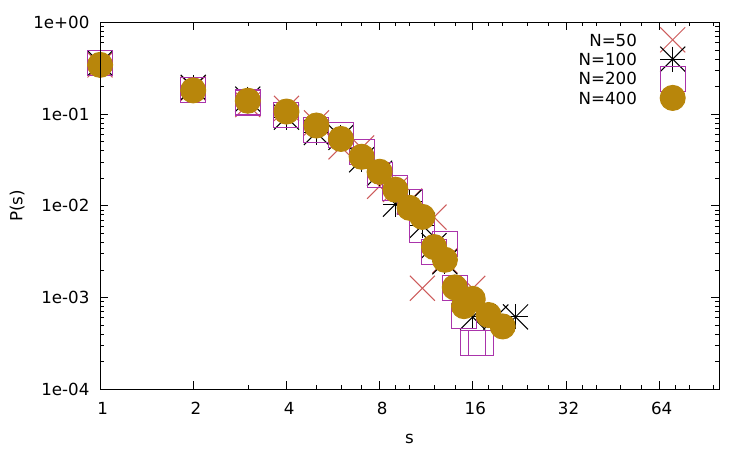}
\includegraphics[]{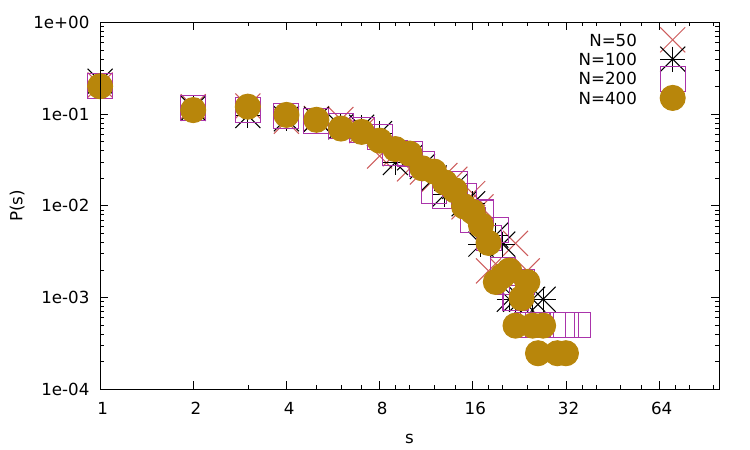}
\includegraphics[]{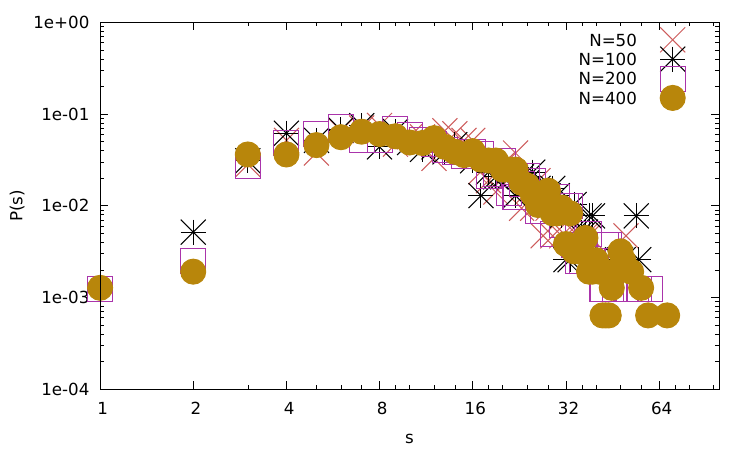}
}
\vspace{0.2cm}
\resizebox{1.1\textwidth}{0.25\textheight}{
\includegraphics[]{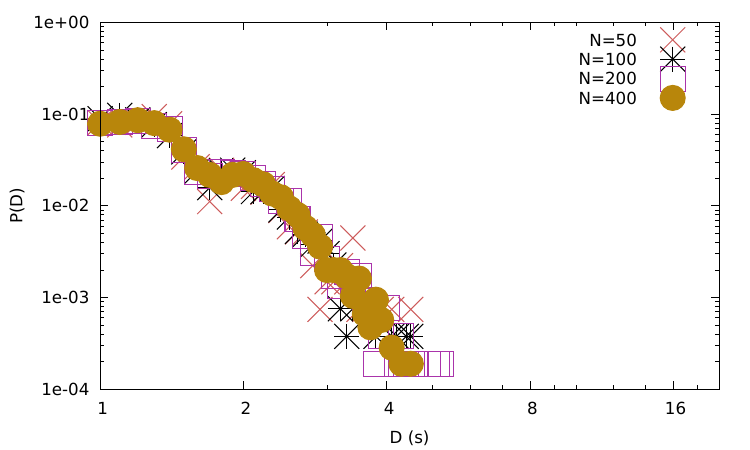}
\includegraphics[]{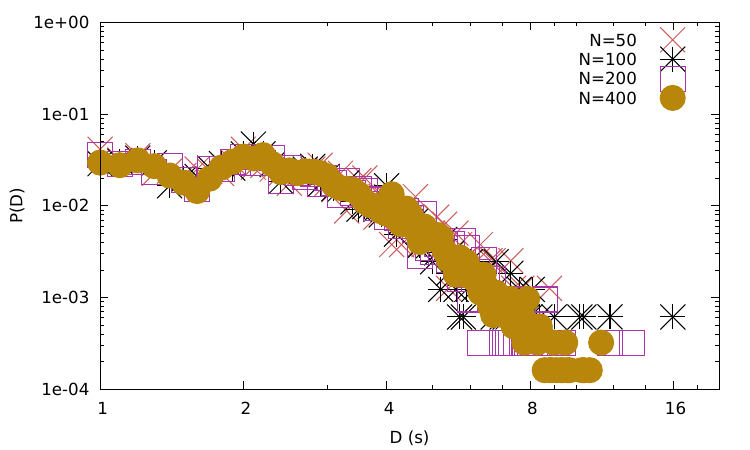}
\includegraphics[]{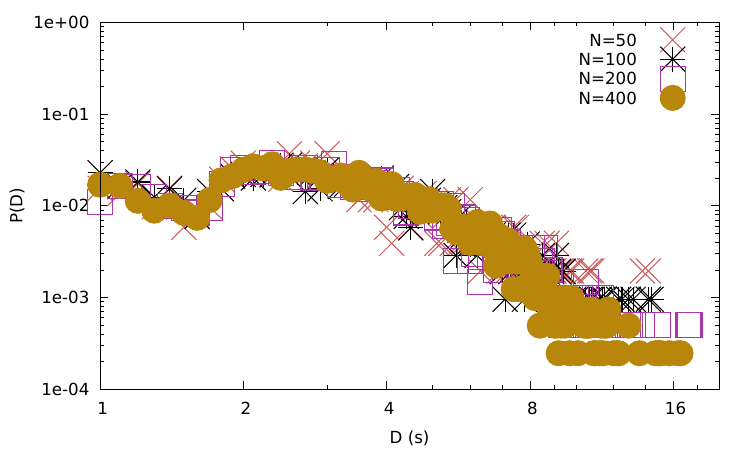}
\includegraphics[]{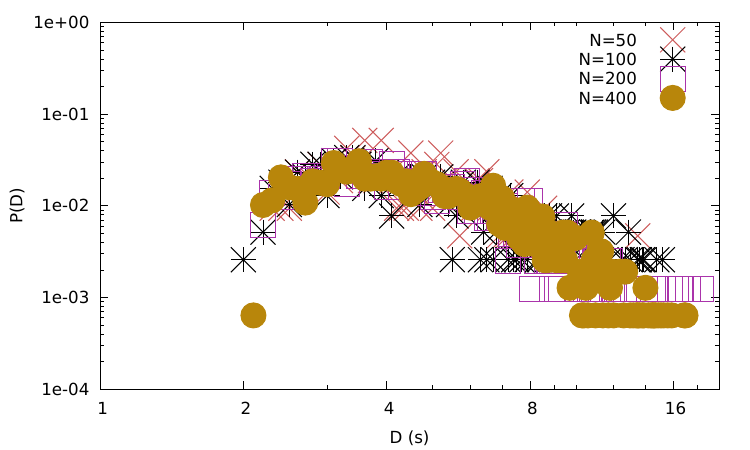}
}
\vspace{0.2cm}
\resizebox{1.1\textwidth}{0.25\textheight}{
\includegraphics[]{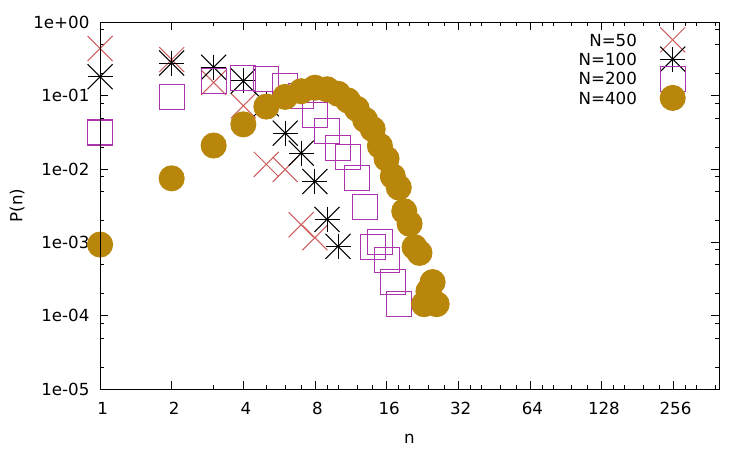}
\includegraphics[]{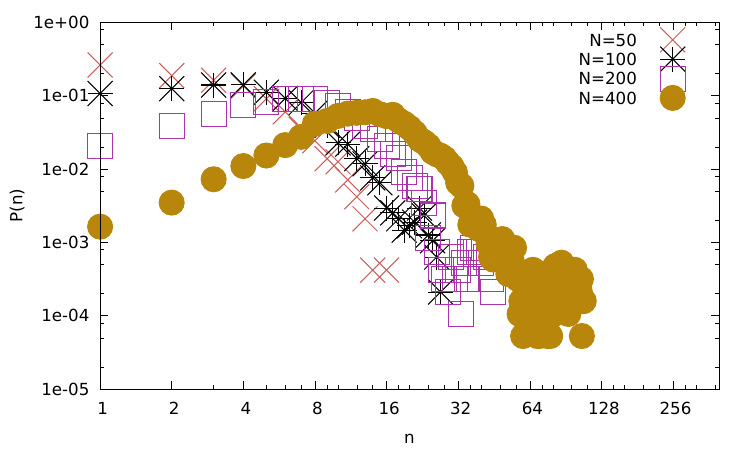}
\includegraphics[]{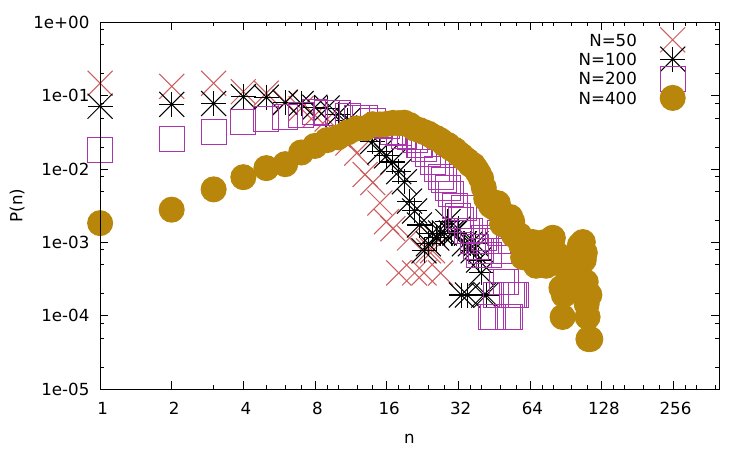}
\includegraphics[]{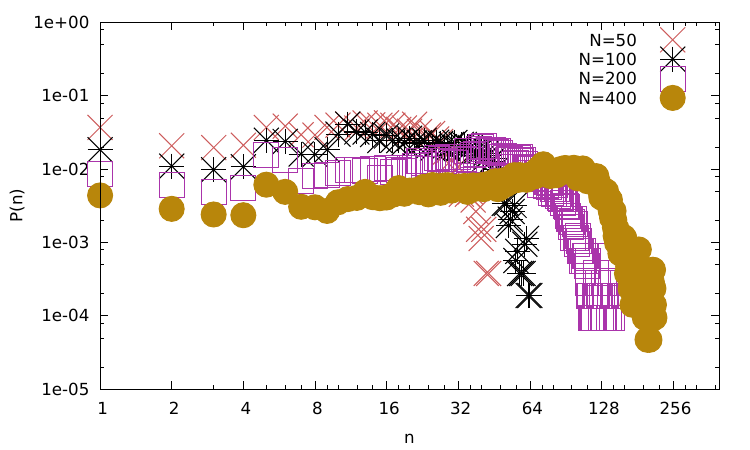}
}
\caption{\textbf{Probability distributions} of waves quantities for different values of $N$ in the $3$ regime corresponding to points I, IIa, IIb, III in Fig. \ref{Fig:Regimes} (from left to right).  \textbf{Top.} Sizes distribution; \textbf{Middle.} Durations distributions; \textbf{Bottom.} Activities distributions. 
\label{Fig:Scaling}
}
\end{figure}

A synthesis of the observed waves statistics is shown in Fig. \ref{Fig:Scaling}, where we plot the probability distribution of $s$ (top row), $D$ (middle row), $n$ (bottom row), for $N=50,100,200,400$, in the $3$ regimes. As before, each column corresponds to a regime ($I$, $II_1$, $II_2$, $III$ from left to right) The plots are in log-log scale to investigate potential power laws. We observe first that the distribution of $s$ and $D$ do not depend on $N$, as expected from waves interactions. In contrast, $P(n)$ does. One also remarks that the shape of these probabilities evolves with $g_A$. Let us interpret what we observe thanks to the previous dynamical analysis and especially Fig. \ref{Fig:Trajectory_GS_GA}. 

In regime I, the bursting of cells is only induced by noise. When bursting, a cell produces Ach and thereby increases the probability that a neighbour cell bursts, due to noise. 
As a consequence, there is an important probability  to observe waves of size $1$ and quite a smaller probability to observe waves of larger size. The probability distribution of waves duration $D$ is more interesting. After a flat maximum, for $D \leq 1.2$ s, this probability  decreases to a local minimum, at $D \sim 1.8$ s before increasing to a secondary maximum, at $D \sim 2$ s. The first maximum corresponds to waves of size $1$. Their duration probability decreases as the duration increases. However, waves of size $1$ with a sufficiently large duration produce more Ach (in a time scale of the order $1-2$  s, a few $\mu_A^{-1}=0.53$ s) thereby increasing the probability to have a neighbouring cell that starts to burst, thanks to noise. But, from Fig. \ref{Fig:AchProfile_NIB_2cells} we know this neighbouring bursting cell will increase the burst duration of the initial cell. This explains the second peak. In other words, the local minimum in the distribution marks a threshold above which the mutual coupling between two bursting neighbours prolongs the (size 2) waves duration. Finally, the probability distribution of the global activity has a bell shape with a maximum $\propto N \av{s}$.  Indeed, the probability of this variable is constrained, on one hand, by the number of coexisting waves and on the other hand by the joint probability that each of these waves has a certain size and that these sizes sum up to $n$. This explains the bimodal shape of the distribution\footnote{The probability of $n$ is related to the probability of sizes by $P(n)= \sum_{m=1}^{N} \sum_{s_1 + s_2 + \dots + s_m=n} P(s_1, \dots, s_m)$ where the sum $ \sum_{s_1 + s_2 + \dots + s_m=n} $ means the sum over all possible configurations with $m$ sets of connected active cells (waves),  each of size $s_i$, $i=1 \dots m$ with $s_1 + s_2 + \dots + s_m=n$, and with the constraint that the total number of cells is $N$. As, in regime $I$, waves are small and not interacting,  $P(n)= \sum_{m=1}^{N} \sum_{s_1 + s_2 + \dots + s_m=n} P(s_1) \dots P(s_m) $.Assuming, for the simplicity of the argumentation, that $P(s) \sim K \gamma^{-s}$ we obtain $P(n)=  \gamma^{-n} \sum_{m=1}^{N} K^m S_N(m,n)$ where $S_N(m,n)$ is the number of possibilities to make, over the set $\set{1 \dots N}$, m connected group with $s_i$ elements and with $\sum_{i=1}^m S_i=n$. We therefore see that $P(n)$ is a combination of a term decaying with $n$ multiplied by a combinatorial term $S_N(m,n)$ which first increases with $n$, then decays. \label{foot:P(n)}}. Actually, this bimodal shape is observed in each regime \footnote{In the other regimes the argument developed in the footnote \ref{foot:P(n)}, extends as we essentially need that the joint probability $P(s_1, \dots, s_m)$ decays sufficiently fast (power low or exponential) with $n$.}. 
%
%

In regime II, as illustrated in Fig. \ref{Fig:Trajectory_GS_GA}, there is a small probability for a cell to penetrate in region C where it bursts without requiring a noise kick. 
Waves propagate until they rich a sAHP barrier, limiting their size and generating a long tail in the distribution. There are still waves of size $1$ though, whose probability decreases with $g_A$. For the duration we still see the local minimum and second maximum, reminiscent of regime I, but beyond the second maximum waves duration probability has, like sizes probability, a longer and longer tail as $g_A$ increases. The global activity probability still has a bell shape, but it looks as a mix of two regimes where it growths linearly (in log-log scale) up to the maximum and decreases also linearly. The maximum appears therefore as a crossover point between two power laws regimes.

In regime III, the probability to have waves of size $1$ is very small. A bursting cell triggers a wave which propagates through the lattice until it meets another wave. 
The distribution of size is bell-shaped, with a maximum corresponding to our definition of waves size: although, in regime III, waves cooperate, when two waves meet their size does not increase. The durations distribution has a similar bell shape, starting, for $g_A=2$, around $2$ s with a maximum around $3$ s (the average duration being $5.6$ s).

There is no clear evidence that the probabilities of connected waves observables ($s$ and $D$) have a power law tail, although this tail is large, especially for duration which goes from $1$ to $20$ s. 
Actually, probability distributions might depend on the dimension and the connectivity, as developed in section \ref{Sec:SOC}.

%

\subsubsection{Transport}\label{Sec:sTransport}

Finally, we analysed the wave speed as a function of $g_A$. First, we remarked that, due to waves interactions and sAHP lanscape, waves are not necessarily propagating ballistically, that is $r(t) =  c.t^z$, where $r(t)$ is the wave radius, $z$ the anomalous diffusion exponent and $c$ the wave "speed" (which actually is the classical speed for $z=1$, whereas, it is e.g. the diffusion coefficient if $z=\frac{1}{2}$). Anomalous diffusion is further analysed in section \ref{Sec:Transport}. 

\begin{figure}[H]
\centerline{
\resizebox{0.8\textwidth}{0.3\textheight}{
\includegraphics[]{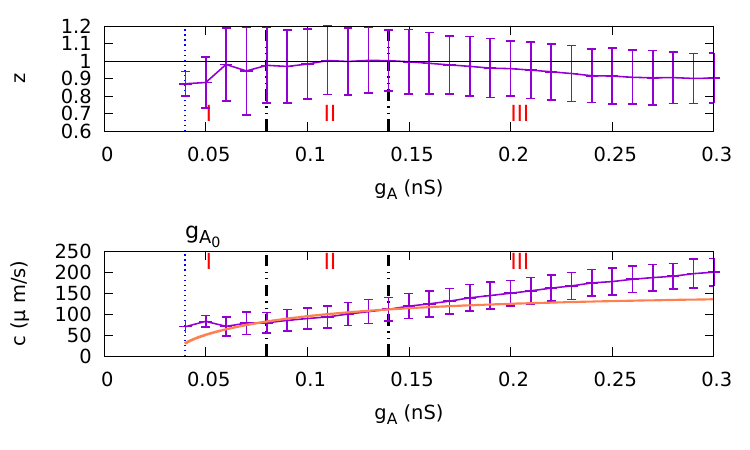}
}
}
 \caption{\scriptsize\textbf{Anomalous propagation} of retinal waves. The three regimes are separated by a black dashed arrow and are indicated with labels. The blue dashed line corresponds to $g_{A_0}=0.04$, the minimal value of $g_A$ below which waves cannot propagate (eq. \eqref{eq:g_0} below). \textbf{Top.} Anomalous diffusion exponent $z$ versus $g_A$ with error bars. The exponent in regime I is lower than $1$ corresponding to anomalous diffusion where propagation is mainly driven by noise. In regime II, $z=1$ corresponding to ballistic transport. It decreases in regime III illustrating direct waves interactions. Here, transport is ballistic too but the radius of a wave stops to grow, in our definition, when two waves collide, whereas duration is incremented. This explains why $z<1$.   \textbf{Bottom.} "Speed" $c$ i.e. proportionality coefficient in the relation  $x(t) = c \, t^z$ (see text). The orange line corresponds to the theoretical derivation \eqref{eq:speed}. The deviation from the theoretical curve in regime I and III is explained in the text. 
\label{Fig:Transport}
}
\end{figure}

We show a plot of $c,z$ as a function of $g_A$ in Fig. \ref{Fig:Transport}. Note that the estimation of these quanties is difficult in regime I as waves are very small. 
We observe that $z$ is (on average, with large error bars) smaller than $1$ in regime I and III corresponding to anomalous transport. This is due to different effects though. In regime I, waves are small and propagation is due to the effect of noise for cells essentially located in region $D$ (see next section and Fig.\ref{Fig:Trajectory_GS_GA}). Thus, wave propagation is slower than ballistic. In contrast, in regime III, waves interact together. When a wave intersects another wave its radius does not evolve any more whereas its duration is incremented. Thus, the effective exponent $z$ is lower than $1$. Although it might look as an artefact of our wave definition, this criterion actually clearly shows when direct wave interactions becomes significant. To illustrate this point we have added in the figure the lines separating the $3$ regimes where we clearly see how propagation is slowed down in regime I and III. 

We also plot, at the bottom of Fig. \ref{Fig:Transport}, the speed $c$, together with the theoretical computation, eq. \eqref{eq:speed} developed in section 
\ref{Sec:Transport}. 
Considering that SACs are spaced by $a \simeq 50 \, \mu m$ our speed would correspond to real waves speed in the range $\bra{50,200}$ $\mu m/s$ in agreement with experimentally observed ranges \cite{singer-mirotznik-etal:01,maccione-hennig-etal:14,sernagor-eglen-etal:00}.

\section{Transport equations}\label{Sec:Transport}

Up to now, we have shown how the non linear coupling between bursting cells and the recurrent return in region $D$
give rise to a rather rich dynamics with different regimes of waves. Still, the numerical illustrations have been done in one spatial dimension, with nearest neighbours interactions. So, one may ask whether a larger dimensionality or a different connectivity could notably impact the global dynamical picture. To avoid multiplying tedious numerical simulations with a larger range of possibilities when considering connectivity patterns \footnote{Even if the real SACs connectivity is known at maturation \cite{tauchi-masland:84}, we don't know about papers reporting how connectivity evolves during development.}, an alternative idea is to derive canonical transport equations, holding in $1$ and $2$ dimensions. \\
\indent In this section, we derive such a transport equation characterising  the retinal waves propagation in a landscape of sAHP. These equations are constructed in the spirit of the bifurcation diagram exposed in section  \ref{Sec:BifAnalysis}, with a main idea: waves transport Ach and sAHP \textit{conductances}, i.e. a wave is mechanism where $G_S$ and $G_A$, considered as \textit{fields}, propagate, obeying a transport equation that we derive. This is done taking into account the 
time scales separation between $G_S$ and $G_A$. From this approach we are able to compute a few important characteristics of waves dynamics, like speed in the ballistic regime.  
We also compare our transport equation to known equations in the non linear and statistical physics literature.

\subsection{Ach conductance transport} \label{Sec:AchCondTransport}

\subsubsection{Approximations} \label{Sec:Approximations}

The equation for Ach production of neuron $i$ is given by \eqref{eq:Achprod}.
We introduce the variable:
\begin{equation}\label{eq:Gamma}
\Gamma_i = \displaystyle{\sum_{k \in {\cal B}_{i}}} U(A_k)
\end{equation}
with $U(A)=\frac{A^2}{\gamma_{A}+A^2}$. Then, the Ach conductance of neuron $i$ at time $t$ 
is $G_A(i,t)=g_A \Gamma_i(t)$.

Differentiating $\Gamma_i$ with respect to time gives:
$$
\frac{d\Gamma_i}{dt} = \displaystyle{\sum_{k \in {\cal B}_{i}}} U'(A_k) \bra{- \mu_A A_{k}+\beta_{A} T_{A}(V_{k})}.
$$
The goal here is to obtain a closed-form equations for the time evolution of $\Gamma_i$, thus for $G_A(i,t)$. For this we are going to make several approximation based on the preliminary analysis of dynamics. 

\begin{enumerate}[(i)]
\item  \textbf{Piecewise linear approximation for $U$.} We remark that $U(A)$ is a sigmoid that can be approximated by the piecewise linear function (Fig. \ref{Fig:Omega}, left, top):
\begin{equation}\label{eq:AppU}
U(A) \sim 
\left\{
\begin{array}{llll}
&\frac{A}{2 \sqrt{\gamma_{A}}}, \, &0 \leq A \leq 2 \sqrt{\gamma_A};\\
&&\\
&1, \, &A \geq 2 \sqrt{\gamma_A}.
\end{array}
\right.
\end{equation}
In this approximation a cell $k$ such that $A_k > 2 \sqrt{\gamma_A}$ does not contribute to the variation of $\Gamma_i$ because $U' \sim 0$. We observe that in our simulations $A < 2 \sqrt{\gamma_A}=2$. Note that the form of $T$ is inspired from biophysics but its shape (sigmoid) matters more than its detailed mathematical form. Under the approximation \eqref{eq:AppU} $U(A) \sim U'(A)A$ so that
$\frac{d\Gamma_i}{dt} = - \mu_A \Gamma_i  +   \frac{\beta_{A}}{2 \sqrt{\gamma_A }} \, \displaystyle{\sum_{k \in {\cal B}_{i}}} T_{A}(V_{k})
$.

\item \textbf{Mean-field approximation for $T_A$.} The time scale of Ach evolution ($\sim 2s$) is quite fast compared to the time scale of evolution of $V$ (a few ms). As we are interested in transport when $G_S,G_A$ vary it is  relevant to replace $T_A(V)$ by its time average on the fast time scale, $\av{T_A(V)}=\lim_{T \to +\infty} \frac{1}{T} \int_{0}^{T} T_A(V(s)) \, ds$ where $G_S,G_A$ are kept constant. We then make the assumption that the time average of Ach production can be approximated as $\av{T_A(V)} \equiv \Omega(G_S,G_A) \in \bra{0,1}$. On the time scales of $G_S,G_A$ $\Omega(G_S,G_A)$ evolves along the pathway described by $G_S,G_A$ in the bifurcation map (Fig. \ref{Fig:Trajectory_GS_GA}). The function $\Omega(G_S,G_A)$ is shown in Fig. \ref{Fig:Omega} right top. It has a sigmoidal shape, with a maximal slope in region $D$. 
In this figure, bottom, we also show a plot of the gradient $\frac{\partial \Omega}{\partial G_S}, \frac{\partial \Omega}{\partial G_A}$ and Hessian of $\Omega$. 

\item \textbf{Continuous space limit.} Cells are located at the nodes of a $\setZ^d$ lattice with small lattice spacing $a \sim 50$ $\mu m$, $d=1,2$ (see section \ref{Sec:Biophys}). We now consider $G_S,G_A$ as continuous fields in space and time, $G_A \equiv G_A(x,y,t)$. 
Then, using the smallness of $a$ (compared to the waves characteristic size):
$$
\frac{\partial \Gamma}{\partial t} = - \mu_A \Gamma+  \frac{d \,\beta_{A}}{ \sqrt{\gamma_A }}  \, \Omega(G_S,G_A) + D_A \, \Delta \, \Omega(G_S,G_A).
$$
where we have set:
\begin{equation}\label{eq:Diff_A}
D_A = a^2 \frac{\beta_{A}}{2 \sqrt{\gamma_A }},
\end{equation}
the diffusion coefficient ($\sim 62.500 \, \mu m^2 s^{-1}$),
$\Delta$ being the Laplacian operator. 
The first term of the right hand side comes from Ach degradation, the second comes from the local Ach production, and the third corresponds to Ach transport by cells interactions, as we develop now. From now on we will write $\Omega$ instead  of $\Omega(G_{S},G_{A})$ to alleviate notations whenever it makes no confusion.
\end{enumerate}

\subsubsection{Non linear diffusion}

Using $G_A=g_A \, \Gamma$, for obtain a transport equation for Ach conductance:
\begin{equation} \label{eq:Transport_A_field}
 \frac{\partial G_A}{\partial t} = g_A  D_A \, \Delta \, \Omega +  S.
\end{equation}
with:
\begin{equation}\label{eq:Source}
 S = - \mu_A G_A+  g_A \,\frac{d  \,\beta_{A}}{ \sqrt{\gamma_A }}  \, \Omega,
\end{equation}
The transport equation is a non linear diffusion equation, because $\Omega$
depends non linearly on the fields $G_S,G_A$ via eq. \eqref{eq:nu}. Note that this is not a reaction-diffusion equation, in contrast e.g. to Lansdell et al approach \cite{lansdell-ford-etal:14}. Indeed, in their case Ach freely diffuse in the medium and non linearity is taken into account in a reaction term whereas, in our case, non linearity is in the diffusion coefficient, characterizing how waves non linearly interact with the sAHP landscape. The equation we obtain is also different from transport equation in calcium fire diffuse models \cite{dawson-keizer-etal:99,thul-smith-etal:08} although propagation and underlying mechanisms are very similar (see section \ref{Sec:ForestFires}). 

\begin{figure}
\resizebox{1.2\textwidth}{0.25\textheight}{
\includegraphics[]{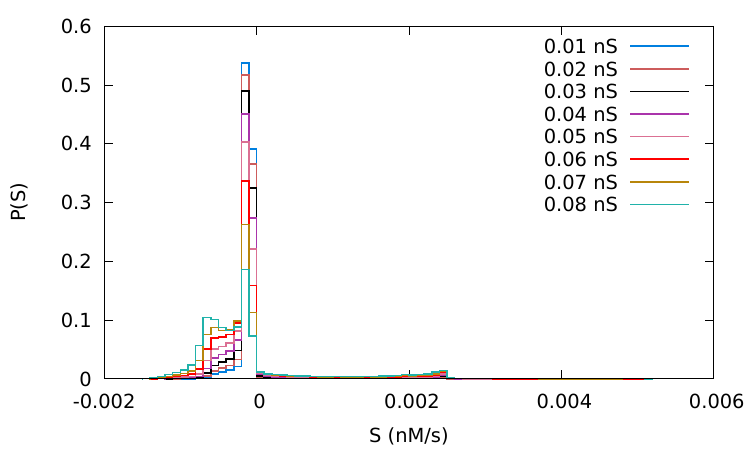}

\vspace{0.2cm}
\includegraphics[]{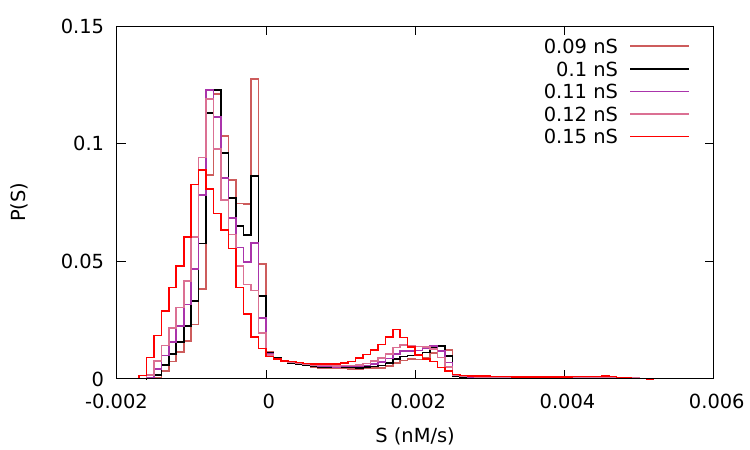}

\vspace{0.2cm}
\includegraphics[]{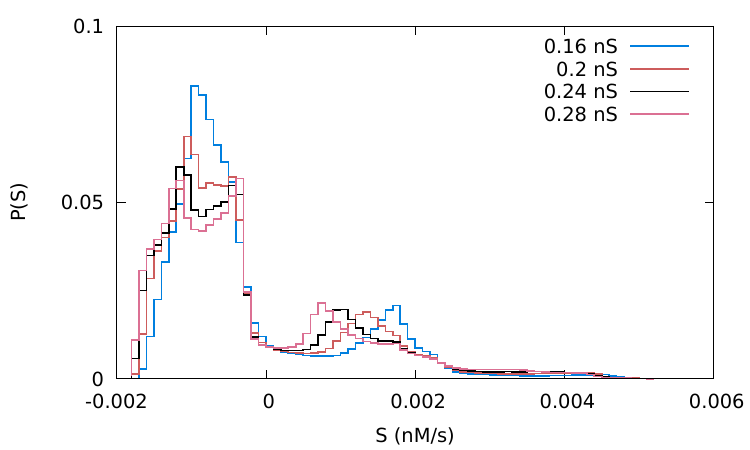}

}
 \caption{
 \textbf{Histogram} of the source term, $S$, in equation \eqref{eq:Transport_A_field} in the $3$ regimes (from left to right).
%
  \label{Fig:HistoSource}}
\end{figure}

The equation \eqref{eq:Transport_A_field} contains also  a "source" 
containing both the degradation term and a local production term. It can therefore be either positive or negative and depends on space-time via $G_S,G_A$. This source term is shown in Fig. \ref{Fig:HistoSource}. Although it has a very different shape in the different regimes, it is slightly negative on average so that dissipation dominates. The source term is purely deterministic here, although it should contain as well a stochastic term that mimics noise induced bursting, at the time scale of Ach evolution. This is briefly discussed in section \ref{Sec:NoiseGA}. The almost-vanishing of the source term, in regime I, expresses that  the local degradation of acetylcholine  is compensed by the local production coming from SACs neighbours. In this regime dynamics is ruled by fluctuations around the threshold, not by the leading source term.

\begin{figure}
\resizebox{\textwidth}{0.25\textheight}{
\includegraphics[width=7cm,height=7cm]{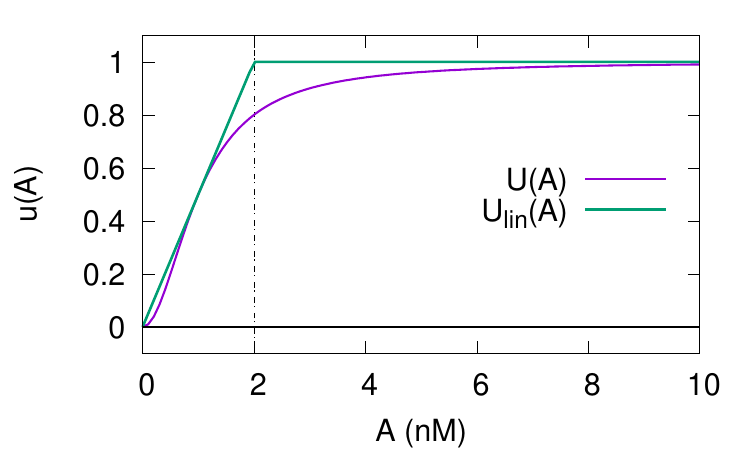}
\hspace{0.5cm}
\includegraphics[]{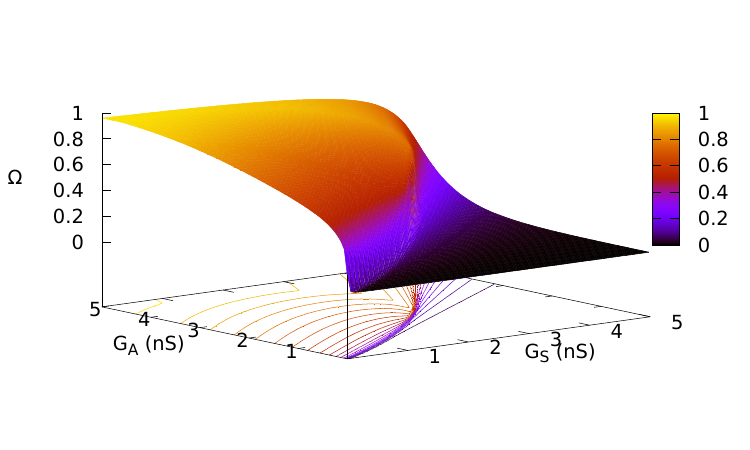}
}
\resizebox{\textwidth}{0.25\textheight}{
\includegraphics[width=6cm,height=5cm]{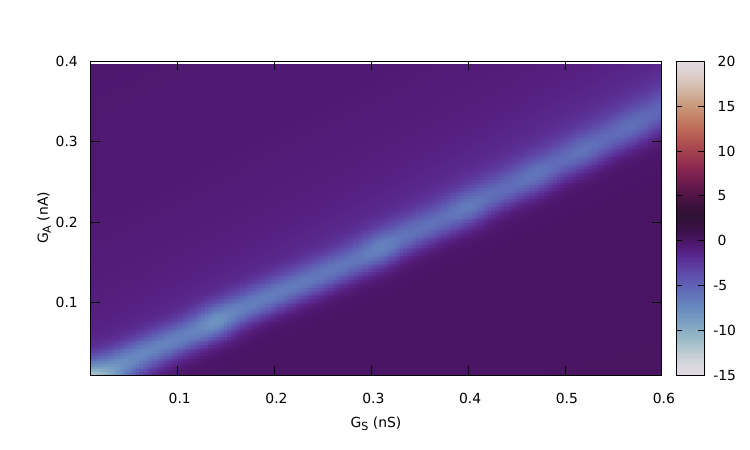}
\hspace{1cm}
\includegraphics[width=6cm,height=5cm]{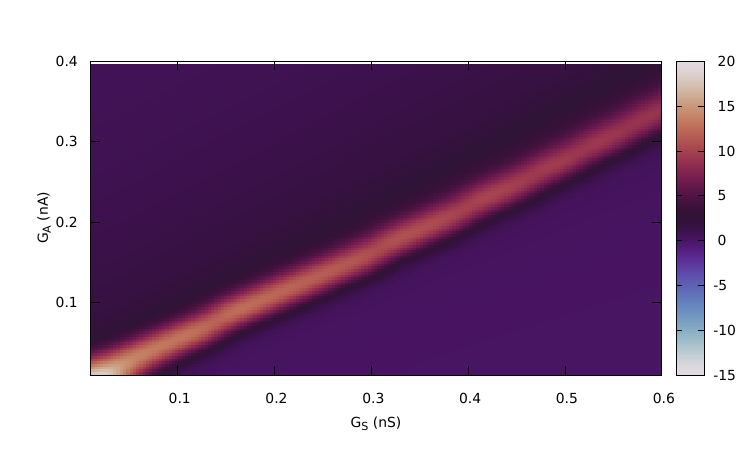}
}

\resizebox{\textwidth}{0.25\textheight}{
\includegraphics[width=6cm,height=5cm]{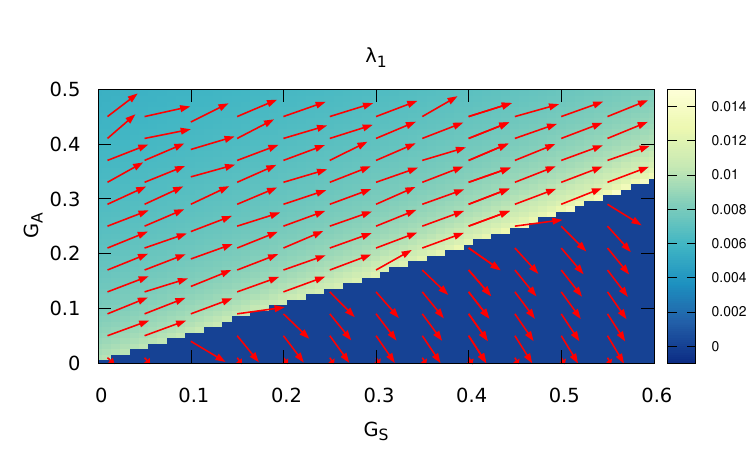}
\hspace{1cm}
\includegraphics[width=6cm,height=5cm]{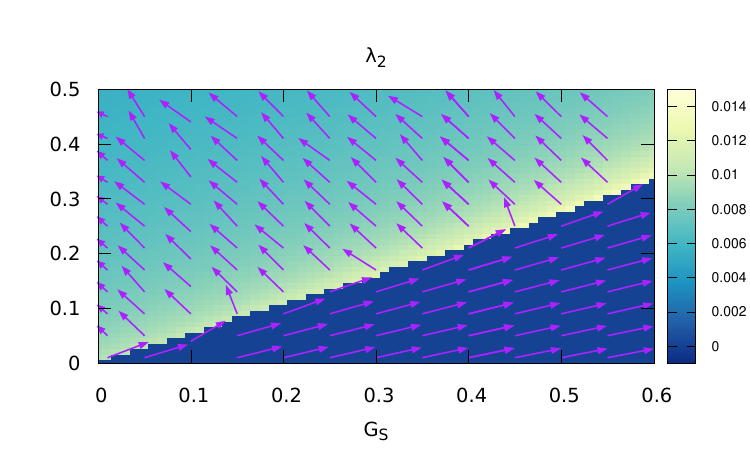}
}

%
\caption{\scriptsize\textbf{Top, left.} Piecewise linear approximation for $U$ with $\gamma_A=1$ $nM^{\frac{1}{2}}$. The dashed line corresponds to $2 \sqrt{\gamma_A}$. \textbf{Top, right.} Average value $\Omega \equiv \av{T_A}$ as a function of $G_S, G_A$. Note the level lines, at the bottom, and compare to the bifurcation diagram Fig. \ref{Fig:BifurcationDiagram_GS_GA}. 
\textbf{Middle row, left.} Gradient of $\Omega$, $\frac{\partial \Omega}{\partial G_S}$.  \textbf{Middle row, right.} Gradient of $\Omega$, $\frac{\partial \Omega}{\partial G_A}$. Gradients are essentially non zero in region $D$ where  $\frac{\partial \Omega}{\partial G_S}  < 0$ and $\frac{\partial \Omega}{\partial G_A} > 0$.
\textbf{Bottom row.} Eigenvalues and eigenvectors of the Hessian of $\Omega$. \textbf{Left.} First eigenvalue $\lambda_1$ as a function of $G_S,G_A$. The color bar, on the right, displays the amplitude of $\lambda_1$. 
The arrows correspond to the normalized eigenvector of the Hessian (depending on $G_S,G_A$) associated to $\lambda_1$.  \textbf{Right.} Same representation for the second eigenvalue, $\lambda_2$. $\lambda_1$ and $\lambda_2$ have an abrupt variation in between the bifurcation lines delimiting region $D$. The motion in the plane $\set{G_S,G_A}$, as illustrated in Fig. \ref{Fig:Trajectory_GS_GA} up, is driven by the transport equation with a second order term, depending on $H_\Omega$, where the trajectory is tangent to a linear combination of these two eigenvectors. It also depends on the gradients. 
\label{Fig:Omega}}
\end{figure}

Let us comment more on the effects of this non linear diffusion.
It is actually useful to let the gradients of $G_A$ and $G_S$ appear in eq. \eqref{eq:Transport_A_field}. We also introduce the Hessian of $\Omega$:
\begin{equation}\label{eq:H2nu}
H_\Omega =
\pare{
\begin{array}{lll}
&\frac{\partial^2 \Omega}{\partial G_S^2}&\frac{\partial^2 \Omega}{\partial G_S \partial G_A } \\
&\frac{\partial^2 \Omega}{\partial G_A \partial G_S}&\frac{\partial^2 \Omega}{\partial G_A^2}\,
\end{array}
}.
\end{equation}
This symmetric matrix characterizes the local convexity of $\Omega$. It also defines a metric. Indeed the transport equation\eqref{eq:Transport_A_field}
can now be written:
\begin{equation} \label{eq:Transport_A_Grad}
 \frac{\partial G_A}{\partial t} =   g_A  D_A \bra{\pare{\nabla G_S, \nabla G_A}  H_\Omega 
\pare{
\begin{array}{lll}
\nabla G_S\\
\nabla G_A
\end{array}
}
+  \frac{\partial \Omega}{\partial G_S}\Delta \, G_S
+
\frac{\partial \Omega}{\partial G_A}\Delta \, G_A } + S.
\end{equation}
This a Laplace-Beltrami diffusion on a manifold whose metric is constrained by \eqref{eq:H2nu}. In other words, the diagram shown in Fig. \ref{Fig:Trajectory_GS_GA} has now to be thought on top of the $\Omega$ manifold, Fig. \ref{Fig:Omega} top, right, whose gradient and curvature, shown in figure \ref{Fig:Omega} middle and bottom row, constrain the diffusion.   We see that the transport of $G_A$ depends on the spatial profile of sAHP. In particular, the extra-diagonal terms of $H_\Omega$ couples $ \nabla G_S$ and $\nabla G_A$, the gradients of sAHP conductance and Ach conductance. 
As expected, the first and second partial derivatives are non vanishing essentially near the bifurcation lines  delimiting region $D$. Interestingly, $\frac{\partial \Omega}{\partial G_S}<0$ in this region corresponding to an anti diffusion (focalisation) mechanism whereas $\frac{\partial \Omega}{\partial G_A}>0$. 

In view of the shape of $\Omega$ and, more specifically, on its derivatives, one sees that there are $3$ main  phases for cells, corresponding to specific regions in the plane $\set{G_S,G_A}$. At a given time, the cells are in one of these $3$ phases, with spatial continuity due to the diffusion term.
%
%
\begin{enumerate}[(a)]
\item \textbf{Quiescent phase.} In regions where $\Omega$ is small and its derivative vanish \eqref{eq:Transport_A_field} reduces to $ \frac{\partial G_A}{\partial t} = - \mu_A G_A + g_A \,\frac{d  \,\beta_{A}}{ \sqrt{\gamma_A }}  \, \Omega$, so that $G_A \sim g_A \,\frac{d  \,\beta_{A}}{ \mu_A \, \sqrt{\gamma_A }}  \, \Omega$ is small. These cells do not participate to propagation.

\item \textbf{Bursting phase.} For cells which are located in region $C$, $\Omega$ is large but its derivatives are small thus the Laplacian term in \eqref{eq:Transport_A_field} vanishes too. There is stationary solution of the same form as above, but $G_A = g_A \,\frac{d  \,\beta_{A}}{ \sqrt{\gamma_A } \, \mu_A}  \, \Omega$ is now large. These cells are the bursting cells participating to waves.

\item \textbf{Intermediate phase}, for cells in the relative refractory period, in region $D$ starting to burst or close to burst. Here, $\Omega$ depends sharply on $G_S,G_A$ and its variations impact dramatically the wave propagation, according to eq.  \eqref{eq:Transport_A_field}. These cells constitute waves boundaries (fronts).

\end{enumerate}

The physical picture which emerges from eq. \eqref{eq:Transport_A_Grad} and Fig. \ref{Fig:Omega} is actually very close to the description made by Bantay and Janosi \cite{bantay-janosi:92},
in a beautiful paper about Self-Organized Criticality  that we essentially rephrase,  although the situation we have here is a bit more complex due to the dependence in sAHP.     
At a given time, one may distinguish
three different spatial regions corresponding to cells in one of the $3$ phases above. In the quiescent and bursting phase the diffusion is very small, while at the interface between these two phases, the diffusion grows steeply so that there is a large flux which enters the intermediate region. In the intermediate region the
diffusion is very large, thus any spatial inhomogeneity disappears quite fast. Consequently, the inflow flux is transported through the intermediate region
to the quiescent phase boundary increasing the level of activity of quiescent cells until they become excitable. As a result, the boundary of the quiescent phases moves towards the bursting region, until this later is completely absorbed by the intermediate phase. 
The behaviour of the other boundary is completely similar, the only difference is that the direction of the flux is opposite. Now, the motion of boundaries is, in our case, strongly constrained by sAHP.

\subsubsection{sAHP conductance} \label{Sec:GSCondTransport}

\begin{figure}
\centerline{
\includegraphics[width=6cm, height=6cm]{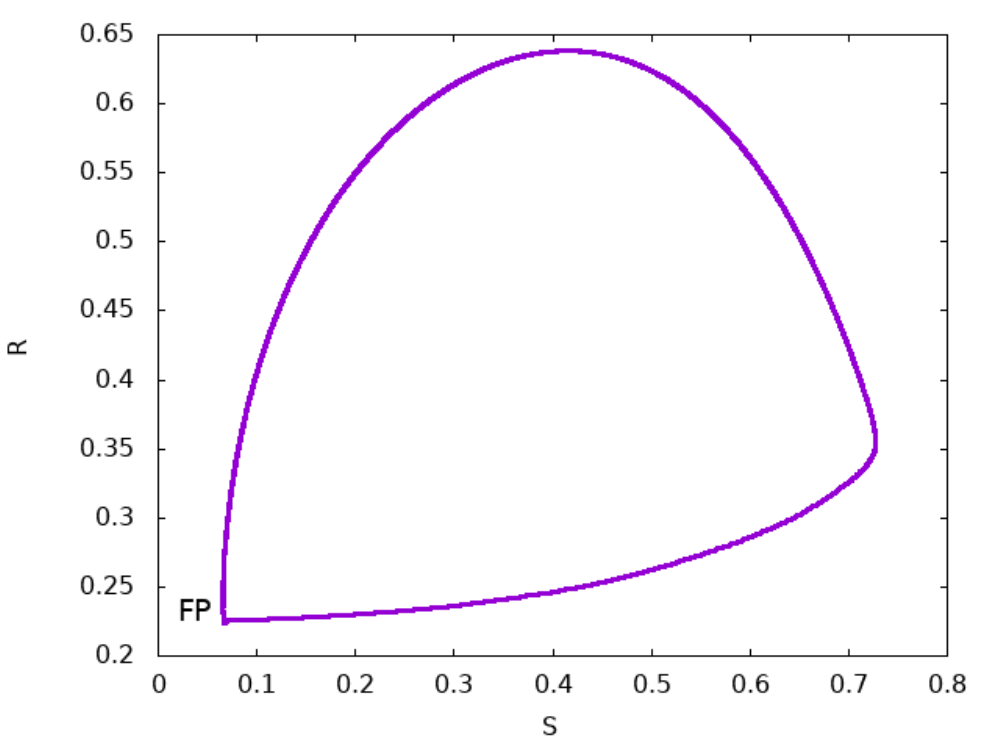}
\hspace{1cm}
\includegraphics[width=6cm, height=6cm]{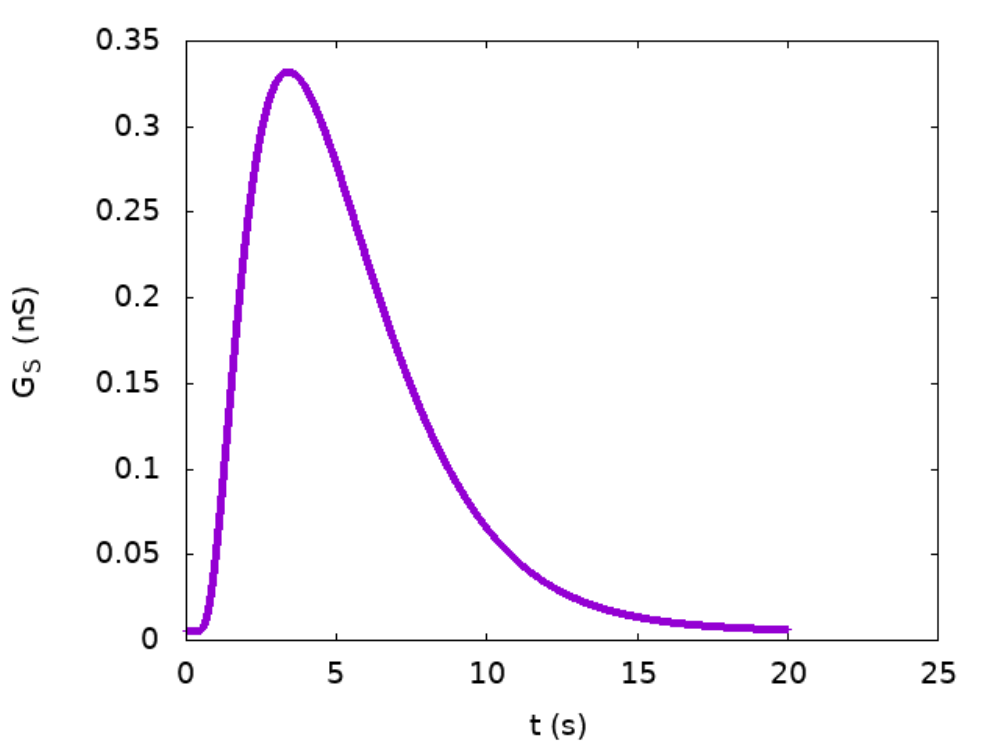}
}
 \caption{\footnotesize\textbf{Left.} Trajectory of $R,S$ during the cycle bursting, hyperpolarization, repolarisation. "FP" means "Fixed Point". \textbf{Right.} Trajectory of $G_S$ ($\tau_S, \tau_R=8.3$ s) in the same conditions. Note that the shape can be more complex depending on parameters such as $g_{sAHP}$.
  \label{Fig:GS_vs_t}
}
 \end{figure}

Equation \eqref{eq:Transport_A_Grad} illustrates how dynamics depends on the dynamics of the sAHP conductance $G_S$, a very slow variable (of order a minute). Using slow-fast analysis  one can see that $V,N,C$ and $A$ evolve, at the time scale of $R,S$, along a slow manifold parametrized by $R,S$. In particular, at this time scale, the calcium concentration is a function $C(R,S)$ of $R,S$, whose shape can be guessed from Fig. \ref{Fig:CRS}, with 4 phases. It results that the coupled dynamics of $S,R$ at the slow time scale, $t_s$, has the form $\frac{dS}{dt_s} = \alpha_S C(R,S)(1-S)-S; \frac{dR}{dt_s}= \alpha_R S(1-R) -R$ and is deterministic. This generates an invariant closed curve represented in Fig. \ref{Fig:GS_vs_t}, left, while, on the right, we see the evolution of $G_S$ as a function of $t$. This is, actually, a deterministic curve of the form $G_S(t)=u(t-t_b)$ where $t_b$ is the time when a cell starts to burst. This means that we may write $G_S(X,t) \equiv u\pare{t-t_{last}(X,t)}$ where $t_{last}(X,t)$ is the last bursting time anterior to $t$, for the cell located at $X$. Thus, $t_{last}(X,t)$ is a stochastic process, whose probability depends on $G_S,G_A$ and therefore evolves along the waves propagation landscape. Fig. \ref{Fig:GS_vs_t} is only illustrative though as the shape of $G_S(t)$ can be quite more complex depending on parameters such as $g_{sAHP}$. This complex dependence on $G_S$, which makes retinal waves dynamics extraordinary rich, lifts the general analysis of transport equations out of our technical reach and we are considering simplified situations in the next sections.   

$G_S$ depends therefore on the network \textit{history} via $t_B(x,t)$. This is the mathematical expression of what was anticipated earlier (section \ref{Sec:WavesProp}). When a retinal wave propagates, transporting Ach conductance, it leaves, behind it a refractory trace in which subsequent waves have to propagate. This induces complex interactions between waves, history depend, as depicted e.g. in Fig. \ref{Fig:Propagation}.

\subsubsection{Noise at the medium time scale}\label{Sec:NoiseGA}

The transport equations are deterministic. They lack a "noise" term that mimics the random initialisation of waves. The detailed form is this "noise" is not completely clear though.
It can be obtained by adding, in the equation of Ach production, a stochastic term $\delta V$ to the voltage, as we did in section \ref{Sec:BurstingProbabilityMap} and then integrate over time $T_A\pare{V+\delta V(t)}$, e.g. by expanding $T_A$ in series, generating terms in powers of $\kappa_A \delta_V$. One obtains a term proportional to $g_A \, \kappa_A \, \sigma \frac{\beta_A}{\sqrt{\gamma_A}}$ which is the time integral of a Gaussian process passing through a highly non-linear function ($T_A$). The resulting process is therefore neither Gaussian nor Brownian.  In addition it must be proportional to $\nu$, the probability \eqref{eq:nu} that a cell burst. A possible spatio-temporal scaling for such a noise has been proposed in \cite{volchenkov-blanchard-etal:02}. Its main interest is to afford a renormalisation group analysis of the dynamics, which could be helpful for the characterization of retinal waves statistics in $1$ and $2$ D.

\subsection{Consequences}\label{Sec:Consequences}

\subsubsection{Threshold value of $g_A$} \label{Sec:ThresholdgA}

A first consequence of these approximations is that there exists a threshold value of $g_A$,
$g_{A_0}$, such that for $g_A < g_{A_0}$ no wave can propagate. 

Consider the situation shown in Fig. \ref{Fig:AchProfile_NIB_2cells}. Cell $2$ is at rest while cell $1$ starts to burst at time $t=0$. Under which condition is cell $2$, initially in the quiescent phase, going to burst upon the excitation of cell $1$ ? As cell $1$ starts to burst at time $0$, from \eqref{eq:Achprod}, the concentration of Ach it produces is $A_1(t)=A^-_{1} e^{- \mu_A t} + \beta_A \int_{0}^{t} T_A(V_1(s)) e^{-\mu_A(t-s)} ds$, where $A^-_{1}=\frac{\beta_A T_A(V_{Si_1})}{\mu_A}$  is the Ach concentration for cell $1$ at rest (note that $V_{Si_1}$ depends on $G_{S_1},G_{A_1}$ via \eqref{eq:VSi}). Now, using the mean-field approximation (ii) above we can write $\int_{0}^{t} T_A(V_1(s)) e^{-\mu_A(t-s)} ds \sim \frac{\Omega(G_{S_1},G_{A_1})}{\mu_A} \bra{1-e^{-\mu_A t}}$ so that:
\begin{equation}\label{eq:A_Burst}
A_1(t) = A^-_{1} e^{- \mu_A t} + \frac{\beta_A \Omega(G_{S_1},G_{A_1})}{\mu_A} \bra{1-e^{- \mu_A t}}.
\end{equation}
Note that $G_{S_1}$ depends on cell $1$'s state whereas $G_{A_1}$ depends on the state of its neighbours, here cell $2$, which is initially in the quiescent phase, so that $G_{A_1} \sim 0$. 

From \eqref{eq:A_Burst} $A_1$ increases, thereby increasing the Ach current seen by cell $2$.
 Without noise, the condition for cell $2$ to burst is given by eq. \eqref{eq:SN2}
characterizing the crossing from region $D$ to region $C$. Indeed, without noise a cell at rest in $D$ cannot reach the limit cycle. Now, the Ach conductance $G_{A_2}$ seen by cell $2$ is, under the approximation (i) above, $G_{A_2}(t)=\frac{g_A}{2 \,\sqrt{\gamma_A}} \, A_1(t)$. Thus, eq. \eqref{eq:SN2} reads:
$$
-\frac{g_A}{2 \,\sqrt{\gamma_A}} \, A_1(t_{B_2}) \, \pare{V_{Si_2}-V_A}  - G_{S_2} \pare{V_{Si_2}-V_K}=I_{SN}
$$
where $t_{B_2}$  is the time where cell $2$ starts to burst by crossing the bifurcation line. 
%
Assuming that $A^-_{1}$ is so small \footnote{Compared to $\frac{g_A}{2 \,\sqrt{\gamma_A}}\, \frac{\beta_A \, \Omega}{\mu_A \, \pare{V_{Si_2}-V_A}}$} that we can neglect it, gives:
%
\begin{equation}\label{eq:tB}
t_{B_2} = -\frac{1}{\mu_A} \, \log\bra{1+  \frac{1}{g_A} \, \frac{2 \, \mu_A \, \sqrt{\gamma_A}}{\beta_A \, \Omega} \, \, \frac{I_{SN} + G_{S_2}  \,  (V_{Si_2}-V_K)}{V_{Si_2}-V_A}}.
\end{equation}

This equation is sufficient to characterize the wave propagation in the geometry we are considering in this paper: one dimensional and nearest neighbours interactions. Note however that is generalizes  to a cell at rest surrounded by $n$ bursting neighbours whose Ach concentration increases according to \eqref{eq:A_Burst} provided these neighbours started to burst at the same time (synchrony). 
%
Let us now comment on this result.
\begin{enumerate}
\item This equation is actually implicit. Indeed, $\Omega$ depends on $G_{S_1},G_{A_1}$ which depend on $t$. Thus, $\Omega$ actually depends on $t_{B_2}$ via $G_{S_1},G_{A_1}$. Likewise, $V_{Si_2}$ depends on $G_{A_1}$.  However, using the methodology developed in this paper, we may consider that  $t_{B_2}$ is a function of the parameters $G_{S_1},G_{A_1}$ which slowly move with time in the bifurcation map as shown in Fig. \ref{Fig:Trajectory_GS_GA} (growing of Ach conductance upon bursting of neighbours).
\item Likewise, $t_{B_2}$ depends on $G_{S_2}$, characterizing the level of hyperpolarization of cell $2$. The higher $G_{S_2}$ the longer it is for cell $2$ to eventually burst, until a value of $G_{S_2}$ where $t_{B_2}$ becomes infinite.
\item More precisely, for eq. \eqref{eq:tB} to have a solution with a positive $t_{B_2}$ one needs that 
$ -1 <\frac{1}{g_A} \, \frac{2 \, \mu_A \, \sqrt{\gamma_A}}{\beta_A \, \Omega} \, \, \frac{I_{SN} + G_{S_2}  \,  (V_{Si_2}-V_K)}{V_{Si_2}-V_A} < 0$ with $V_{Si_2}-V_A<0$.
%
%
 As $I_{SN} + G_{S_2}  \,  (V_{Si_2}-V_K)>0$, this implies that there is a minimal value of $g_A$, 
\begin{equation}\label{eq:g_0}
g_{A_0} \equiv  \frac{2 \, \mu_A \, \sqrt{\gamma_A}}{\beta_A \, \Omega} \, \, \frac{I_{SN} + G_{S_2}  \,  (V_{Si_2}-V_K)}{V_A - V_{Si_2}},
\end{equation}
such that the cell $2$ is excited by cell $1$ only if $g_A > g_{A_0}$. 
This value depends on $G_{A_1},G_{S_1}$ via $\Omega$, and also on $G_{S_2}$. 
%
The most important here is that $g_{A_0}$ depends on $G_{S_2}$, the level of hyperpolarisation of cell $2$. As $G_{S_2}$ increases $g_{A_0}$ increases so that the conductance $g_A$ necessary to trigger excitation propagation to a quiescent cell depends of its level of hyper-polarisation, as expected. For a cell at rest, we found, with our parameters value (Table \ref{TabParameters}) $g_{A_0} \sim 0.04$ nS. This corresponds to the blue dashed line in Figs. \ref{Fig:AchProfile_NIB_2cells} right, \ref{Fig:Transport}.
\end{enumerate}

\subsubsection{Propagation speed}\label{Sec:Speed}

The time $t_{B_2}$ gives the time for a rest cell, excited by its neighbour, to burst. Now, to numerically compute the wave propagation speed, as in Fig. \ref{Fig:Transport}, bottom, we actually consider waves of calcium activity i.e. a cell is considered as bursting if its calcium concentration reaches a certain threshold (see section \ref{Sec:NumSim}). The time to reach this threshold from the moment where cell $2$ starts to burst is a time $t_C$ weakly depend on $g_A$ but that can be estimated numerically to a few hundreds of milliseconds. 
The inverse time $t_C+t_{B_2}$ gives therefore the speed of propagation, in this specific geometry, and in the deterministic case:
\begin{equation}\label{eq:speed}
c=\frac{a}{t_C-\frac{1}{\mu_A} \, \log\bra{1+  \frac{1}{g_A} \, \frac{2 \, \mu_A \, \sqrt{\gamma_A}}{n \, \beta_A \, \Omega} \, \, \frac{I_{SN} + G_{S_2}  \,  (V_{Si_2}-V_K)}{V_{Si_2}-V_A}}}.
\end{equation}
This corresponds to the theoretical curve displayed in Fig. \ref{Fig:Transport} bottom.
Note that this equation is valid only in the ballistic case, because we have assumed a deterministic propagation without wave interactions. This corresponds to regime II. 

%

\subsection{Specific propagation regimes}\label{Sec:SpecificPropagationRegimes}
Let us now consider several interesting limits for the transport equation.

\subsubsection{Flat sAHP landscape and Kardar-Parisi-Zhang equation}\label{Sec:KPZ}

 We consider the situation where $G_S$ is homogeneous in space (flat sAHP landscape). This characterizes e.g. the propagation of a wave in a region which has not seen a previous propagation since a time larger than the characteristic time of return to rest. The terms $\nabla G_S$ and $\Delta \, G_S$ vanish and the transport equation  \eqref{eq:Transport_A_Grad} reads now:
\begin{equation}\label{eq:NonLinearDiffGamma}
 \frac{\partial G_A}{\partial t} =    g_A D_A \, \frac{\partial^2 \Omega}{\partial G_A^2}  \| \nabla G_A \|^2  
+
g_A D_A \, \frac{\partial \Omega}{\partial G_A}\Delta \, G_A  + S(G_S,G_A)\\
\end{equation}
Here $G_S$ acts therefore as a parameter changing the characteristics of the transport coefficients. Especially it constrains the diffusion coefficient:
\begin{equation}\label{eq:CoefDifSingular}
 D \equiv D(G_S,G_A) = D_A \,  \frac{\partial \Omega}{\partial G_A}
\end{equation}

Note that we may write:
\begin{equation}\label{eq:Transport_Gamma_flat_psi}
 \frac{\partial G_A}{\partial t}
\begin{array}{lll}
&=& 
 g_A \, \frac{\partial D}{\partial G_A}  \| \nabla G_A \|^2  
+  g_A \, D \Delta G_A + S.
\end{array}
\end{equation}
which has the form of the Kardar-Parisi-Zhang equation (KPZ) \cite{kardar-parisi-etal:86}.The KPZ equation models interface growth: the first term mimics surface growth in a direction normal to the surface whereas the second term corresponds to diffusive motion along the surface.
However, in contrast to KPZ the  terms  $D(G_S,G_A), \frac{\partial D(G_S,G_A)}{\partial G_A}$ are not constant. Additionally, the source term is white noise in KPZ whereas, in our case this is a function of $G_A, G_S$. This suggests interesting questions that could be addressed in further studies. Are there scale invariant solutions  ? Assume that $G_S$ has such a scale invariance, what are the consequences for $G_A$ ? Is it possible to characterize (anomalous) tranport in this context ?

\subsubsection{Transition between regime I and II}
In regime I waves are weakly interacting with the sAHP landscape so that one can consider it as uniform along the path of the waves. This assumption leads to an explicit $g_A$ value for the transition between regime I and II.

Let us  approximate, in region $D$, $\Omega(G_S,G_A)$ by a linear function of $G_A$, 
$\Omega(G_S,G_A)=\Omega_0+ \Omega_1(G_S) \, G_A$ with $\Omega_1(G_S) \equiv \frac{\partial \Omega}{\partial G_A}(G_S)>0$. 
As $G_S$ is assumed spatially uniform here, the transport equation \eqref{eq:Transport_A_field} reads, using \eqref{eq:Diff_A}: 
%
$$
\frac{\partial G_A}{\partial t} = P - \lambda \, G_A + \kappa \, \Delta G_A
$$
with $P=g_A \,\frac{d  \,\beta_{A}}{ \sqrt{\gamma_A }}  \, \Omega_0$, $\lambda = \mu_A -  g_A \,\frac{d  \,\beta_{A}}{ \sqrt{\gamma_A }} \, \Omega_1$, $\kappa=g_A   \frac{\beta_{A}\, \Omega_1}{2 \sqrt{\gamma_A }}  \, a^2 $.
Taking the spatial Fourier transform gives:
%
%
$$
\frac{\partial \hat{G}_A}{\partial t} = P \delta(k) -  \sigma(k) \, \hat{G}_A
$$
with $\sigma(k)=\lambda \, + \, \kappa \, k^2$. Solutions are stable if $\sigma(k) >0$
and marginally stable if $\sigma(k^\ast)=0$. 

This last condition reads:
%
%
\begin{equation}\label{eq:k2}
k^2 = \frac{2d}{a^2} \pare{1 -\frac{1}{g_A} \, \frac{\mu_A \, \sqrt{\gamma_A }}{d \, \beta_{A}} \,\frac{1} {\Omega_1}}
\end{equation}
and has a solution if $g_A \geq \frac{\mu_A \, \sqrt{\gamma_A}}{d \, \beta_A} \, \frac{1}{\Omega_1}$. Let us interpret this condition. If $g_A < \frac{\mu_A \, \sqrt{\gamma_A}}{d \, \beta_A} \, \frac{1}{\Omega_1}$ then $\sigma(k)>0$, $\forall k$, so that linear solutions are damped
and converge to $\frac{P}{\sigma(0)}=g_A \,\frac{d  \,\beta_{A}}{\mu_A \, \sqrt{\gamma_A }}  \, \Omega_0$. This corresponds to a uniform quiescent phase. In this range of value of $g_A$ there is no wave propagation so that $\frac{\mu_A \, \sqrt{\gamma_A}}{d \, \beta_A} \, \frac{1}{\Omega_1}$ corresponds to the critical value $g_{A_0}$, eq. \eqref{eq:g_0} found in section \ref{Sec:ThresholdgA} in the case of two cells. There are slight differences though. While \eqref{eq:g_0} was holding for two cells 
without linear approximation on $\Omega$, here, we find the critical value without constraint on dimension, but assuming that $\Omega$ is linear in $G_A$. 
%

Using this we can rewrite the positive solution \eqref{eq:k2} as $k^\ast \equiv k^\ast(g_A)=\frac{\sqrt{2d}}{a} \, \sqrt{1-\frac{g_{A_0}}{g_A}}$. 
This defines a characteristic length:
\begin{equation}\label{eq:CharLength}
\Lambda(g_A)=\frac{a}{\sqrt{2d}} \,\frac{1}{\sqrt{1-\frac{g_{A_0}}{g_A}}},
\end{equation}
which interprets as follows. Consider cells in the quiescent phase and excite a connected cluster of cells, with radius $r$. If $r < \Lambda(g_A)$ this perturbation is damped, whereas for $r \geq \Lambda(g_A)$ it is amplified, getting rapidly (exponentially fast) out of the range of $G_A$ where the linear approximation on $\Omega$ holds. This corresponds to drawing cells to region $C$. Thus, $\Lambda(g_A)$ is a critical size for a cluster of excited cells to nucleate a wave.

As a consequence, the value of $g_A$ such that $\Lambda(g_A)=a$ corresponds to a regime where exciting one cell is enough to propagate a wave in a uniform landscape of $G_S$. This value is $g_A=\frac{g_{A_0}}{1-\frac{1}{2d}}$, so that, in one dimension, $g_A=2 \,g_{A_0}$. This is the value drawn in Figs. \ref{Fig:AchProfile_NIB_2cells}, \ref{Fig:Regimes}, \ref{Fig:Transport}, at the vertical dashed line separating regime I and regime II. Note that this theoretical value is obtained in a purely deterministic setting (without noise).  

\subsubsection{"Self-Organized" Criticality and beyond}\label{Sec:SOC}

Let us now explore another regime. In the zero noise limit, the transition to bursting can be only achieved when a cell switches from region $D$ to region $C$. This corresponds to a sharp threshold condition where $\Omega(G_S,G_A)$ takes the form $\pi_A H\bra{G_A-G_{A_c}(G_S)}$, $H$ being the Heaviside function and $G_{A_c}$ the bifurcation value given by eq. \eqref{eq:SN2}. Then, eq. \eqref{eq:Transport_A_field} has the form of a continuous state sandpile, proposed by several authors \cite{zhang:89,bantay-janosi:92} in the context of Self-Organised Criticality \cite{bak-tang-etal:87,bak:97,jensen:98,christensen-moloney:05,markovic-gros:14,buendia-santo-etal:20}. More generally, the presence of noise smooths the Heaviside function leading to a sigmoid of the form displayed in Fig. \ref{Fig:Omega}, allowing, in SOC models, power expansions and dynamical renormalisation group analysis \cite{diaz-guilera:94,volchenkov-blanchard-etal:02}. 

In comparison with a SOC model there are some important differences though. First, we have, in our case, an extra term $S$ violating the local conservation required to reach a Self-Organized Critical regime \cite{hwa-kardar:89,jensen:98,vespignani-zapperi:98,cessac-blanchard-etal:04}. Actually, $S$ behaves in a different way in the $3$ regimes as shown in Fig. \ref{Fig:HistoSource}.
Second, in contrast to early SOC models the "dissipation" does not occur at the boundary but in the bulk, similarly to extended SOC models studied in \cite{bak:92,hwa-kardar:89,hwa-kardar:92,grinstein:95,vespignani-dickman-etal:98,drossel:02,cessac-blanchard-etal:04,malcai-shilo-etal:06}.  Indeed, $S$ is negative on average, leading to a small bulk dissipation. 
Finally, the wave activation in our case is not "adiabatic" as in SOC model where one waits that an avalanche stops before activating a new one, thereby preventing avalanches interactions. Thus, we don't have the infinite time scale separation necessary to reach SOC \cite{grinstein-lee-etal:90,socolar-grinstein-etal:93,grinstein:95}. 
In contrast, several retinal waves can propagate simultaneously and strongly interact, as illustrated in Fig. \ref{Fig:Propagation}. These interactions limit the size of each wave in the regime II.

In a series of recent papers, Buendia et al report on the different scenarios leading to SOC-like behaviours \cite{buendia-santo-etal:20,buendia-villegas-etal:21}. Even if "perfect" SOC is not reached there exist observed regimes close to it, such as "self-organized bistability" (SOB), "self-organized quasi-criticality" (SOqC), "self-organized collective oscillations" (SOCO). In addition, the paper \cite{buendia-villegas-etal:21} exhibits, in Fig. 1, a bifurcation diagram quite similar to ours although it deals with a different model (Kuramoto) and different parameters. Also, the paper  \cite{buendia-santo-etal:20} shows, in Fig. 2, a behaviour for the equivalent of our source term quite similar to ours. This suggests that our model, and especially equations \eqref{eq:Transport_A_field}, \eqref{eq:Transport_A_Grad},  may exhibit different regimes corresponding to different modalities of transport equations reported in \cite{buendia-santo-etal:20} especially "SOC like" behaviours.  The fact that we do not observe anything striking in this direction with our simulations may just signify that one has to consider other settings. That is:
\begin{enumerate}
\item \textbf{Dimension.} We have only considered 1 dimensional simulations here. One could happen in $2$ dimensions ? We observed that the slow hyperpolarization current  following the bursting phase generates a landscape in which subsequent waves have to propagate. In particular, a cell in the absolute refractory state cannot be excited by its bursting neighbours. This way, absolute refractory cells constitute temporary forbidden zone where waves cannot go through, thereby compartmentalizing space
with slowly varying boundaries. Clearly, these boundaries are quite different in one and in two dimensions. In 1D, these boundaries are just points; in 2D these are lines with complex geometry. As the topological structure of active and refractory regions could play an important role in shaping the visual system, it is important to consider them in detail. This actually means, as outlined in this paper, that it is not sufficient to consider indicators for the \textit{global}
activity of the network, it is also essential to consider the shape and size of \textit{connected}  regions where activity holds. Unfortunately, this is easy to do in 1D, but is complex in 2D. (Fortunately, SACs are physiologically organized in a two dimensional layer, so, a priori, there is no need to consider $3$ dimensional lattices). 
The detailed study of this question deserves thorough investigations beyond the scope of this paper. Yet, we want to make some comments coming out from the present analysis.

First, a large part of the mechanism of retinal waves can be explained by local dynamics and does not depend on dimensionality. Yet, the synchronisation mechanism is a non local effect, depending, as we saw, on the non linear coupling intensity $g_A$, sAHP conductance $g_S$ and noise $\eta$, fixing the characteristic size of active and refractory regions. It is also constrained by boundary conditions. Especially, zero boundary conditions imposes quiescent cells at the boundaries, thereby constraining the way how waves propagate and interact and their characteristic size. Here, we can make a formal analogy with statistical mechanics and Ising model. SACs dendritic tree is rather short compared to the size of the retina ($2$ mm in the mouse, compared to $50 \, \mu m$ for the dentritic tree radius of a SAC), so SACs interaction can be considered as local. Yet, Ising model, taught us that local interactions can induce long range correlations at a phase transition point.  
However, phase transitions in Ising model, and more generally, models with short range interactions do not appear in 1D. In this sense too, the behaviour of retinal waves could be quite different in 1D and 2D.

\item \textbf{Potential phases transition and connectivity.} The formal analogy with Ising model actually raises further comments and questions. First,
dynamics here is quite more complex than an Ising model (or classical forest fires) as  couplings are slowly evolving in time, and depend non linearly in dynamics. Refractory domains constitute somewhat a glassy phase, evolving on a very slow time scale, so an analogy with spin glasses could be more relevant \cite{binder-young:86,mezard-parisi-etal:87}. In particular, it would be interesting to investigate if evidences of ageing appear in the model when taking the thermodynamic limit. This might be a pure academic question, however,
as the meaning of a "thermodynamic limit" in a retina remains to be clarified. Another question coming out from the analogy with an Ising model is "what could play the role of the external field". Photoreceptors become more and more active during the first weeks of development and it could be interesting to investigate if they could play a role, in stage II, by acting on Bipolar cells and thereby on
Starburst Amacrine cells.    

Finally, we considered here nearest neighbours interaction. Although, this coupling shouldn't be taken "a la lettre" as a real connectivity but an effective one \cite{lansdell-ford-etal:14} 
one may be willing to consider more realistic connectivities. In particular, using random graphs (Erdos-Reyni) dramatically impacts the phase diagram of the Kuramoto model \cite{buendia-villegas-etal:21}, exhibiting a range of parameters with a Griffith phase. Griffith phase is characterized by power-laws extending over broad regions in parameters space. In magnetic systems, it was shown that Griffith phases may have many different origins, e.g., phase separation, occurrence of clusters of sizes ranging from nano-meters to micrometers, competing intra- and interlayer magnetic interaction. In general, it stems from the presence of quenched disorder (e.g. structural heterogeneity) in
classical, quantum, and non-equilibrium disordered systems \cite{vojta:06} with local regions characterized by parameter values which
differ significantly from their corresponding system averages. This is very close to what we observe in our model. In the neuroscience field, Griffith phases were observed by 
 Moretti - Munoz \cite{moretti-munoz:13} for the whole brain, and Girardi-Schappo et al \cite{girardi-schappo-bortolotto-etal:16} for the V1 cortex. It was argued by these authors
 that Griffith phase yield enhanced functionality in a generic way, facilitating the task of self-organizing, adaptive and evolutionary mechanisms selecting for criticality.
 As the retina is the entry of V1 and as retinal waves conduct the development of the visual system it might interesting to relate the potential existence of a Griffith phase in retinal waves to the observation of such a phase in V1.
 
 \item \textbf{Percolation.} In their paper \cite{hennig-adams-etal:09} Hennig et al.  conclude that "the network of SACs participating in stage II retinal waves is capable of operating at a transition point between purely local and global functional connectedness, corresponding to a percolation phase transition", where waves of activity - often referred to as "avalanches"- are distributed according to power laws (see Fig. 4 of \cite{hennig-adams-etal:09}, although it is not completely clear if this is a power law). They interpret this regime as an indication that "early spontaneous activity in the developing retina is regulated according to the following principle: maximize randomness and variability in the resulting activity patterns". A similar point of view is defended in \cite{lansdell-ford-etal:14}. The equivalent of a percolation transition in our model is at the transition between regime II and regime III. One switches from a \textit{competitive} regime where waves interact with the sAHP landscape, \textit{preventing} them to propagate, to a \textit{cooperative} regime where waves collides, with the effect of non linearly \textit{enhancing} their activity, as shown in Fig. \ref{Fig:AchProfile_NIB_2cells} and activity spreads through the whole lattice. Thus, in contrast to classical percolation, here cells interaction is active, potentially leading to a different "critical" behaviour (critical exponents) than in directed percolation or forest fires. Note also that the transition corresponds to a determined value of $g_A$ whereas $g_A$ is evolving during development, so that the "percolation" phase is a brief instant. Thus, regime II, extending on a whole interval (Griffith phase), might be more relevant for visual system development. 

\item \textbf{Self-organization.} Assume that their exist some notion of criticality in retinal waves how could the retina manage to maintain a critical state as $g_A$ is evolving ? Several biological mechanisms have been proposed in the literature to explain how the brain could achieve to maintain a critical state: synaptic adaptation \cite{levina-herrmann-etal:07,levina-herrmann-etal:09,bonachela-franciscis-etal:10,wang-zhou:12,di-santo-villegas-etal:18,kinouchi-brochini-etal:19} , dynamic neuronal gains \cite{brochini-costa-etal:16,kinouchi-brochini-etal:19}, adaptive firing thresholds \cite{girardi-schappo-brochini-etal:20}, topological self-organization \cite{bornholdt-rohlf:00} (see \cite{kinouchi-pazzini-etal:20} and references therein for a recent review), homeostasy \cite{menesse-marin-etal:21}. We first would like to remark that most of these models where made for cortical, spiking neurons. In contrast, most neurons in the retina, including SACs, do not spike \footnote{Neural spikes are produced by sodium and potassium fluxes, whereas it was shown by Zhou et al \cite{zheng-lee-etal:04} that sodium plays no role in SACs bursting.}. Also, there is no evidence of long term plasticity although short term plasticity has been reported \cite{kastner-ozuysal-etal:19} (dealing with non spiking neurons). It is therefore not clear to us which mechanism could apply. Note however that the most important aspect revealed by our paper, making dynamics so rich, is that region $D$ is a recurrent region close to bifurcations.  Dynamics naturally drives the cell back to this region where it is quite sensitive to perturbations thanks to basic mechanisms observed in experiments. In this way, the mechanism of "critical stability"  could be close to the scenario called "Mapping Self-Organized Criticality onto Criticality" introduced by D. Sornette  and co-workers in \cite{sornette-johansen-etal:97}. In the past, this line of thought has collected a broad consensus and resulted in numerous interesting results \cite{sornette-johansen-etal:97,levina-herrmann-etal:07,levina-herrmann-etal:09,di-santo-villegas-etal:18,rubinov-sporns-etal:11,shew-clawson-etal:15}. Yet, the recurrence to region $D$ is not sufficient to ensure criticality, requiring also long range correlations via Ach coupling. A thorough investigation of criticality in our model would reveal whether the dynamics we propose is sufficient or if we need extra mechanisms such as the one proposed in  \cite{kossio-goedeke-etal:18} where the feedback of the control parameter to the order parameter is carried out through the dynamics of the synaptic tree radius.
\end{enumerate}

\section{Discussion}\label{Sec:Discussion}

In this paper, following \cite{karvouniari-gil-etal:19}, we answered several questions with respect to our understanding of retinal waves. Which is the biophysical mechanism that generates sustained periodically bursting in immature SACs and which are the parameters that control it? Is there a potential, universal, mechanism of waves generation  yet accounting for the observed waves variability ? How do SACs synchronize in order to produce propagating waves? How do waves propagate ? How do they stop? How do the characteristics of waves depend on biophysical parameters  ? 

The scenario that we proposed relies on a detailed modeling of Starbust Amacrine Cells, based on the experimental literature and taking into account the role of calcium gated potassium channels, intracellular calcium dynamics, and cholinergic coupling between cells. The model is based on a variant of the Morris-Lecar model, with a fast potassium current, calcium dynamics, slow after hyperpolarization current and cholinergic interactions. 
There are $3$ times scales and the fast dynamics is characterized by a two dimensional model of Morris-Lecar type with additional parameters controlling the sAHP conductance ($G_S$) and the cholinergic conductance ($G_A$). Most parameters have been tuned from the experimental literature (see \cite{karvouniari:18} for detail).

A thorough bifurcation analysis of the fast dynamics, in the plane $\set{G_S,G_A}$ reveals the existence of $4$ regions, $A,B,C,D$. The most important in our analysis is the tiniest, region $D$. Indeed, as we show: (1) Dynamics returns to that region in a recurrent way; (2) in this region, the cell is bistable (type 1 excitability) and a small amount of noise allows it to switch from rest state to fast oscillations. This transition triggers waves generation and propagation.
In region $D$ the transition probability is quite sensitive to the actual value of $G_S$ and $G_A$ resulting, at the cell level, to a wide variability in the time of transition, and, at the network level, in a wide variability in waves duration and size. 
This mechanism is purely dynamical and biophysically grounded. Especially, the  power $R^4$ in the sAHP conductance, playing a central role here, is due to the mechanism of calcium-gating in slow potassium channels. 

All in all, what makes retinal waves dynamics of our model so rich is the closeness of bifurcations arising in the tight, recurrent zone $D$.
In this spirit, we would like to briefly comment on the work of Ford and Feller
\cite{ford-feller:12} where they insisted on the "\textit{considerable variability in the current underlying the slow AHP  and in
depolarization of individual SACs in subsequent waves [...] Variability in the proportion of nonrefractory SACs is necessary to produce waves of finite size. With a high rate of spontaneous depolarization, as in rabbit retina and previous models, this variability is
introduced by spontaneous depolarizations and subsequent hyperpolarizations between waves"}.  
This remark was motivated by experimental observations. 
They modelled it using Hennig et al model where, in addition, the characteristic times $\tau_R, \tau_S$ are random distribution, introducing this way a source of variability in the sAHP. 
As we have shown in this paper, there is no need to introduce this extra variability. Dynamics makes the job.\\


On the basis of numerical simulations and theoretical arguments we were able to exhibit $3$ regimes of wave propagation as $g_A$, the maximal Ach conductance, increases. 
When $g_A$ is smaller than a value $g_{A_0}$, analytically computed in one dimension, no wave propagation is possible. Then as $g_A$ increases one enters in regime I where one observes small waves whose propagation is ruled by noise fluctuations, corresponding to cells in region $D$. This results in an anomalous transport. In this regime, the probability $\rho$ that a cell is bursting at time $t$ increases sharply as well as waves characteristics (size, duration) and network activity ($n$). The separation between regime I and II holds at the unique point where $\frac{\partial \rho}{\partial g_A}$ is maximum, that is the point where the probability $\rho$ that a cell bursts at a given time has the maximum sensitivity to variations of $g_A$. The transition corresponds to a characteristic scale of wave nucleation equal to the SACs spacing.
As $g_A$ further increases, in regime II, waves are essentially propagating deterministically, with a decreasing influence of noise, corresponding to cells in-between region $D$ and $C$. Transport is ballistic there.
What mainly limits the wave size is the interaction with the sAHP landscape left by previous retinal waves. This is therefore a regime of competition between waves. The probability $\rho$ that a cell is bursting at time $t$ as well as waves characteristics are increasing essentially linearly with $g_A$, with a steep slope.
In the transition between regime II and regime III one switches from a regime of competition to a regime of cooperation (non linear feedback) prolonging burst durations and sAHP. At the transition, one typically have several cooperating waves with a global activity spreading through the whole lattice. Hence, the transition between regime II and III resembles a percolation transition, as formerly pointed out by several authors \cite{hennig-adams-etal:09,lansdell-ford-etal:14}. However, in contrast to classical percolation, non linear coupling cells enhances cooperative effects. 
In regime III, waves size and duration saturate and $\rho$ increases linearly, with a small slope, in the range of $g_A$ values we investigated. For larger $g_A$ we expect $\rho$ to saturate and to converge to the fraction of time a cell can be excited, dictated by sAHP parameters. 
We didn't explore this regime though as we were interested in ranges of $g_A$ values
compatible with retinal development. In this perspective, actually, $g_A$ is decreasing, presumably leading from regime III to regime I, which, according to our estimation of table \ref{Tab:AchEVolution} could coincide with the switch from cholinergic-SAC (stage II) with glutamatergic-AII (stage III) retinal waves.\\

In this perspective, transition between regimes such as the percolation-like transition from regime III to regime II is a brief instant in the evolution and one does not see why it should play an important role. Nevertheless, the question of critical behaviour in retinal waves is important. Since the seminal work of Beggs and Plen \cite{beggs-plenz:03} - reporting that neocortical activity in rat slices occur in the form of neural avalanches with power law distributions close to a critical branching  process - there have been numerous papers suggesting that the brain as a dynamical system fluctuates around a critical point. Scale-free neural avalanches have been found to occur in a wide range of neural tissues and species \cite{shew-plenz:13}. From a theoretical point of view, it has been suggested \cite{haldeman-beggs:05,shew-plenz:13} that such a scale-free organisation could favor information storage and transfer, improvement of the computational capabilities \cite{bertschinger-natschlager:04}, information transmission \cite{beggs-plenz:03,bertschinger-natschlager:04,shew-yang-etal:11}, sensitivity  to  stimuli  and  enlargement  of  dynamic  range  \cite{kinouchi-copelli:06,shew-yang-etal:09,gautam-hoang-etal:15,girardi-schappo-bortolotto-etal:16}. 

It has been proposed by Hennig et al \cite{hennig-adams-etal:09} that such a critical regime could also be helpful for the visual system developmental phase triggered by  
stage II retinal waves. In such a regime the retina would maximize randomness and variability in the resulting activity patterns. However, they do not propose a mechanism explaining how this regime can be robust to parameters variations during development, especially cholinergic coupling. As we discussed in the paper, such a structural stability could correspond to a Griffith phase (also observed in V1 \cite{girardi-schappo-bortolotto-etal:16}) instead of a classical critical regime. This interesting possibility would however deserve extensive simulations in 2D, with a realistic connectivity, and a thorough analysis data to cleanly identify a critical signature. There is a wide literature dealing with the precise characterization of a critical state, where scaling exponents obey relations fixed by renormalization group theory \cite{ma:76,hohenberg-halperin:77,kadanoff-nagel-etal:89,sethna:95}. In the field of neuroscience 
this has been widely discussed by authors such as Touboul and Destexhe
\cite{touboul-destexhe:10,touboul-destexhe:17} and many authors have looked for solid evidences of criticality \cite{girardi-schappo-kinouchi-etal:13,moretti-munoz:13,girardi-schappo-bortolotto-etal:16}. In particular, it is believed that the so-called crackling noise relationship (from Sethna's terminology \cite{sethna:95}) can be used to discriminate between true critical systems and "fake" ones.
Recently, Kanders et al \cite{kanders-hyungsub-etal:20} have shown that developing neural cultures in specific conditions satisfy the crackling noise relationship. The same type of investigations could be done with our model.\\

We would like now to address a few further open questions related to this work.

\subsection{Spontaneous Bursting}\label{Sec:SB}

It is possible to move regions $C,D$ downward or upward by a slight change in the leak potential $V_L$. Experimentally, is it possible to modify $V_L$ by changing the outside potassium concentration. Note, however, that the change in a reversal potential is logarithmic in the concentration. Nevertheless, in our case, a change of $V_L$ from $-72$ mV, the value used in the paper, to $V_L=-70$ mV has the effect to move region $D$ to negative values for  $G_A$. This means that, in this case, when the cell returns back from its sAHP excursion it eventually penetrates in region $C$, instead of region $D$. Then, it starts to oscillate again (stable limit cycle in $C$).
Therefore, in this situation, no noise is required to restart bursting. It occurs spontaneously. 

\begin{figure}
\centerline{
\resizebox{0.5\textwidth}{0.2\textheight}{
\includegraphics{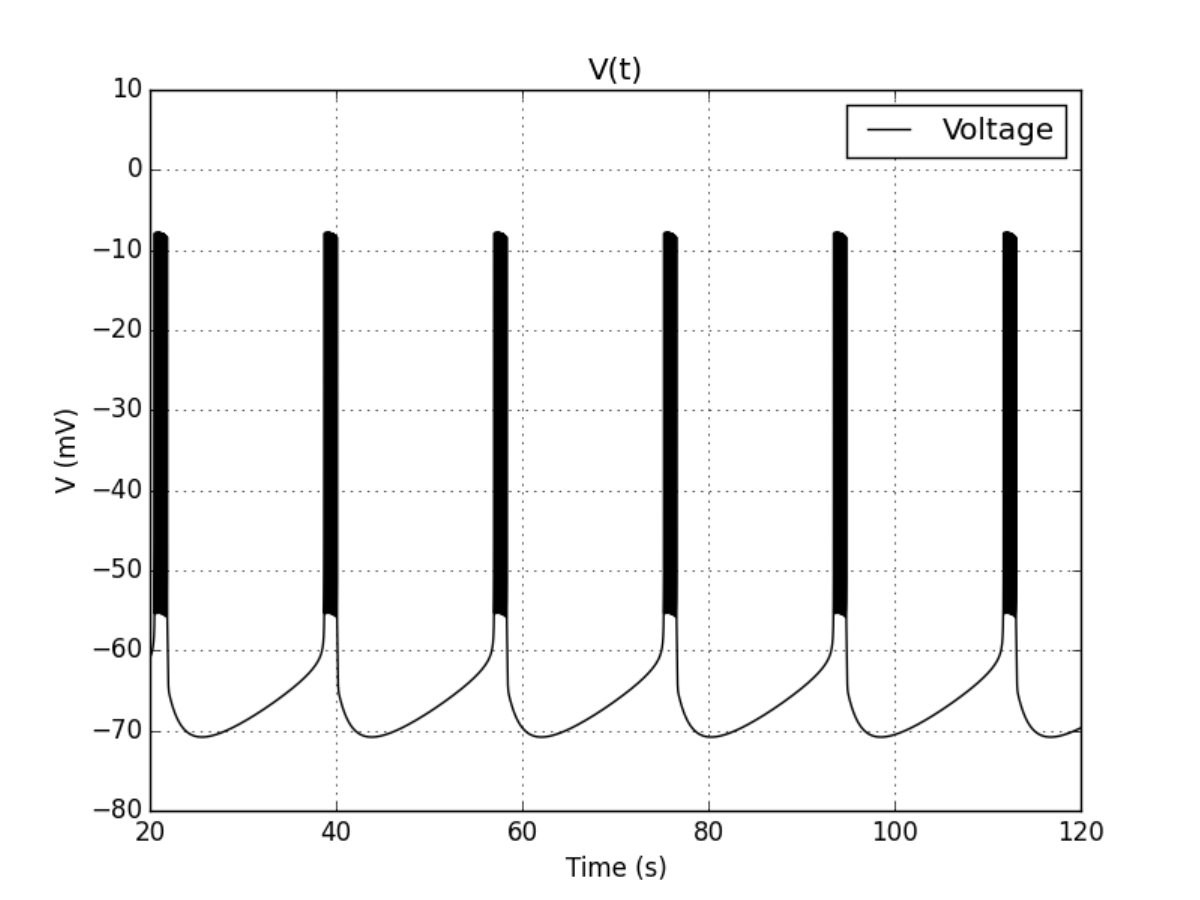}
}
}
 \caption{\textbf{Spontaneous bursting (SB) regime}. Voltage Rhythmic bursting for one isolated SAC as a function of time. The interburst depends on $\tau_S,\tau_R$ \label{Fig:SB}.}
\end{figure}

In this scenario, cells burst therefore periodically, with a frequency controlled by $\tau_R, \tau_S$. An example is shown in Fig. \ref{Fig:SB}. Here, cells have a tendency to synchronise. However, introducing a bit of randomness in initial conditions one can obtain, depending on the maximal conductances $g_S,g_A$, a regime with spatio-temporal disorder where waves interacts and leads to random regions similar to the Noise Induced Bursting regime \cite{karvouniari-gil-etal:16}.

The cholinergic interactions of SACs in a quenched disordered sAHP landscape breaks down the spontaneous bursting periodicity and leads to a spatio-temporal chaotic dynamics quite similar to the noise induced bursting for waves dynamics \cite{karvouniari-gil-etal:16}. We didn't address in detail the spontaneous bursting regime in this paper though
leaving it for further studies.

\subsection{Understanding the functional role of retinal waves}\label{Sec:FuncRole}

It would be interesting to explore, at a modelling level, what could be the functional impact of the different waves regimes we observe in the development of the retino-thalamico-cortical pathways, elucidating why early neural networks would choose to maintain their activity in such regimes. As we have seen in section \ref{Sec:AchDecays} experimental evidences show that $g_A$ decays during development so that, according to our results, the retina ought to go successively through regimes III (extended waves), II (wide distribution of waves and strong interaction with the sAHP landscape), and I (localized waves) although this step in evolution coincide with the arousal of stage III waves.  More particularly, we reported the apparition of sAHP patterns that spatially bound the propagating waves. Those localized bounding patches are found to persist more than a sAHP timescale, indicating that in the network scale, they introduce a type of spatial memory of previous activity. In other words, they create an heterogeneous landscape where waves could propagate, inducing an effect, first on waves characteristics, and second on the selective synaptic shaping of the network through this spatial bias \footnote{This spatial heterogeneity might also have an impact on blood vessels organisation with, thus, a very long term influence.}  For example, late waves (stage III) appear to be spatially bounded and more localised than stage II waves \cite{maccione-hennig-etal:14}, suggesting that probably late stage II waves could prepare the landscape for the transition to more spatially confined activity. Hence, those localized activity patterns, could potentially have a link with how receptive fields are formed before vision becomes functional.

A next step would therefore be to link the retinal waves model presented here to a cortical V1 model such as the one developed in \cite{souihel-chavane-etal:19} so as to investigate the role of synaptic shaping under retinal wave stimulation to drive the cortical response during development.

This work has also possible future outcomes with respect to retina therapy as well. Understanding how retinal waves are initiated and propagate in the retina could enable one to define protocols to trigger such retinal waves in the in vivo adult retina. Inducing such waves is expected to reintroduce some plasticity in the retinal tissue and the projections in the brain. This induced plasticity could have important therapeutic applications to treat patients or stimulate regeneration of retinal ganglion cell axons following optic nerve crush.


\subsection{Acknowledgements}
We thank the reviewers for insightful remarks and constructive criticism.
D.M.K. was supported by a doctoral fellowship from Ecole Doctorale des  Sciences et Technologies de l'Information et de la Communication de Nice-Sophia-Antipolis (EDSTIC) and by the RENVISION European Union Project No. 600847.  This work also benefited from the support of the Neuromod institute of the University C\^ote d'Azur and the Doeblin federation. We warmly acknowledge Lionel Gil, Matthias Hennig,  Evgenia Kartsaki, Evelyne Sernagor, Olivier Marre and Serge Picaud for their help. 

\vfill\eject
\clearpage

\section*{Parameters Value}

\label{AppendixParameter}
\begin{table}[!h]
\begin{center}
    \begin{tabular}{ | p{2cm} | p{2.6cm} | }       			   \hline
    Parameter 	& 	Physical value 					\\ \hline
    $C_m$		& 	$22\, pF$  					\\ \hline
    $g_L$ 		& 	$2\, nS$ 	 					\\ \hline
    $g_{C}$	& 	$12 \, nS$						\\ \hline
    $g_{K}$		& 	$10 \, nS$						\\ \hline   
    $g_{S}$	& 	$\left[2,12\right] \, nS$ 			\\ \hline    
    $V_L$		& 	$-72 \, mV$					\\ \hline
    $V_{C}$	& 	$50 \, mV$ 					\\ \hline
    $V_K$		& 	$-90  \, mV$   					\\ \hline
    $V_1$		& 	$-20  \, mV$  					\\ \hline
    $V_2$		& 	$20 \, mV$ 					\\ \hline
    $V_3$		& 	$-25 \, mV$					\\ \hline
    $V_4$		& 	$7 \, mV$ 						\\ \hline
    $\tau_{N}$	&	$5 \, ms$    					\\ \hline
    $\tau_{R}$	&	$8300 \, ms$ 					\\ \hline
    $\tau_{S}$	&	$8300 \, ms$  					\\ \hline
    $\tau_{C}$	&	$2000 \, ms$  					\\ \hline
    $\delta_{C}$	&	$10.503 $ $nM \,\, pA^{-1}$ 		\\ \hline
    $\alpha_{S}$	&	$\frac{1}{200^4}$ $nM^{-4}$  		\\ \hline
    $\alpha_{C}$&	$4865$\, nM  					\\ \hline
    $\alpha_{R}$&	$4.25$  						\\ \hline
    $H_X$		&	$1800$ $nM$					\\ \hline
    $C^{(0)}$		&	$88$ $nM$					 \\ \hline
    $\eta$	    &	$6 \,pA ms^{1/2}$	     \\ \hline
    \end{tabular}
\end{center}
\caption{Typical parameters values used in the model.}
\label{TabParameters}
\end{table}

\begin{table}[!h]
\begin{center}
    \begin{tabular}{ | p{2cm} | p{2.6cm} | }       			   \hline
    Parameter 		& 	Physical value 					\\ \hline
    $\mu_A$			& 	$ 1.86 \,s^{-1}$  					\\ \hline
    $K_{A}$		&	$200 \, V^{-1}$					\\ \hline
    $V_{A}$		& 	$0 \, mV$						\\ \hline
    $V_{0}$			& 	$-40 \, mV$					\\ \hline
    $\gamma_{A}$ 	& 	$1 \,nM^2$					\\ \hline
    $\beta_{A}$ 	& 	$5 \,nM \,s^{-1}$				\\ \hline
    \end{tabular}
\end{center}
\caption{Typical parameters values related to the acetylcholine neurotransmitter (Ach).}
\label{TabParametersNext}
\end{table}

\section*{References}
\bibliographystyle{abbrv,unsrt}
\bibliography{biblio,odyssee}

\end{document}